\pdfoutput=1
\documentclass[superscriptaddress,amsmath,amssymb,twocolumn]{revtex4-1}
\usepackage[utf8]{inputenc}
\usepackage{graphicx}
\usepackage[colorlinks,allcolors=blue]{hyperref}
\usepackage{soul}
\usepackage{tabularx}
\usepackage[figuresleft]{rotating}

\usepackage{lipsum}

\newcommand{\avg}[1]{\left\langle #1 \right\rangle}

\begin{document}

\title{Relationship between fitness and heterogeneity  in exponentially growing microbial populations}




\author{Anna Paola Muntoni}
\affiliation{Politecnico di Torino, Corso Duca degli Abruzzi, 24, I-10129, Torino, Italy}
\affiliation{Italian Institute for Genomic Medicine, IRCCS Candiolo, SP-142, I-10060 Candiolo (TO), Italy}

\author{Alfredo Braunstein}
\affiliation{Politecnico di Torino, Corso Duca degli Abruzzi, 24, I-10129, Torino, Italy}
\affiliation{Italian Institute for Genomic Medicine, IRCCS Candiolo, SP-142, I-10060 Candiolo (TO), Italy}
\affiliation{INFN, Sezione di Torino, Torino, Italy}

\author{Andrea Pagnani}
\affiliation{Politecnico di Torino, Corso Duca degli Abruzzi, 24, I-10129, Torino, Italy}
\affiliation{Italian Institute for Genomic Medicine, IRCCS Candiolo, SP-142, I-10060 Candiolo (TO), Italy}
\affiliation{INFN, Sezione di Torino, Torino, Italy}

\author{Daniele De Martino}
\affiliation{Biofisika Institute (CSIC,UPV-EHU) and Ikerbasque Basque Foundation for Science, Bilbao 48013, Spain}
\email{daniele.demartino@ehu.eus}

\author{Andrea De Martino}
\affiliation{Politecnico di Torino, Corso Duca degli Abruzzi, 24, I-10129, Torino, Italy}
\affiliation{Italian Institute for Genomic Medicine, IRCCS Candiolo, SP-142, I-10060 Candiolo (TO), Italy}
\affiliation{Istituto di Nanotecnologia (CNR-NANOTEC), Consiglio Nazionale delle Ricerche, I-00185 Roma, Italy}
\email{andrea.demartino@polito.it}


\begin{abstract}
Despite major environmental and genetic differences, microbial metabolic networks are known to generate consistent physiological outcomes across vastly different organisms. This remarkable robustness suggests that, at least in bacteria, metabolic activity may be guided by universal principles. The constrained optimization of evolutionarily-motivated objective functions like the growth rate has emerged as the key theoretical assumption for the study of bacterial metabolism. While conceptually and practically useful in many situations, the idea that certain functions are optimized is hard to validate in data. Moreover, it is not always clear how optimality can be reconciled with the high degree of single-cell variability observed in experiments within microbial populations. To shed light on these issues, we develop an inverse modeling framework that connects the fitness of a population of cells (represented by the mean single-cell growth rate) to the underlying metabolic variability through the Maximum-Entropy inference of the distribution of metabolic phenotypes from data. While no clear objective function emerges, we find that, as the medium gets richer, the fitness and inferred variability for {\it Escherichia coli} populations follow and slowly approach the theoretically optimal bound defined by minimal reduction of variability at given fitness. These results suggest that bacterial metabolism may be crucially shaped by a population-level trade-off between growth and heterogeneity.

~

Evolutionary reasoning suggests that growth rate maximization may be the key organization principle of microbial metabolism. While appealing, optimality is hard to validate by directly extracting objective functions from data. Using a Maximum Entropy framework to infer metabolic phenotypes from population-level experiments, we show here that, as growth conditions improve, {\it Escherichia coli} cells approach a theoretical limit that connects their average growth rate to cell-to-cell variability in metabolic phenotypes. Specifically, as the former increases, the reduction in variability stays close to a minimum. This suggests that the organization of microbial metabolism (and some of the trade-offs that characterize it) may result from the need to preserve large metabolic heterogeneity in any growth condition.
\end{abstract}

\maketitle

\section*{Introduction}

A standard assumption of theoretical models of microbial metabolism is that cells regulate the fluxes through metabolic reactions so as to maximize their growth rate (i.e. their biomass output) \cite{lewis12,feist10,dourado20,bruggeman20}. While intuitive and highly successful in many applications \cite{bordbar14}, this idea is not easy to validate in exponentially growing microbial populations. Quantitative studies of the interplay between metabolism and gene expression suggest for instance that microbial fitness is strongly gauged by regulatory constraints \cite{scott10,hui15}, biosynthetic costs \cite{flamholz13,basan15} and the ability to respond to changing environments \cite{utrilla16,mori17,towbin17,erickson17, basan20}. As a consequence, the trade-offs arising from a complex multi-objective optimization often give a more accurate description of microbial growth than straightforward biomass maximization \cite{schuetz12,shoval12,mori19}. Moreover, experiments characterizing bacterial growth at single-cell resolution have shown tight links between fitness and cell-to-cell variability  \cite{kiviet14,taheri15,kennard16}. Growth rate distributions and metabolic fluxes indeed appear to be best captured by modeling such variability rather than assuming growth rate maximization \cite{demartino16,demartino18}, with the implication that trade-offs between metabolism and gene expression may affect not only bulk (average) properties but also the overall structure of a microbial population. While genome-scale models of metabolic networks can account for some of these facts \cite{obrien13,goelzer15,mori16,reimers17}, the maximization of biomass output remains a key conceptual premise. 

Addressing the question of ``what cells actually want" \cite{feist16} requires in essence to reverse the usual theoretical pipeline and infer from empirical data (reaction fluxes, growth rates, nutrient intake rates, etc.) (a) how the flow of metabolites through the metabolic network is organized, and (b) whether some objective function is optimized. In this work we develop a framework to learn the probability distribution of metabolic phenotypes (namely, whole-network flux configurations) using mass-spectrometry data \cite{dai17} to inform a constraint-based model of {\it E. coli}'s metabolism. This approach differs significantly from previous inference studies, specifically those of Refs \cite{demartino16,demartino18}, where the probability to observe a certain phenotype was effectively assumed to be a Boltzmann-like exponential function of its growth rate. No such assumption is made here. Rather, for each experimental sample, we compute the most likely distribution of phenotypes compatible with flux data resorting to the Maximum Entropy (MaxEnt) principle. In a nutshell, this approach prescribes that, if one has to infer a probability distribution subject to a given set of constraints, the distribution having the largest entropy provides the best guess, in the sense of being closest to uniform (i.e. minimizing its divergence from the uniform distribution), thus avoiding the introduction of biases that are not needed to accommodate constraints (see \cite{demartino18h} for a simple introduction to this idea). Each inferred distribution is then characterized via (i) its mean biomass output (a proxy for the fitness), and (ii) its `information content', a global measure of cell-to-cell variability introduced in \cite{demartino16} that in essence quantifies the ``volume'' of the space of allowed phenotypes over which the inferred distribution is spread, with high information content corresponding to small ``volume''. Fitness and information values inferred at different glucose levels appear to draw a well-defined curve in the fitness-information plane, supporting the idea of a tight link between growth and (inferred) variability. As a benchmark, we compare this curve against a purely theoretical bound obtained by maximizing, in each condition, the mean biomass output at fixed information content (similar to a `rate-distortion curve' in information theory  \cite{mackay03}). We found that empirical populations qualitatively follow and slowly approach the theoretical limit as the growth medium gets richer. In other words, as the fitness (mean biomass output) increases, the inferred phenotypic variability tends to remain as large as possible. This quantitatively supports the idea that heterogeneity plays a key role in shaping the fitness of a microbial population.

\section*{Materials and Methods}

\subsection*{Constraint-based model of the metabolic network \label{sub:feasible}} 

Given a network reconstruction defined by the $M\times N$ matrix $\mathbf{S}$ of the stoichiometric coefficients of metabolic reactions (with $N$ the number of reactions and $M$ that of chemical species, including exchange fluxes between the cell and the medium), feasible flux vectors $\mathbf{v}=\{v_i\}$ are assumed to satisfy the non-equilibrium steady-state (NESS) mass-balance conditions  $\mathbf{Sv=0}$ \cite{bordbar14} (Fig. \ref{uno}a). Once ranges of variability of the form $v_i\in[v_{i,\min},v_{i,\max}]$ are supplied for each flux, solutions span a convex polytope of dimension at least equal to $N-\mathrm{rank}(\mathbf{S})$ \cite{shrijver98}. This represents the `feasible space' $\mathcal{F}$ of the metabolic network. In principle, all vectors $\mathbf{v}\in\mathcal{F}$ are viable network states (phenotypes). Flux Balance Analysis (FBA) and related approaches typically focus on the optimal phenotype, defined as the flux vector that maximizes a specific objective, usually the ``biomass synthesis rate'' $v_{\rm bm}$ included in the network reconstructions \cite{feist10}, which quantifies the rate at which biomass precursors are produced, in the correct proportions, in state $\mathbf{v}$. It follows that the vector $\mathbf{v}\in\mathcal{F}$ that maximizes $v_{\rm bm}$ can be found by Linear Programming. We shall hereafter write $v_{\rm bm}^{\rm max}\equiv\max_{\mathbf{v}\in\mathcal{F}} v_{\rm bm}$. 
Here we aim at using experimental data on fluxes to infer a probability density on $\mathcal{F}$ that most efficiently represents our empirical knowledge. 

Since the biomass synthetic rate models the growth rate according to the metabolic model, the variable $v_{\rm bm}$ has the dimension of a rate, i.e. $h^{-1}$. In the following we will make use of a heavy notation to denote the biomass synthetic rate under different conditions and approximations. For sake of clarity, we summarize in Supplemental Information Sec. J the notation associated with this quantity.

\subsection*{Experimental data and network reconstruction}

We have considered the 17 experiments described in \cite{nanchen06} and the 16 from \cite{schuetz12}, which study glucose-limited {\it E. coli} growth and employ the same network reconstruction for flux estimation. Taken together, data cover $33$ values of the growth rate, from ca. $0.05$/h to $1$/h (a range that includes the key phenotypic crossover marked by the onset of acetate overflow \cite{basan15,wolfe05}) and provide, for each condition, estimates for the population-averaged fluxes through a small set of reactions from the central carbon pathways \cite{buescher15}. The former dataset yields the expectation values of 26 fluxes at various growth rates and glucose intakes below the acetate onset point. The latter quantifies instead 25 fluxes in a nutrient-rich medium, with acetate excretion observed in 11 experiments. As fluxes were measured relative to the glucose uptake in \cite{schuetz12}, we converted them to net fluxes using the glucose uptake values reported in Table S5 of \cite{schuetz12}. The data we used finally comprised $33$ vectors $\mathbf{v}^{\mathrm{xp}}$ of average fluxes and the vector $\boldsymbol{\sigma}^{\mathrm{xp}}$ of their experimental errors. To define the feasible space $\mathcal{F}$, we derived the stoichiometric matrix relative to the network reconstruction given in Table S1 in \cite{schuetz12}, after mapping these fluxes to those measured in \cite{nanchen06} via their chemical equations (Supplemental Information, Sec. A). To define reaction reversibility, we assigned bounds $[-1000,0]$ or $[0,1000]$ to irreversible reactions, and $[-1000,1000]$ to reversible ones. We then implemented two modifications. First, we turned reaction \textit{udhA} from positive irreversible to reversible in order to allow for negative values (measured). Next, using the algorithm given in \cite{demartino13}, we found that reaction \textit{sdhABCD} was responsible for a thermodynamically infeasible loop in the network. To prevent it, we changed it from reversible to positive irreversible, so that all flux configurations we consider are thermodynamically consistent. We finally performed Flux Variability Analysis \cite{gudmundsson10} to restrict the range of variability of each flux. To define the growth medium, the glucose import flux was set to the value reported in each experiment. The corresponding flux can then be encoded in an experiment-dependent vector $\mathbf{b}$, so that the NESS conditions take the form $\mathbf{S}\mathbf{v} = \mathbf{b}$. The network we consider is ultimately composed of $N = 73$ fluxes and $M = 49$ metabolites, while the feasible space $\mathcal{F}$ is a convex polytope of dimension 26, i.e. with 26 independent degrees of freedom. Fig. \ref{uno}b shows, for all experiments, the measured biomass rate $v_{\rm bm}^{\rm xp}$ (markers) together with the maximum value of the biomass  $v_{\rm bm}^{\rm max}$ predicted by FBA for the network just described (black line).

\subsection*{Maximum Entropy distribution}

To infer the probability density of flux configurations ($p(\mathbf{v})$), we use empirical data to constrain the space of probability densities on $\mathcal{F}$. Specifically, we would like to impose that the mean flux $\avg{v_j}=\int_{\mathcal{F}} v_j\,p(\mathbf{v})\,d\mathbf{v}$ of every reaction $j$ that has been experimentally quantified matches its experimental estimate $v_j^{\mathrm{xp}}$. We denote by $\mathcal{E}$ the set of experimentally measured fluxes. According to the Maximum Entropy principle, the least biased guess for  $p(\mathbf{v})$ compatible with these constraints is given by the solution of \begin{equation}\label{const}
\max_{p(\mathbf{v})}\,H[p]~~~\text{subject  to}~\avg{v_j}=v_j^{\mathrm{xp}}\quad\text{$\forall j\in \cal{E}$}\quad,
\end{equation}
where $H[p]=-\int_{\mathcal{F}}p(\mathbf{v})\log_2 p(\mathbf{v})\,d\mathbf{v}$ is the entropy and $\cal{E}$ stands for the set of measured fluxes. This yields 
\begin{equation}\label{dist}
p(\mathbf{v};\mathbf{c})=\frac{1}{Z(\mathbf{c})}\, \exp\left[\sum_{j\in\cal{E}}c_j v_j\right]~~,
\end{equation}
where $\mathbf{c}=\{c_j\}$ is the vector of Lagrange multipliers (``fields'' for short) enforcing the constraints (\ref{const}) and $Z(\mathbf{c})$ is a factor ensuring proper normalization ($\int_{\mathcal{F}} p(\mathbf{v};\mathbf{c})\,d\mathbf{v}=1$). The values of the fields $c_j$ must be determined from the conditions
\begin{equation}\label{constr}
\avg{v_j}\equiv\int_{\mathcal{F}} v_j \, p(\mathbf{v};\mathbf{c}) \, d\mathbf{v}=v_j^{\mathrm{xp}}\qquad\text{$\forall j\in \cal{E}$}\quad.
\end{equation} 
To solve Eqs (\ref{constr}) in a general setting one can neither rely on the Monte Carlo schemes employed in \cite{demartino16} nor to Boltzmann learning (Supplemental Information, Sec. F) due to exceeding computational costs.
Furthermore, we observed that, due to experimental uncertainties, some empirical means $v_j^{{\rm xp}}$ given in \cite{nanchen06,schuetz12} lie outside the feasible polytope $\mathcal{F}$ defined by the network reconstruction employed in those studies. In other words, there is no configuration $\mathbf{v} \in \mathcal{F}$ satisfying $v_{j} = v_{j}^{\rm xp}$ exactly, therefore the above-mentioned approach cannot directly be applied.
To account for the last issue, we introduce in the next section a slightly modified MaxEnt model, whose parameters are determined by Expectation Propagation (EP), a highly efficient algorithm for approximate Bayesian inference with broad applicability \cite{opper00,minka01, braunstein17}.

\subsection*{Computation of the MaxEnt distribution via Expectation Propagation}

We define for each measured flux $j\in\mathcal{E}$, the auxiliary variable $v_j^{e}=v_j^{{\rm xp}}+\eta_j$, where $\eta_j$ is a Gaussian random variable with zero mean and standard deviation $\gamma^{-1/2}$ (no cross-correlations between different fluxes are assumed). We then aimed at determining the distribution of phenotypes $\mathbf{v}$ such that (i) $\mathbf{v}$ lies in $\mathcal{F}$, (ii) fluxes in $\mathcal{E}$ take on values as close as possible (within  experimental errors) to those encoded in the auxiliary vector $\mathbf{v}^{e}$, and (iii) averages of $\mathbf{v}^{e}$ match empirical averages $\mathbf{v}^{\mathrm{xp}}$. The (MaxEnt) distribution satisfying these constraints reads 
\begin{equation}
\label{eq:jointdist}
p\left(\mathbf{v},\mathbf{v}^{e};\mathbf{c}\right)= \frac{1}{Z_{e}(\mathbf{c})}\prod_{j\in\mathcal{E}}e^{-\frac{\gamma}{2}\left(v_{j}^{e}-v_j\right)^{2}+c_j v_{j}^{e}}~~,
\end{equation}
where $Z_{e}(\mathbf{c}) = \int_{\mathbb{R}^{n}} d\mathbf{v}^{e} \int_{\mathcal{F}} d\mathbf{v}\, p\left(\mathbf{v},\mathbf{v}^{e};\mathbf{c}\right) $ is the normalization constant, and $\mathbf{c}$ is the vector of Lagrange multipliers ensuring that 
\begin{equation}
\label{eq:constr_means}
\left\langle v_{j}^{e}\right\rangle _{p\left(\mathbf{v},\mathbf{v}^{e};\mathbf{c}\right)}=v_{j}^{\mathrm{xp}}\qquad\mathrm{for\,each}\,\,j\in\mathcal{E}~~.
\end{equation}
In the limit of large $\gamma$ the distribution will concentrate on the point (or face) that is ``closest'' to the empirical vector $\mathbf{v}^{\mathrm{xp}}$. To set a specific value for $\gamma$, however, we imposed that the average variances of auxiliary fluxes $\mathbf{v}^{e}$ match the average variance of experimental errors, i.e.
\begin{equation}
\label{eq:constr_av_noise}
\frac{1}{|\mathcal{E}|}\sum_{j\in\mathcal{E}}\avg{\left(v_j^{e}-v_j^{\mathrm{xp}}\right)^{2}}_{p\left(\mathbf{v},\mathbf{v}^{e};\mathbf{c}\right)}=\frac{1}{|\mathcal{E}|}\sum_{j\in\mathcal{E}}(\sigma_j^{\mathrm{xp}})^{2}
\end{equation}
(see Supplemental Information, Sec. B for details). Note that (\ref{eq:jointdist}) is tightly related to (\ref{dist}), since
\begin{eqnarray}
\label{eq:P-marginaliz}
\int_{\mathbb{R}^n} p\left(\mathbf{v},\mathbf{v}^{e};\mathbf{c}\right) d\mathbf{v}^{e}\equiv p\left(\mathbf{v};\mathbf{c}\right) ~~.
\end{eqnarray}
To avoid the high computation cost of calculating the vector $\mathbf{c}$ and the scalar $\gamma$ through Boltzmann learning, 
we propose to employ a variant of Expectation Propagation (EP) \cite{opper00,minka01,braunstein17}, an algorithm computing a multivariate Gaussian approximation $q\left(\mathbf{v},\mathbf{v}^{e};\mathbf{c}\right)$ of $p\left(\mathbf{v},\mathbf{v}^{e};\mathbf{c}\right)$. Note that thanks to the relation in (\ref{eq:P-marginaliz}) and the Gaussian approximation provided by EP, we can characterize the fluxes $\mathbf{v} \in \mathcal{F}$ by means of a marginal density $q\left(\mathbf{v}; \mathbf{c}\right)$, fully parametrized by a vector of means $\boldsymbol{\mu}$ and a covariance matrix $\boldsymbol{\Sigma}$. Details are given in Supplemental Information, Sec. B. After applying EP to each experiment $a$ to compute the distribution $q\left(\mathbf{v},\mathbf{v}^{e};\mathbf{c}_{a}\right)$ approximating $p\left(\mathbf{v},\mathbf{v}^{e};\mathbf{c}_{a}\right)$, expectation values of constrained fluxes coincide within error bars with empirical means as well as with results obtained by (potentially more accurate but less efficient) Monte Carlo Hit-and-Run calculations \cite{ell} (Supplemental Information Secs. F and G, Supplementary Figs. S3, S4 and S5). The script performing EP to compute the joint distribution (\ref{eq:jointdist}) along with its single-flux marginals using constraints derived from the datasets considered in this paper is available at  \url{https://github.com/infernet-h2020/MetaME}.

\subsection*{Inferred fitness and information content}

To characterize the inferred distributions (\ref{eq:jointdist}) (one per experimental population), we use two quantities (Fig. \ref{uno}c): (i) the mean inferred biomass output 
\begin{gather}\label{otto}
v_{\rm bm}^{\rm inf} = \avg{v_{\rm bm}}_{q\left(\mathbf{v};\mathbf{c}\right)}=\int_{\mathcal{F}}v_{\rm bm}\,q(\mathbf{v;c})d\mathbf{v}~~,
\end{gather}
and (ii) the `information content' per degree of freedom defined as
\begin{gather}\label{nove}
I^{\rm inf}=\frac{H[p(\mathbf{v;0})]-H[p(\mathbf{v;c})]}{\mathrm{dim}(\mathcal{F})\,\ln 2}~~~~~{\rm [bits]}~~.
\end{gather}
$v_{\rm bm}^{\rm inf}$ is a proxy for the population growth rate (fitness) when cell-to-cell variability is sufficiently small \cite{taheri15}.  $I^{\rm inf}$, namely the entropy loss relative to the uniform density on $\mathcal{F}$, quantifies instead how ``spread'' over $\mathcal{F}$ is $p(\mathbf{v;c})$. Low entropy or low phenotypic variability implies high information content and vice-versa. We calculated the information content per degree of freedom from the Gaussian approximation to (\ref{eq:jointdist}), i.e.
\begin{gather}
H_{a}\left[q\right]=\frac{1}{2}\log\det (2\pi e\boldsymbol{\Sigma}_{a})~~,\\
I_{a}^{\rm inf}=\frac{H_{\mathbf{0}}-H_{a}}{\mathrm{dim}(\mathcal{F})\, \ln 2}~~,
\end{gather}
where $H_{\mathbf{0}}$ is the entropy of the multivariate Gaussian distribution approximating the uniform distribution on $\mathcal{F}$.

\subsection*{Fitness-Information bound}

To benchmark inferred distributions, we reasoned that a large population of cells can be described by a probability density $p(\mathbf{v})$ over $\mathcal{F}$ as long as the feasible space can be taken to be the same for all cells in the population (i.e. if all cells obey the same set of physico-chemical, regulatory and environmental constraints). Clearly, the same fitness (mean biomass output $\avg{v_{\rm bm}}=\int_{\mathcal{F}} v_{\rm bm}\,p(\mathbf{v})d\mathbf{v}$) can be achieved by different probability densities $p(\mathbf{v})$. Among equally fit populations, however, those with the largest entropy encode less information into $p(\mathbf{v})$ and are hence likely to face smaller costs associated with the regulation of fluxes. It is therefore reasonable to define the optimal population for any given information content as the one solving
\begin{equation}
\max_{p(\mathbf{v})}\,\avg{v_{\rm bm}}~~~\text{subject to}~I[p]=\bar{I}~~.
\end{equation}
Equivalently, the optimal population is the one achieving a given fitness at the smallest reduction of flux variability, i.e. the solution of
\begin{equation}\label{maxent}
\min_{p(\mathbf{v})}\,I[p]~~~\text{subject to}~\avg{v_{\rm bm}}=\bar{v}_{\rm bm}~~.
\end{equation}
The solution of (\ref{maxent}) has the form
\begin{equation}\label{star}
p^\star(\mathbf{v})=\frac{1}{Z(\beta)}\,e^{\beta v_{\rm bm}}~~,
\end{equation}
where $\beta$ is the Lagrange multiplier enforcing the constraint $\avg{v_{\rm bm}}=\bar{v}_{\rm bm}$ and $Z(\beta)$ is a normalization constant. The distribution (\ref{star}) depends on the single parameter $\beta$. As $\beta$ increases, $\bar{v}_{\rm bm}$ increases and the values of $\avg{v_{\rm bm}}$ and $I$ associated with (\ref{star}) will change, drawing a curve in the $(I,\avg{v_{\rm bm}})$ plane that only depends on the specifics of $\mathcal{F}$. We call this curve the {\it Fitness-Information (F-I) bound}. By definition, optimal populations (fastest-growing if $\beta>0$, slowest-growing if $\beta<0$) have fitness and information values that lie on this curve. 

\begin{figure*}[t!]
\begin{center}
\includegraphics[width=17cm]{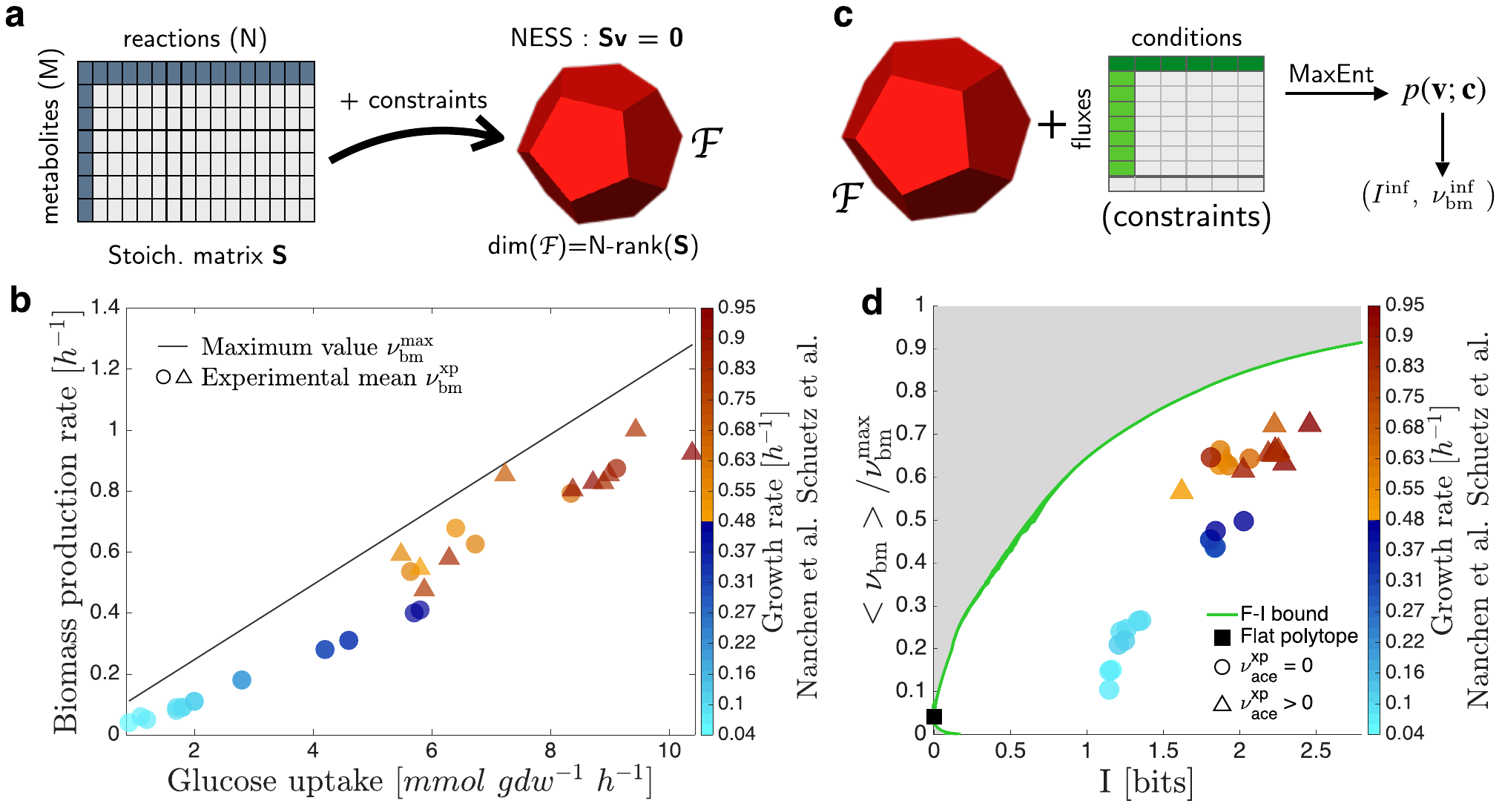}
\end{center}
\caption{\label{uno}(a) Constraint-based models use large-scale reconstructions of cellular metabolic networks, encoded in a stoichiometric matrix $\mathbf{S}$,  together with biochemical or regulatory constraints on fluxes. Non-equilibrium steady states (NESS) of the network satisfy the conditions $\mathbf{Sv} = \mathbf{0}$. The (high-dimensional, as $N\gg M$) polytope of solutions is the feasible space $\mathcal{F}$ of the system. The biomass output associated with each feasible flux vector (phenotype) represents its fitness. (b) Empirical mean biomass rate $v_{\rm bm}^{\rm xp}$ (markers, from \cite{nanchen06,schuetz12}) and maximum biomass output $v_{\rm bm}^{\rm max}$ (continuous line) predicted by Flux Balance Analysis for the metabolic network and glucose uptakes from \cite{nanchen06,schuetz12}. (c) Experiment-derived averages for a small set of fluxes and the bounds defined by the feasible space can be used to inform a Maximum Entropy inference procedure that allows us to determine the most likely distribution of phenotypes for the entire network. The distribution inferred for each experiment yields a point in the (fitness, information) plane. (d) Inferred fitness-information values for a set of 33 experiments probing {\it E. coli} growth in glucose-minimal media. Green-line: fitness-information (F-I) bounds numerically computed by EP for the 33 experiments (all curves perfectly overlap). In each condition, fitness values $\avg{v}_{\rm bm} = v^{\rm optm}_{\rm bm}$ have been re-scaled by the corresponding value of $v^{\max}_{\rm bm}$. Markers: fitness is computed according to the inferred distribution, i.e. $\avg{v}_{\rm bm} = v^{\rm inf}_{\rm bm}$. The black marker denotes the re-scaled fitness  $\avg{v_{\rm bm}}_{q\left(\mathbf{v}; \beta = 0\right)}/v^{\max}_{\rm bm}$ corresponding to a uniform distribution on $\mathcal{F}$ ($I=0$, $\beta=0$), which separates the upper branch of the F-I bound (growth faster than $\avg{v_{\rm bm}}_{q\left(\mathbf{v}; 0\right)}$, $\beta>0$) from the lower one (growth slower than $\avg{v_{\rm bm}}_{q\left(\mathbf{v}; 0\right)}$, $\beta<0$). The grey area represents the infeasible region. In color bars in (b) and (d), orange and blue shades are used for data coming from \cite{schuetz12} and \cite{nanchen06}, respectively.}
\end{figure*}

Given a feasible space $\mathcal{F}$, the F-I bound can be computed by EP, as the latter yields a multivariate Gaussian approximation $q\left(\mathbf{v};\beta\right)$ to the distribution (\ref{star}). The mean biomass synthesis rate increases as $\beta$ increases. We therefore selected $10^3$ values of $\beta$ in the range $\left[0,10^{4}\right]$ and, for each of these, computed $q\left(\mathbf{v};\beta\right)$, and the corresponding mean biomass $v_{\rm bm}^{\rm optm} = \avg{v_{\rm bm}}_{q\left(\mathbf{v};\beta\right)}$ and information content $I^{\rm optm}$. The latter quantities give the F-I bound (the green line in Fig. \ref{uno}d). (The same curve can also be computed thorough the Monte Carlo protocols described in \cite{demartino16,demartino18}.) Note that in principle one obtains different F-I bounds in each condition, since both $\mathcal{F}$ and $v_{\rm bm}^{\rm max}$ change with the glucose level. However, when the values of $v_{\rm bm}^{\rm optm}$ are re-scaled by $v_{\rm bm}^{\rm max}$, all curves collapse on the green line due to inherent linear scaling of fluxes with the glucose uptake. In this respect, the F-I bound shown in Fig. \ref{uno}d provides a highly robust characterization of the cell's metabolic capabilities.

\section*{Results}

\subsection*{Inferred fitness-information relationship}

Inferred values of (rescaled) fitness and information $\left( v_{\rm bm}^{\rm inf}/v_{\rm bm}^{\rm max},\, I^{\rm inf}\right)$ for the 33 experiments used to inform the MaxEnt problem (Materials and Methods) are shown by the markers in Fig. \ref{uno}d. Different color groups (orange vs blue) are used for data coming from \cite{schuetz12} and \cite{nanchen06}, respectively, with shades corresponding to different growth conditions, while triangles vs circles denote the presence ($\nu_{{\rm ace}}^{{\rm xp}}>0$) or absence ($\nu_{{\rm ace}}^{{\rm xp}}=0$) of acetate excretion. As the growth medium gets richer both the mean biomass output and the information encoded in inferred distributions increase, following a remarkably well-defined behavior. In other terms, faster-growing populations tend to spread over smaller and smaller portions of the feasible space. Such a relationship represents the trade-off between static (instantaneous) fitness and cell-to-cell variability in bacterial populations starting from reaction fluxes rather than through direct quantification of elongation rates or interdivision times (see e.g. \cite{taheri15}).

\subsection*{Empirical vs optimal fitness-information relationship}

\begin{figure*}[t!]
\begin{center}
\includegraphics[width=16.5  cm]{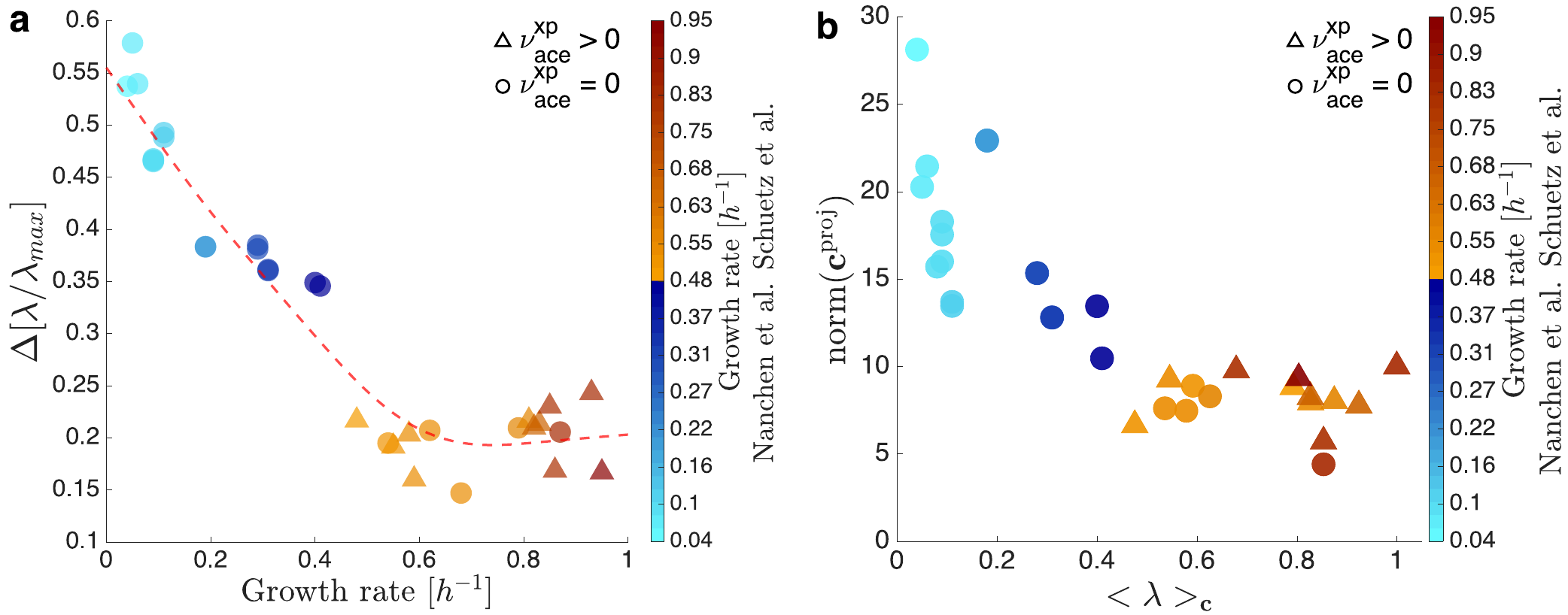}
\end{center}
\caption{(a) Vertical distance between the F-I bound and inferred F-I pairs versus empirical growth rate. The dashed line is a guide for the eye. (b) Norms of the projections of the fields $\mathbf{c}$ onto the feasible space $\mathcal{F}$ (Supplemental Information, Sec. C) as a function of the experimentally measured biomass. In color bars, orange and blue shades are used for data coming from \cite{schuetz12} and \cite{nanchen06}, respectively.
\label{due}}
\end{figure*}

The theoretically optimal fitness-information relationship is encoded by the F-I bound describing the maximal mean biomass output achievable in $\mathcal{F}$ at any given information content (or, vice versa, the minimum information content required to achieve a given mean biomass) (Materials and Methods). The bound obtained for network reconstruction of \cite{nanchen06,schuetz12} is shown by the green line in Fig. \ref{uno}d. The top (resp. bottom) branch of the line  corresponds to optimal states with $\beta>0$ (resp. $\beta<0$) in (\ref{star}), where the biomass synthetic rate is higher (resp. lower) than the value obtained by an unbiased uniform sampling of $\mathcal{F}$ ($\beta=0$ in (\ref{star})). Such a value is displayed as a black square in Fig. \ref{uno}d. F-I pairs lying within the two branches of the F-I bound (white area) are feasible, while pairs in the gray area are forbidden. Inferred F-I pairs consistently lie in the feasible region. A comparison between inferred and optimal scenarios reveals however two qualitatively different regimes (Fig. \ref{due}a). 
In poorer media, populations appear to be significantly sub-optimal but rapidly close the gap with the theoretically optimal mean biomass production rate as the medium improves. Here, inferred values of $I$ are larger than optimal ones by about 1 bit across the whole range of growth rates (Fig. \ref{uno}d). This suggests that a re-modeling of the protein repertoire leading to faster growth at roughly the same regulatory costs is likely the key driver of the organization of flux patterns, as described e.g. in \cite{hui15}. In richer media (faster growth), the fitness remains instead at a roughly constant (small) distance to the optimum, suggesting that growth is mainly information-limited: increases in fitness require fine tuning metabolic fluxes to encode more information into $p(\mathbf{v;c})$. Noticeably, the crossover from one regime to the other occurs around the growth rate where acetate overflow sets in \cite{basan15}.

\subsection*{Physical meaning of the inferred Lagrange multipliers}

Qualitative changes in the flux distributions are also reflected in the behavior of the inferred fields $\mathbf{c}=\{c_j\}$ (see (\ref{dist})). Roughly speaking, these quantities can be interpreted as `forces' acting on fluxes: the larger $c_j$, the more $p(\mathbf{v};\mathbf{c})$ is deformed in the direction of flux $j$ with respect to the uniform distribution on $\mathcal{F}$ with the prescribed glucose uptake in order for $\avg{v^{e}_j}$ to match $v_j^{\mathrm{xp}}$ (Materials and Methods). The projection $\mathbf{c}^{\mathrm{proj}}$ of the fields vector $\mathbf{c}$ onto the feasible space $\mathcal{F}$ (or, more precisely, its norm $||\mathbf{c}^{\mathrm{proj}}||$, see Supplemental Information, Sec. C) therefore quantifies the overall deformation required to reproduce data once the glucose uptake rate is given. One sees (Fig. \ref{due}b) that $||\mathbf{c}^{\mathrm{proj}}||$ decreases approximately as $v_{\rm bm}^{-0.4}$ as the growth rate increases. In other words, as the glucose uptake increases inferred distributions get closer to being as broad as possible given the glucose uptake. Because optimal populations are the least constrained at any given fitness, one may think that experimental populations also get globally less constrained as the medium gets richer. This is however not the case. To show it, one must compute, for every experiment, the fitness-information pairs obtained by varying only the biomass output (i.e. the coefficient associated with the biomass reaction) at fixed fields $\mathbf{c}$. For any given experiment (i.e. for any given $\mathbf{c}$), this  procedure returns a line in the fitness-information plane describing the values of $I$ that would correspond to a population subject to the same fields, but carrying a higher fitness. By construction, populations lying to the left of this line at any given growth rate encode less information into the flux distribution than the reference population and are therefore globally less constrained. A representative example in Supplementary Fig. S8 shows the isofield lines for two experiments from \cite{nanchen06} and one from \cite{schuetz12}. One sees that populations tend to lie to the right of these lines. Hence, contrary to intuition, faster-growing populations are slightly more constrained than slower ones despite being closer to the optimal fitness and information content.

Finally, the projections of $\mathbf{c}$ along the directions of individual degrees of freedom of the feasible space $\mathcal{F}$ provide information about the degree of regulation of individual reactions (Supplementary Fig. S10). At slower growth, reactions belonging to glycolysis, the Entner-Doudoroff pathway and the glyoxylate shunt get more and more downregulated compared to the unbiased mean as the glucose level is limited, while fluxes through the pentose phosphate pathway, the TCA cycle and oxidative phosphorylation are mostly upregulated. This picture effectively recapitulates  known patterns of proteome allocation in {\it E. coli} \cite{hui15}. The projection of $\mathbf{c}$ on the biomass production rate is relatively large at faster growth but behaves erratically in poor media. Marginal flux distributions (Supplemental Information Sec. I and Supplementary Fig. S7) confirm this picture in greater detail.

\begin{figure*}
\begin{center}
\includegraphics[width=0.9\textwidth]{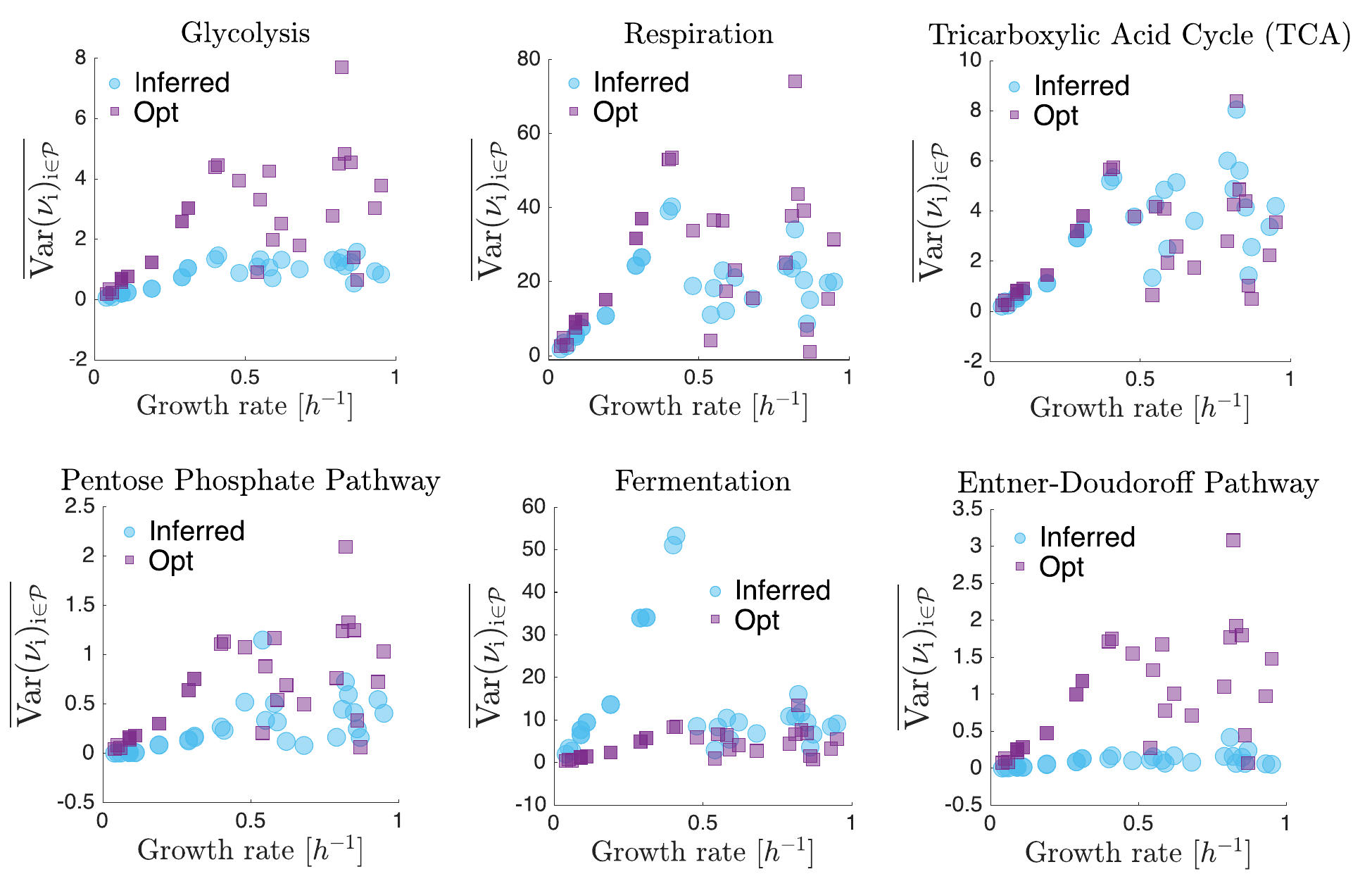}
\end{center}
\caption{\label{tre}Mean variance  of fluxes (Eq. (\ref{mv})) through six metabolic pathways as a function of the growth rate (empirical values) in optimal (purple) and inferred (cyan) distributions. (Note that markers appear darker at points where many of them overlap. This is mainly due to replicated experiments in \cite{nanchen06})}
\end{figure*}

\subsection*{Inferred vs optimal patterns of pathway regulation}

Besides quantifying the distortion of the inferred phenotype distribution (\ref{dist}) with respect to a uniform distribution (at given nutrient intake), the fields $\mathbf{c}$ also provide information about how different metabolic pathways are used in different conditions. At the simplest level, Principal Component Analysis (PCA) performed on the ensemble of the 33 inferred $\mathbf{c}$ vectors (one per experiment) shows that experiments are classified by the projection on the first component in two clusters characterized by distinct acetate excretion profiles (Supplemental Information, Sec. D and Supplementary Fig. S1; note that PCA correctly assigns the data from \cite{schuetz12} to the two clusters). This confirms carbon overflow as the key separator of phenotypic behavior in carbon-limited {\it E. coli} growth.


At a more refined level, one can consider the average variance of fluxes in each pathway $\mathcal{P}$ defined according to the network reconstruction provided in Table S1 from \cite{schuetz12} defined as
\begin{equation}\label{mv}
\overline{\mathrm{Var}(v_{i})_{i\in\mathcal{P}}} = \frac{1}{|\mathcal{P}|} \sum_{i\in \mathcal{P}} \Sigma_{ii}.
\end{equation}
A larger (resp. smaller) variance indicates that the pathway is globally less (resp. more) regulated, as its fluxes are allowed larger variability on average. Fig. \ref{tre} reports a comparison between optimal and inferred values for six key pathways. While glycolysis, Entner-Doudoroff, and pentose phosphate pathways appear to be tightly controlled in all conditions (both in  optimal and inferred distributions), respiration and the TCA cycle display a significant (and remarkably similar) pattern of variability in both cases.  Noticeably, their variability undergoes strong modulations as the growth rate changes. The major difference between the two cases is seen in the fermentation pathway, which appears to be much more variable in empirical population than it is at optimality. Finally, in both  optimal and inferred distributions the overall variability of fluxes within pathways appears to decrease close to the acetate onset point (ca. 0.6/h), again pointing to the occurrence of a major regulatory transition. Detailed results for all pathways included in the network model are given in Supplementary Fig. S9. Taken together, changes in the variability of pathways (Fig. \ref{tre}) and in the regulation of different reactions (Supplementary Fig. S10)  suggest that some of the known empirical facts regarding the use of cellular resources may result from the need to preserve a sufficiently large metabolic variability in any growth condition.

\begin{figure}[t!]
\begin{center}
\includegraphics[width=8cm]{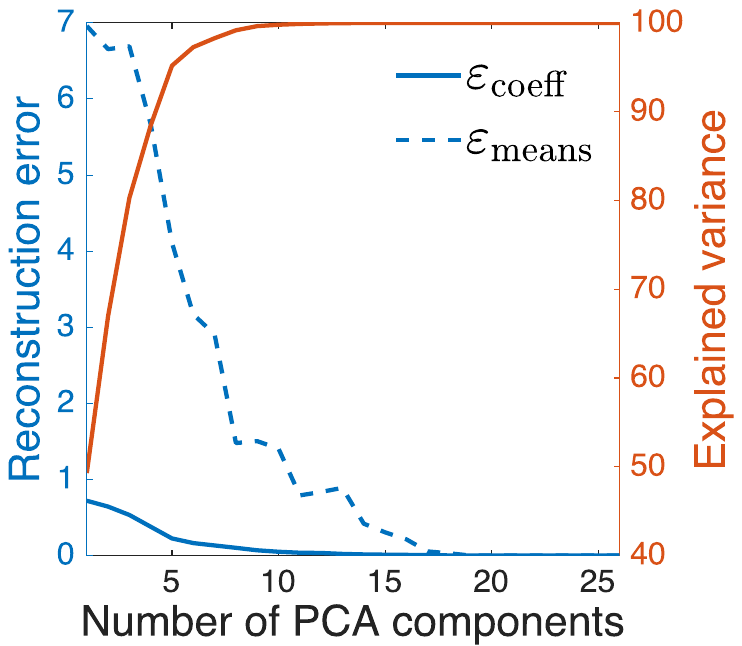}
\end{center}
\caption{\label{four}Explained variance of the coefficients (in percentage, red line) and reconstruction error of the original fields (continuous blue line) and phenotypes (dashed blue line) as a function of the number of PCA components employed to compute their low-rank counterpart.}
\end{figure}

\subsection*{Reduced representations of metabolic activity are not sparse}

\begin{figure}[t!]
\begin{center}
\includegraphics[width=\columnwidth]{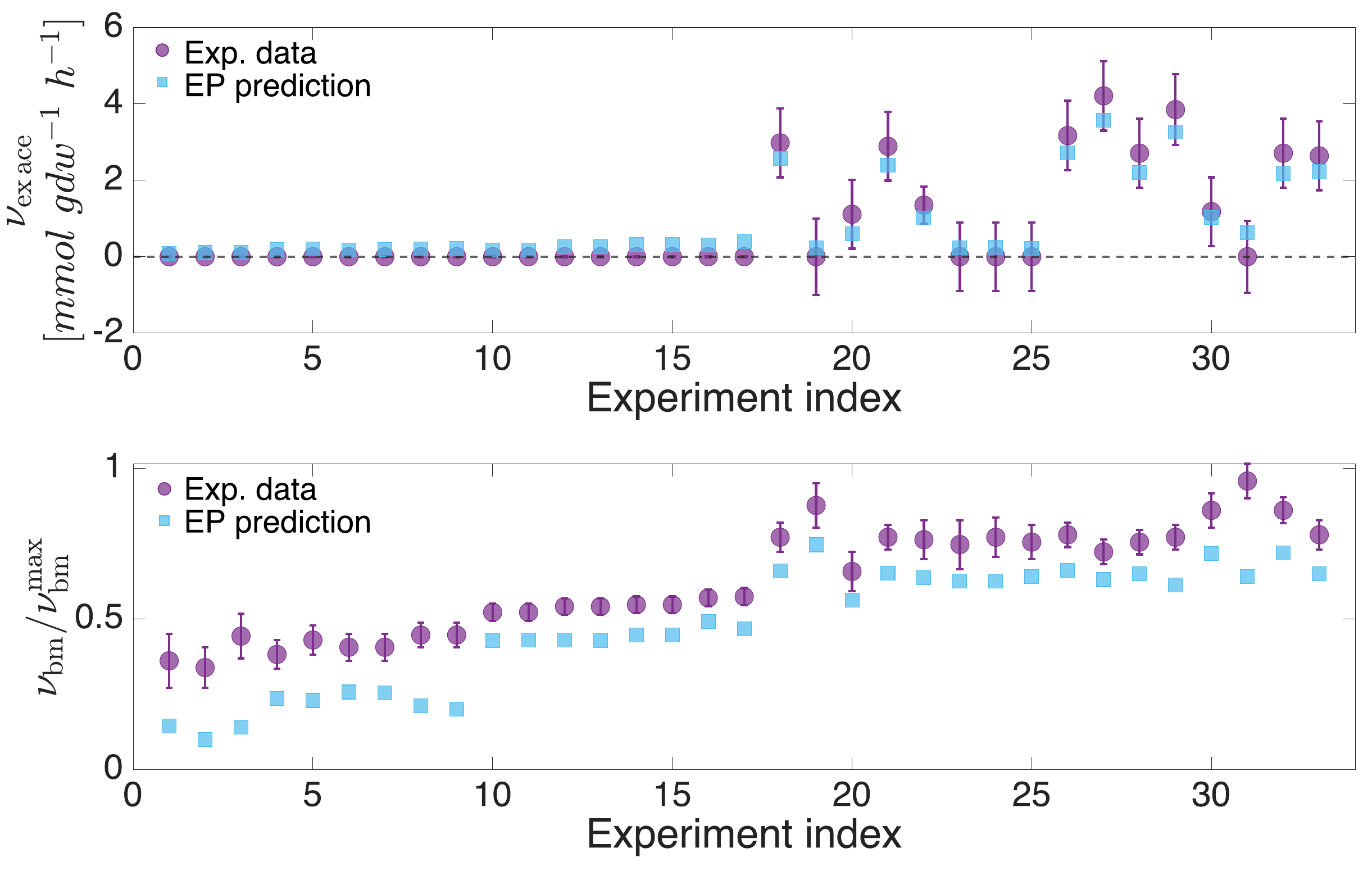}
\end{center}
\caption{\label{five}Mean values of the acetate excretion rate (top) and biomass production rate (bottom) measured (purple) and predicted from the inferred distribution (cyan) for the different experiments from \cite{nanchen06,schuetz12}. For comparison, bare FBA predictions correspond to $v_{\mathrm{ex\,ace}}=0$ and $v_{\rm bm}/v^{\max}_{\rm bm}=1$, respectively.}
\end{figure}

An important problem arising in the analysis of metabolic networks concerns the possibility that whole-network flux configurations might be efficiently represented by a small number of collective variables (`pathways'), whose control would be the central task of metabolic regulation. Experiments probing these variables would effectively allow us to reconstruct the activity across the whole network. Fig. \ref{four} indeed shows that about 90\% of the empirical variance of inferred fields $\mathbf{c}$ (red line) is explained by the first 5 principal components, suggesting that a significant reduction of dimensionality might be possible (as found e.g. in \cite{Kaneko}). Encouraged by this, we then estimated how well one can reconstruct inferred Lagrange coefficients and mean fluxes from PCA coefficients as a function of the number of PCA components included in the calculation. Ideally, an efficient representation would only require a small number of components. Denoting by $k$ the number of PCA components included, as quality indicators we used the quantities
\begin{gather}
\varepsilon_{\mathrm{coeff}}(k) = \overline{||\mathbf{c}_{a}^{\mathrm{proj}} - \mathbf{c}_{a}^{\mathrm{PCA}(k)}||},\\
\varepsilon_{\mathrm{means}}(k) = \overline{|| \langle \mathbf{v}_{\mathcal{E}} \rangle_{\mathbf{c}_{a}} - \langle \mathbf{v}_{\mathcal{E}} \rangle_{\mathbf{c}_{a}^{\mathrm{PCA}(k)}}||}~~,
\label{eq:PCAest}
\end{gather}
where $\mathbf{c}_{a}^{\mathrm{proj}}$ (resp. $\mathbf{c}_{a}^{\mathrm{PCA(k)}}$) is the projected vector of Lagrange multipliers for experiment $a = 1,\ldots,33$ from the full inference problem (resp. obtained by including only the first $k$ PCA components), while $\langle \mathbf{v}_{\mathcal{E}} \rangle_{(.)}$ collects the mean values of the measured fluxes $\mathcal{E}$ according to a MaxEnt distribution parametrized by the external fields $(.)$. The over-bar denotes the average over the 33 experiments. Fig. \ref{four} shows that $\varepsilon_{\mathrm{coeff}}$ is generically small and decreases fast with $k$, in line with the fact that the first 5 principal components explain most of the variability of the original coefficients. $\varepsilon_{\rm means}$, however, decreases much more slowly. In practice, lossless inference of the coefficients $\mathbf{c}$ (and hence the reconstruction of mean fluxes)  requires the inclusion of the first 18 principal components (Supplemental Information, Sec. E and Supplementary Fig. S2). Incidentally, this number coincides with the number of fluxes measured in \cite{nanchen06,schuetz12} that are linearly independent. Note that, besides the achievable effectiveness in reproducing measured fluxes, inferred distributions also yields predictions for metabolic fluxes that are inaccessible to labeling experiments, stored in the marginal densities of individual fluxes (Supplemental Information, Sec. I and Supplementary Fig. S7). Likewise, they allow for accurate predictions of the values of physiological parameters quantified independently of fluxes, like growth and acetate excretion rates when such quantities are excluded from the inference procedure (Fig.  (\ref{five})).

\subsection*{Metabolic control coefficients}

The fact that the complexity of metabolic activity is not captured by a few effective variables, indicates that the moments of  (\ref{dist}) are highly sensitive to the values of the inferred fields $\mathbf{c}$: small changes in the latter can induce large re-arrangements in the  distribution. Metabolism, in other words, forms a system of globally coupled processes. Within the theoretical framework employed here, this aspect is fully quantified by the flux-flux correlation matrices. Indeed, starting from the definition of $Z(\mathbf{c})$, i.e. (see (\ref{dist}))
\begin{gather}
Z(\mathbf{c})=\int_{\mathcal{F}}\exp\left[\sum_{j\in\mathcal{E}}c_j v_j\right]\,d\mathbf{v}~~,
\end{gather}
one can easily show by a direct calculation that
\begin{gather}
\frac{\partial\ln Z(\mathbf{c})}{\partial c_i}=\frac{1}{Z(\mathbf{c})}\int_{\mathcal{F}} v_i \,\exp\left[\sum_{j\in\mathcal{E}}c_j v_j\right]\,d\mathbf{v}\equiv\avg{v_i}_{\mathbf{c}}\\
\frac{\partial^2\ln Z(\mathbf{c})}{\partial c_i\partial c_j}=\avg{v_i v_j}_{\mathbf{c}}-
\avg{v_i}_{\mathbf{c}}\avg{v_j}_{\mathbf{c}}\equiv C_{ij}\\
\frac{\partial\avg{v_i}_{\mathbf{c}}}{\partial c_j}=\frac{\partial\avg{v_j}_{\mathbf{c}}}{\partial c_i}=C_{ij}~~,\label{venti}
\end{gather}
where $\avg{\cdots}_{\mathbf{c}}=\int\cdots p(\mathbf{v;c})d\mathbf{v}$. Hence (see (\ref{venti})), flux-flux correlations $C_{ij}$ (computable by EP, see Supplemental Information, Sec. B) can be immediately interpreted as `metabolic control coefficients' relating changes in flux $i$ to changes in field $j$ (or vice-versa). For instance,  the correlation matrix computed for a representative experiment from \cite{nanchen06} shown in Supplemental Information H and Supplementary Fig. S6 suggests that glycolitic reactions are positively coupled to other glycolytic reactions (e.g. an upregulation in one reaction, quantified by a change in the corresponding field, will increase the mean flux through the other), whereas they are mostly negatively coupled to reactions in the pentose phosphate pathway (upregulations in the latter will decrease the flux through the former). These couplings span across the entire metabolic network. Indeed one sees that changes in one field typically propagate to a large number of reactions, supporting the idea that the cross-talk between metabolic reactions is significantly non-local.

\section*{Discussion}

\subsection*{Biological significance of the information content}

While the idea that fitness and information content of flux distributions are interrelated seems rather natural, the physiological meaning of the latter is not obvious. Technically, $I$ quantifies the deviation of the inferred distribution $p(\mathbf{v;c})$ from uniformity in a given medium. Small values of $I$ imply that experiment-derived constraints do not significantly modify our prior knowledge of the flux distribution, corresponding to all flux vectors in the feasible space defined by the given uptake rates being equally likely. On the other hand, large values of $I$ imply that the inferred likelihood has a small overlap with the uniform distribution in the same medium. In this sense, one can say that $I$ provides a proxy for the amount of metabolic regulation required to grow in a given medium. We have seen that the information content {\it per degree of freedom} more than doubles as the growth rate goes from 0.05/h to about 1/h (Fig.\,\ref{uno}d). Such a gain is mainly due to a systemic fine-tuning of fluxes and correlations rather than to the tightened control of a few pathways, in agreement with evidence suggesting that system-wide rearrangements underlie response to changing carbon levels in {\it E. coli} \cite{hui15}. 

Our analysis also shows that inferred distributions exceed the minimum required information content by roughly 1 bit per degree of freedom in all growth conditions (Fig.\,\ref{uno}d).  
This suggests that the biochemical constraints used to define the feasible space (flux reversibility, ranges of variability, etc.) might be too conservative. Further ingredients affecting the metabolism of single cells, like biosynthetic costs \cite{mori16}, might also reduce the feasible space and bring data closer to the theoretical bound. However, the gap may also indicate that population growth requires a minimum amount of regulatory information, in line with the idea that minimal complexity (as opposed to minimal number of components) is the defining characteristics of cells \cite{xavier14}. That regulatory interactions and mechanical effects are crucial in determining {\it E. coli}'s overall metabolic capabilities is indicated by the fact that they remain substantially unchanged following a large-scale removal of unnecessary genes \cite{posfai06,gerosa11,minton11}. By contrast, they are significantly affected by the selective knock-out of a small number of genes through which specific cellular tasks are optimized \cite{carlson04,trinh06,trinh08}. In this sense, constraint-based models may be missing a substantial amount of regulatory interactions that would effectively reduce the size of the feasible space $\mathcal{F}$. Identifying these constraints could bring empirical populations closer to the F-I bound and provide crucial hints about the nature of optimality in bacterial growth.

It is finally important to remark that the F-I bound we define is fundamentally different from the fitness-information relationship derived in \cite{bialek12}. In that case, one quantifies the information about nutrient availability that has to be encoded in the level of a nutrient-processing enzyme in order to achieve a given fitness. In our case, information is a measure of the high-dimensional space of flux configurations that is effectively accessible to the system.

\subsection*{Limitations of our analysis}

Besides the information encoded in the network structure, the key physical assumption made in our inference is that metabolic networks are at a NESS described by the mass balance conditions alone. This means that we do not account for factors like biosynthetic costs, molecular crowding, membrane occupancy etc. All of these are likely essential for the metabolic behavior single cells. However, including them in an inverse model defined on $\mathcal{F}$ would necessarily require additional assumptions about how they are linked to metabolic fluxes.

On the technical side, our study is limited by two not-easily-avoidable facts. (i) The data sets we used are not homogeneous, so it is {\it a priori} difficult to consider one as a continuation of the other at different growth rates. Carrying out this study on a broader, unique fluxomic data set covering a large enough range of growth rates would likely yield a more clean-cut picture. That a consistent scenario can emerge despite this limitation is in this respect quite remarkable. (ii) In our framework, we implicitly interpret measured flux variances as proxies for the cell-to-cell variability. While this assumption has given consistent results when used in the context of single-cell data \cite{demartino16,demartino18}, a more detailed understanding of the sources of variability and error in fluxomics would allow to fine-tune the application of the MaxEnt scheme for the inverse problem considered here. It is however important to note that, while our approach is capable of efficiently representing the empirical variability, it cannot point to specific causal factors behind it. For this goal, different types of models (e.g. biochemically detailed dynamical models) are necessary.

\subsection*{Relation to other approaches}

The most immediate comparison for our results is given by standard biomass maximization, which corresponds to the limit $\beta\to\infty$ in (\ref{star}).  Previous work has shown that empirical data, including distributions of elongation rates in exponentially growing populations and measured fluxes, are better described using (\ref{star}) with finite $\beta$ rather than its $\beta\to\infty$ limit \cite{demartino16,demartino18}. Here we have effectively quantified how close flux distributions inferred from data are to (\ref{star}) in terms of fitness and information content. Another set of potentially related problems concerns the experiment-guided determination of an objective function for constraint-based models. Different techniques have been proposed in the past to infer objectives or discriminate between various alternatives \cite{burgard03,knorr06,gianchandani08,chiu08,zhao16,yang19}. While the vector $\mathbf{c}$ of Lagrange multipliers does partially align with the biomass output, our analysis does not highlight a clear objective function for constraint-based models. On the contrary, our results support the idea that the growth of bacterial populations is governed by a trade-off between mean single-cell biomass and heterogeneity. Notice that optimizing the mean biomass {\it over time} provides individual cells with an effective way to cope with multiple sources of variability \cite{miotto19}. For instance, bacteria in fluctuating environments may be unable to adjust fluxes to the distribution that maximizes the instantaneous biomass synthetic rate due to the biosynthetic cost of the regulatory machinery implementing the adjustments. Regulatory programs selected over longer time scales would essentially optimize the frequency with which metabolism is adjusted in varying conditions.

Finally, we note that MaxEnt-based models of metabolic networks have been employed in the past for a variety of purposes: to guide the decomposition of flux configurations into physiologically significant modes \cite{zhao09,zhao10}; explain the variability observed in bacterial populations \cite{demartino16,demartino17} and continuous cell cultures \cite{cossio19,pereiro21} (note: the latter article appeared while the present paper was under review); reproduce empirical data on fluxes  \cite{demartino18}; derive dynamic strategies of cellular resource allocation \cite{tourigny20}; or predict response times to changing environments \cite{masoero}. While also based on the Maximum Entropy principle, the work presented here faces the question of heterogeneity from a rather different viewpoint, aiming essentially at bridging the gap between optimization-based methods and empirical results by building an efficient representation of metabolic data using constraint-based models. Our hope is that such an approach will lead to new theoretical insights into the nature and optimality of bacterial growth.

\section*{Conclusion}

We have shown here that, as the growth medium gets richer, phenotype distributions inferred for {\it E. coli} populations appear to follow and slowly approach a theoretical limit that quantitatively relates the mean biomass production rate to the cell-to-cell variability in metabolic phenotypes. Specifically, the mean biomass gets closer to the maximum allowed by the inferred heterogeneity of the population. Despite the fact that large fluctuations affect the activity of metabolic pathways, the scenario we obtain reproduces some of the well-known tradeoffs that characterize {\it E. coli} growth under carbon limitation, including downregulation of glycolysis, upregulation of respiration and the TCA cycle, and the transition to acetate overflow. This suggests that {\it E. coli} populations trade some of their fitness to maintain their metabolic heterogeneity nearly as large as possible in all growth conditions considered.

\section*{Author Contributions}

APM performed research, analysed the data, prepared figures and tables, contributed materials and analysis tools, wrote the paper and critically reviewed the manuscript. AB, AP, DDM and ADM conceived and designed research, contributed materials and analysis tools, wrote the paper and critically reviewed the manuscript.

\section*{Acknowledgments}

AB, ADM, AP, and APM acknowledge financial support from Marie Sk{\l}odowska-Curie, grant agreement No. 734439 (INFERNET).


\begin{widetext}

\appendix

\centerline{\textsf{\large{Relationship between fitness and heterogeneity  in exponentially growing microbial populations}}}

\vspace{0.3cm}

\centerline{\textsf{\large{SUPPLEMENTAL INFORMATION}}}

\vspace{0.3cm}

\centerline{Anna Paola Muntoni$^{1,2}$, Alfredo Braunstein$^{1,2,3}$, Andrea Pagnani$^{1,2,3}$, Daniele De Martino$^{4,*}$, and Andrea De Martino$^{1,2,5,\dag}$}

\vspace{-0.3cm}

\begin{center}\small{
{\it $^1$ Politecnico di Torino, Corso Duca degli Abruzzi, 24, I-10129, Torino, Italy\\
$^2$Italian Institute for Genomic Medicine, IRCCS Candiolo, SP-142, I-10060 Candiolo (TO), Italy\\
$^3$INFN, Sezione di Torino, Torino, Italy\\
$^4$Biofisika Institute (CSIC,UPV-EHU) and Ikerbasque Basque Foundation for Science, Bilbao 48013, Spain\\
$^5$Istituto di Nanotecnologia (CNR-NANOTEC), Consiglio Nazionale delle Ricerche, I-00185 Roma, Italy}\\
$^*$ daniele.demartino@ehu.eus\\
$\dag$ andrea.demartino@polito.it}
\end{center}

\section*{Supporting Text}

\subsection{Reactions mapping}

The reaction network used in this work is that reported in Table S1 of ~\cite{schuetz12}. For the experiments in ~\cite{nanchen06} we need to map each measured flux to one of the model in ~\cite{schuetz12}. Looking at the chemical equations reported in ~\cite{nanchen06} and those associated with the fluxes in Table S1 of ~\cite{schuetz12} we perform the following mapping:
\begin{table}[h]
\centering
\begin{tabular}{c|c}
Measured reaction fluxes in ~\cite{nanchen06}  & Mapping to the reaction network in ~\cite{schuetz12} \tabularnewline
\hline 
GLC+ATP$\rightarrow$G6P  & glk\tabularnewline
\hline 
G6P $\rightarrow$6PG+NADPH  & zwf\tabularnewline
\hline 
6PG$\rightarrow$P5P+CO2+NADPH  & gnd\tabularnewline
\hline 
6PG$\rightarrow$F6P  & pgi\tabularnewline
\hline 
6PG$\rightarrow$T3P+PYR  & eda\tabularnewline
\hline 
F6P+ATP$\rightarrow$2T3P  & pfk\tabularnewline
\hline 
2P5P$\rightarrow$S7P+T3P  & tktA\tabularnewline
\hline 
P5P+E4P$\rightarrow$F6P+T3P  & tktB\tabularnewline
\hline 
S7P+T3P$\rightarrow$E4P+F6P  & talA\tabularnewline
\hline 
T3P$\rightarrow$PGA+ATP+NADH  & gapA\tabularnewline
\hline 
PGA$\rightarrow$PEP  & eno\tabularnewline
\hline 
PEP$\rightarrow$PYR+ATP  & pyk\tabularnewline
\hline 
PYR$\rightarrow$AcCoA+CO2+NADH  & aceEFlpd\tabularnewline
\hline 
OAA+AcCoA$\rightarrow$ICT  & gltA\tabularnewline
\hline 
ICT$\rightarrow$OGA+CO2+NADPH  & icd\tabularnewline
\hline 
OGA$\rightarrow$FUM+CO2+1.5ATP+2NADH  & sucABlpd\tabularnewline
\hline 
FUM$\rightarrow$MAL  & fum\tabularnewline
\hline 
MAL$\rightarrow$OAA+NADH  & mdh\tabularnewline
\hline 
MAL$\rightarrow$PYR+CO2+NADH  & maeA\tabularnewline
\hline 
OAA+ATP$\rightarrow$PEP+CO2  & pck\tabularnewline
\hline 
PEP+CO2$\rightarrow$OAA  & ppc\tabularnewline
\hline 
ICT+AcCoA$\rightarrow$MAL+FUM+NADH & aceA\tabularnewline
\hline 
NADPH$\rightarrow$NADH  & udhA\tabularnewline
\hline 
O2+2NADH$\rightarrow$2P/O x ATP  & ndh\tabularnewline
\hline 
Acetate production rate & ex ace\tabularnewline
\hline 
Glucose consumption rate & ex glc\tabularnewline
\hline 
Biomass & biomass\tabularnewline
\end{tabular}

\end{table}

\subsection{Mathematical details of the Expectation Propagation algorithm}

In this section we report (i) the detailed derivation of the statistical model involving the fluxes $\mathbf{v}$ and $\mathbf{v}^e$ presented in the main text and (ii) the derivation of the Expectation Propagation approximation scheme for the target distribution.

\subsubsection{Modeling the posterior probabilities of the fluxes given experimental evidence}

Let us briefly mention how to probabilistically treat the constrained fluxes, as done in ~\cite{braunstein17}.
Within a Bayesian framework, we can investigate what is the \textit{a posteriori} probability of observing a configuration of fluxes given a vector $\mathbf{b}$, encoding the mass-balance equations into a likelihood function, and the constraints on the range of variability into single-variable priors:
\begin{equation}
p\left(\mathbf{v};\mathbf{b}\right) = \frac{1}{Z_{p}}  \delta\left(\mathbf{S}\mathbf{v} - \mathbf{b}\right)
\prod_{i} \frac{\mathbb{I}_{\left[v_{i}^{max}, v_{i}^{min}\right]}(v_{i})} {v_{i}^{max} - v_{i}^{min}}
\end{equation}
where $\delta\left(\mathbf{S}\mathbf{v} - \mathbf{b}\right)$ is the Dirac delta function equals to 1 for configurations of fluxes satisfying the mass-balance conditions and 0 otherwise, $Z_{p}$ is the normalization constant, or partition function in the statistical mechanics jargon. \\
Here we are interested in the joint probability distribution between the auxiliary `experimental' fluxes (affected by noise) $\mathbf{v}^{e}$, and the `constrained' ones (confined within the polytope described by the stoichiometric constraints and the boundaries of variability) $\mathbf{v}$, shown in Eq. (\ref{eq:jointdist}) of the main text. For sake of simplicity we define as $E = |\mathcal{E}|$ the number of observed fluxes (it varies from 26 to 25 when considering ~\cite{nanchen06} or ~\cite{schuetz12} respectively), and we restrict the analysis on $\mathbf{v}^{e}\in\mathbb{R}^{E}$, hence, at difference with the main text, the latter collects only the values of the noisy and measured fluxes. Furthermore, we recall that the indices of the measured fluxes in $\mathcal{E}$ allow for the mapping between $\mathbf{v}^{e}$ and the corresponding components in $\mathbf{v}$; finally, we can define the joint distribution of the two sets of fluxes as
\begin{equation}
\label{eq:True-p-v-ve}
p\left(\mathbf{v},\mathbf{v}^{e};\mathbf{c},\mathbf{b}\right)\propto 
\delta\left(\mathbf{S}\mathbf{v} - \mathbf{b}\right)
\prod_{i=1}^{N} \psi_{i}(v_{i})
\prod_{i=1}^{E}e^{-\frac{\gamma_{i}}{2}\left(v_{i}^{e}-v_{\mathcal{E}\left(i\right)}\right)^{2}+c_{\mathcal{E}\left(i\right)}v_{i}^{e}}
\end{equation}
being $\psi_{i}(v_{i}) = \frac{\mathbb{I}_{\left[v_{i}^{max}, v_{i}^{min}\right]}(v_{i})} {v_{i}^{max} - v_{i}^{min}}$, $\mathbf{c}$ the vector containing the Lagrange multipliers in correspondence to the measured fluxes and $\gamma_{i}$ the inverse variance of the noise affecting the $i^{\rm th}$ auxiliary flux, i.e. $v^{e}_{i}$.  Notice that estimating any observable from the joint distribution $p\left(\mathbf{v},\mathbf{v}^{e};\mathbf{c},\mathbf{b}\right)$ turns out to be intractable because the computation of the partition function is analytically unfeasible as it corresponds to the calculation of the volume of a convex polytope in high dimensions. We cope with this issue exploiting the approximation scheme provided by Expectation Propagation (EP). EP is an iterative algorithm that provides a Gaussian approximation of intractable probability distributions. In the context of metabolic fluxes reconstruction, it has been shown in ~\cite{braunstein17} its ability in treating the marginal probability density of fluxes satisfying the mass-balance constraints $\mathbf{S}\mathbf{v} = \mathbf{b}$ and bounded intervals of variability, i.e. $v_{i}\in\left[v_{i}^{min}, v_{i}^{max}\right]$. While the likelihood function alone, expressed as a Dirac delta function, would allow for a treatable separation between free and dependent flux, the single variable priors are responsible for the intractability of the partition function. The idea behind EP scheme is to approximate each `hard' term $\psi_{i}\left(v_{i}\right)$ through a univariate Gaussian density having mean $a_{i}$ and variance $d_{i}$, to be determined within the approximation. Following the approximation in ~\cite{braunstein17}, we can easily define a multivariate Gaussian approximation of the joint posterior distribution in Eq. (\ref{eq:True-p-v-ve}) as
\begin{equation}
\label{eq:Gauss-v-ve}
q\left(\mathbf{v},\mathbf{v}^{e};\mathbf{c},\mathbf{b}\right)\propto 
\delta\left(\mathbf{S}\mathbf{v} - \mathbf{b}\right)
\prod_{i=1}^{N}e^{-\frac{\left(v_{i}-a_{i}\right)^{2}}{2d_{i}}}
\prod_{i=1}^{E}e^{-\frac{\gamma_{i}}{2}\left(v_{i}^{e}-v_{\mathcal{E}\left(i\right)}\right)^{2}+c_{\mathcal{E}\left(i\right)}v_{i}^{e}}
\end{equation}
keeping in mind that here, together with the parameters of the overall Gaussian approximation $\mathbf{a}$, $\mathbf{d}$, we have to concurrently determine the values of the unknown Lagrange multipliers $\mathbf{c}$ and $\boldsymbol{\gamma}$.
We show in the following that both measures can be iteratively determined within the EP approximation through a two-steps algorithm: at each iteration $t$, we can update the parameters $(\mathbf{a}^{t+1}$, $\mathbf{d}^{t+1})$ in Eq. (\ref{eq:Gauss-v-ve}), for fixed $(\mathbf{c}^{t}, \boldsymbol{\gamma}^{t})$ using the `standard' EP fixed point equations ~\cite{braunstein17}. Then, we can proceed updating $(\mathbf{c}^{t+1}, \boldsymbol{\gamma}^{t+1})$, for fixed $(\mathbf{a}^{t}$ and $\mathbf{d}^{t})$, requiring that the matching constraints in Eqs. (\ref{eq:constr_means}) and (\ref{eq:constr_av_noise}) of the main text are satisfied. Notice that practically, the update is slightly differently performed, as described in Section \ref{sec:impl}. \\
In the following we will first exploit the linear relationship among fluxes to identify the set of dependent and independent fluxes (notice that the Expectation Propagation scheme for this subdivision of the target variables has been already exploited in ~\cite{braunstein20,saldida20} in different contexts). Then, we will derive the two-steps EP update scheme for a general framework in which each experimental flux $i\in\mathcal{E}$ has its own experimental error and a parameter $\gamma_{i}$ associated with it (the scheme exploiting a unique value for $\gamma$ can be straightforwardly derived from the general one as described in the following). Finally, we give some implementation details of the overall scheme.

\subsubsection{Pre-process of the fluxes}

Applying the Gaussian eliminations on the rows of the stoichiometric matrix (together with the known term $\mathbf{b}$), allows us to get the equivalent row echelon form $\mathbf{A}^{'}$ associated with the system of equations $\mathbf{S}\mathbf{v} = \mathbf{b}$
\begin{equation}
\mathbf{A}^{'}=\left[\begin{array}{ccccccc}
1 & 0 & \ldots & 0 & A_{1,1} & \ldots & A_{1,n'}\\
0 & 1 & \ldots & \ldots & \ldots &  & \ldots\\
\ldots & 0 & \ldots & \ldots &  & \ldots\\
& \ldots & \ldots & 0 & \ldots &  & \ldots\\
\ldots & \ldots & \ldots & 1 & A_{m',1} & \ldots & A_{m',n'}\\
\ldots & \ldots & \ldots & \dots & \ldots & \ldots & \ldots \\
0 & 0 & \ldots & 0 & 0 & \ldots & 0
\end{array}|\begin{array}{c}
y_{1}\\
\ldots\\
\\
\ldots\\
y_{m'}\\
\ldots\\
0
\end{array}\right]
\end{equation}
where $m'$ ($n'$) is the number of dependent (independent) fluxes. Here, we can identify on the left the identity matrix of size $m' \times m'$, in the central part a submatrix $\mathbf{A}\in\mathbb{R}^{m'\times n'}$  encoding the linear relationships among free (hereafter denoted as $\mathbf{v}^{\rm f}$) and dependent fluxes (called in the following $\mathbf{v}^{\rm d}$) and a vector $\mathbf{y}\in\mathbb{R}^{m'}$ containing the transformed constant terms. As a consequence, the set of dependent fluxes can be retrieved from the free ones using
\begin{equation}
\mathbf{v}^{\rm d} = -\mathbf{A} \mathbf{v}^{\rm f} + \mathbf{y}
\end{equation}
As for the fluxes, we can split the vector of the Lagrange multipliers $\mathbf{c}$ in $(\mathbf{c}^{d}, \mathbf{c}^{f})$ where the first (second) set is associated with the dependent (free) fluxes, taking non-zeros values only in correspondence to the measured fluxes. \\
In the following, we will focus on the distributions of the two subsets of fluxes $(\mathbf{v}^{f}, \mathbf{v}^{\rm d})$ and the original stoichiometric equations $\delta\left(\mathbf{S}\mathbf{v} - \mathbf{b}\right)$ will be replaced by the constraints $\delta(\mathbf{v}^{\rm d} + \mathbf{A}\mathbf{v}^{\rm f}-\mathbf{y})$. Besides, we should also differentiate within the full set of auxiliary and noisy fluxes $\mathbf{v}^{e}$ (and the associated coefficients $\mathbf{c}^{e}$), those that belong to the set of dependent or free fluxes; to this end, we define four extra auxiliary vectors, $\mathbf{v}^{e,d} \in \mathbb{R}^{m}$, $\mathbf{c}^{e,d} \in \mathbf{R}^{m}$ and $\mathbf{v}^{e,f}\in\mathbb{R}^{n}$, $\mathbf{c}^{e,f}\in\mathbf{R}^{n}$ being $m$ ($n$) the number of measured and dependent (free) auxiliary fluxes. Obviously, $m \leq m'$, $n \leq n'$, and $m + n = E$.
Therefore, it is possible to re-phrase the target distribution in Eq. (\ref{eq:True-p-v-ve}) and the full Gaussian approximation in Eq. (\ref{eq:Gauss-v-ve}) as:
\begin{eqnarray}
\label{eq:True-p-dep-free}
p\left(\mathbf{v}^{d},\mathbf{v}^{f},\mathbf{v}^{e,f},\mathbf{v}^{e,d};\mathbf{c}^{e,d}, \mathbf{c}^{e,f},\mathbf{y}\right) & \propto &
\delta(\mathbf{v}^{\rm d} + \mathbf{A}\mathbf{v}^{\rm f}-\mathbf{y})
\prod_{i=1}^{m'} \psi_{i}(v^{f}_{i})
\prod_{i=1}^{n'} \psi_{i}(v^{d}_{i}) \times \\
& & \times
\prod_{i=1}^{m}e^{-\frac{\gamma_{i}}{2}\left(v_{i}^{e,d}-v_{\mathcal{E}_{d}\left(i\right)}^{d}\right)^{2}+c^{e,d}_{i}v_{i}^{e,d}} 
\prod_{i=1}^{n}e^{-\frac{\gamma_{i}}{2}\left(v_{i}^{e,f}-v_{\mathcal{E}_{f}\left(i\right)}^{f}\right)^{2}+c^{e,f}_{i}v_{i}^{e,f}} \nonumber 
\end{eqnarray}
\begin{eqnarray}
\label{eq:Gauss-dep-free}
q\left(\mathbf{v}^{d},\mathbf{v}^{f},\mathbf{v}^{e,f},\mathbf{v}^{e,d};\mathbf{c}^{e,d}, \mathbf{c}^{e,f},\mathbf{y}\right) & \propto &
\delta(\mathbf{v}^{\rm d} + \mathbf{A}\mathbf{v}^{\rm f}-\mathbf{y})
\prod_{i=1}^{m'}e^{-\frac{\left(v^{d}_{i}-a^{d}_{i}\right)^{2}}{2d^{d}_{i}}}
\prod_{i=1}^{n'}e^{-\frac{\left(v^{f}_{i}-a^{f}_{i}\right)^{2}}{2d^{f}_{i}}} \times \\
& & \times
\prod_{i=1}^{m}e^{-\frac{\gamma_{i}}{2}\left(v_{i}^{e,d}-v_{\mathcal{E}_{d}\left(i\right)}^{d}\right)^{2}+c^{e,d}_{i}v_{i}^{e,d}} 
\prod_{i=1}^{n}e^{-\frac{\gamma_{i}}{2}\left(v_{i}^{e,f}-v_{\mathcal{E}_{f}\left(i\right)}^{f}\right)^{2}+c^{e,f}_{i}v_{i}^{e,f}} \nonumber
\end{eqnarray}
where $(\mathbf{a}^{d}, \mathbf{d}^{d})$ and $(\mathbf{a}^{f},\mathbf{d}^{f})$ are the set of means and variances of the Expectation Propagation approximation for dependent and free fluxes respectively, while $\mathcal{E}_{d}$ and $\mathcal{E}_{f}$ are the indices of measured dependent and free fluxes within the sets $\mathbf{v}^{d}$ and $\mathbf{v}^{f}$ respectively.

\subsubsection{Determining $\mathbf{a}$ and $\mathbf{d}$, fixed $\mathbf{c}$ and $\gamma$}
Let us first determine the update equations of the means $(\mathbf{a}^{d}, \mathbf{a}^{f})$ and the variances $(\mathbf{d}^{d}, \mathbf{d}^{f})$ of the Gaussian approximation, following the usual scheme of the Expectation Propagation algorithm, for fixed Lagrange multipliers $\mathbf{c}$ and $\boldsymbol{\gamma}$. Since the `hard` single-prior terms involve the constrained fluxes only (and as a consequence the parameters $\mathbf{a}$ and $\mathbf{d}$), we will first proceed marginalizing the approximated joint distribution over the noisy fluxes $\mathbf{v}^{e}$.  \\ 
For sake of simplicity, let us define:
\begin{itemize}
	\item $\mathbf{A}^{\rm exp}_{d}$, a sub-matrix of $\mathbf{A}$ composed of all the columns and only the rows associated with measured (and dependent) fluxes. Similarly, we define $\mathbf{y}_{d}^{\rm exp}$ as the sub-vector of $\mathbf{y}$ aggregating only the components related to the measured dependent fluxes. As a consequence, the dependent fluxes that have been experimentally observed can be expressed in terms of all free ones as $\mathbf{v}_{\mathcal{E}_{d}}^{d}=\mathbf{y}_{d}^{\rm exp}-\mathbf{A}_{d}^{\rm exp} \mathbf{v}^{f}$;
	\item $\boldsymbol{\gamma}_{m\times m}$ and $\boldsymbol{\gamma}_{n \times n}$, the diagonal matrices containing the Lagrange multipliers $\gamma_{i}$ being $i$ a measured flux of the set of the dependent or the free fluxes respectively. For uniform $\gamma$, the two matrices coincide with $\gamma\mathbb{I}_{m\times m}$ and $\gamma\mathbb{I}_{n\times n}$, where $\mathbb{I}_{x \times x}$ is the identify matrix of size $x$;
	\item $\mathbf{D}^{d}$ and $\mathbf{D}^f$, two diagonal matrices containing the inverse variances of the approximation, $\mathbf{d}^d$ and $\mathbf{d}^f$, i.e. $\mathbf{D}^{d} = \textrm{diag}(\frac{1}{d_{1}^{d}},\ldots,\frac{1}{d_{m'}^{d}})$, $\mathbf{D}^{f} = \textrm{diag}(\frac{1}{d_{1}^{f}},\ldots,\frac{1}{d_{n'}^{f}})$.
\end{itemize}
Using the formalism just introduced and the Dirac delta function in Eq. (\ref{eq:Gauss-dep-free}) we explicitly remove the dependent variables $\mathbf{v}^{d}$ in the joint distribution $q\left(\mathbf{v}^{d},\mathbf{v}^{f},\mathbf{v}^{e,f},\mathbf{v}^{e,d};\mathbf{c}^{e,d}, \mathbf{c}^{e,f},\mathbf{y}\right)$ and re-write it as
\begin{eqnarray}
\label{eq:Gauss-free-exp}
q\left(\mathbf{v}^{f},\mathbf{v}^{e,f},\mathbf{v}^{e,d};\mathbf{c}^{e,d}, \mathbf{c}^{e,f},\mathbf{y}\right) & \propto & e^{-\frac{1}{2}\left(\mathbf{v}^{f}-\mathbf{a}^{f}\right)^{T}\mathbf{D}^{f}\left(\mathbf{v}^{f}-\mathbf{a}^{f}\right)-\frac{1}{2}\left(\mathbf{y}-\mathbf{A}\mathbf{v}^{f}-\mathbf{a}^{d}\right)^{T}\mathbf{D}^{d}\left(\mathbf{y}-\mathbf{A}\mathbf{v}^{f}-\mathbf{a}^{d}\right)} e^{\mathbf{c}^{e,d^{T}}\cdot\mathbf{v}^{e,d}+\mathbf{c}^{e,f^{T}}\cdot\mathbf{v}^{e,f}} \times \\
& & \times e^{-\frac{1}{2}\left(\mathbf{y}_{d}^{\mathrm{exp}}-\mathbf{A}_{d}^{\mathrm{exp}}\mathbf{v}^{f}-\mathbf{v}^{e,d}\right)^{T}\boldsymbol{\gamma}_{m\times m}\left(\mathbf{y}_{d}^{\mathrm{exp}}-\mathbf{A}_{d}^{\mathrm{exp}}\mathbf{v}^{f}-\mathbf{v}^{e,d}\right)}e^{-\frac{1}{2}\left(\mathbf{v}^{f}_{\mathcal{E}_{f}}-\mathbf{v}^{e,f}\right)^{T}\boldsymbol{\gamma}_{n\times n}\left(\mathbf{v}^{f}_{\mathcal{E}_{f}}-\mathbf{v}^{e,f}\right)} \nonumber
\end{eqnarray}
Let us briefly re-introduce the auxiliary fluxes, sorted as $\mathbf{v}^{e} = (\mathbf{v}^{e,d}, \mathbf{v}^{e,f})$, and let us marginalize the (approximated) joint distribution over them, i.e. let us compute $q\left(\mathbf{v}^{f};\mathbf{c}^{e,d}, \mathbf{c}^{e,f},\mathbf{y}\right) = \int d\mathbf{v}^{e} q\left(\mathbf{v}^{f},\mathbf{v}^{e};\mathbf{c}^{e,d}, \mathbf{c}^{e,f},\mathbf{y}\right) $. To do this we introduce
\begin{eqnarray}
\boldsymbol{\Sigma}_{e}^{-1} &	= &\left(\begin{array}{cc}
\boldsymbol{\gamma}_{m\times m} & 0\\
0 & \boldsymbol{\gamma}_{n\times n}
\end{array}\right) \\
\boldsymbol{\mu}_{e} &	= &\boldsymbol{\Sigma}_{e}\left[\begin{array}{c}
\mathbf{c}^{e,d}+\boldsymbol{\gamma}_{m\times m}\left(\mathbf{y}_{d}^{\mathrm{exp}}-\mathbf{A}_{d}^{\mathrm{exp}}\mathbf{v}^{f}\right)\\
\mathbf{c}^{e,f}+\boldsymbol{\gamma}_{n\times n}\mathbf{v}^{f}_{\mathcal{E}^{f}}
\end{array}\right]
\end{eqnarray}
which allows us to write the integrand in standard form. Therefore
\begin{eqnarray}
q\left(\mathbf{v}^{f};\mathbf{c}^{e,d},\mathbf{c}^{e,f}, \mathbf{y}\right) &	\propto &	e^{-\frac{1}{2}\left(\mathbf{v}^{f}-\mathbf{a}^{f}\right)^{T}\mathbf{D}^{f}\left(\mathbf{v}^{f}-\mathbf{a}^{f}\right)-\frac{1}{2}\left(\mathbf{y}-\mathbf{A}\mathbf{v}^{f}-\mathbf{a}^{d}\right)^{T}\mathbf{D}^{d}\left(\mathbf{y}-\mathbf{A}\mathbf{v}^{f}-\mathbf{a}^{d}\right)} \times \\ & & \times e^{-\frac{1}{2}\left(\mathbf{y}_{d}^{\mathrm{exp}}-\mathbf{A}_{d}^{\mathrm{exp}}\mathbf{v}^{f}\right)^{T}\boldsymbol{\gamma}_{m\times m}\left(\mathbf{y}_{d}^{\mathrm{exp}}-\mathbf{A}_{d}^{\mathrm{exp}}\mathbf{v}^{f}\right)-\frac{1}{2}\mathbf{v}_{\mathcal{E}^{f}}^{f^{T}}\boldsymbol{\gamma}_{n\times n}\mathbf{v}_{\mathcal{E}^{f}}^{f}} \times \nonumber \\
& & \times  e^{\frac{1}{2}\boldsymbol{\mu}^{e^{T}}\boldsymbol{\Sigma}^{-1}_{e}\boldsymbol{\mu}^{e}}\int d\mathbf{v}^{e}e^{-\frac{1}{2}\left(\mathbf{v}^{e}-\boldsymbol{\mu}_{e}\right)^{T}\boldsymbol{\Sigma}^{-1}\left(\mathbf{v}^{e}-\boldsymbol{\mu}_{e}\right)} \nonumber
\end{eqnarray}

The results of the Gaussian integration gives us a factor which is independent of $\mathbf{v}^f$ and thus it can be absorbed by the normalization constant. The term $e^{\frac{1}{2}\boldsymbol{\mu}^{e^{T}}\boldsymbol{\Sigma}^{-1}_{e}\boldsymbol{\mu}^{e}}$ allows for a further simplification of the expression which, using the equivalence $\mathbf{v}_{\mathcal{E}_{d}}^{d}=\mathbf{y}_{d}^{\mathrm{exp}}-\mathbf{A}_{d}^{\mathrm{exp}}\mathbf{v}^{f}$, can be nicely stated as
\begin{eqnarray}
q\left(\mathbf{v}^{f};\mathbf{c}^{e,d},\mathbf{c}^{e,f}, \mathbf{y}\right)	\propto	e^{-\frac{1}{2}\left(\mathbf{v}^{f}-\mathbf{a}^{f}\right)^{T}\mathbf{D}^{f}\left(\mathbf{v}^{f}-\mathbf{a}^{f}\right)-\frac{1}{2}\left(\mathbf{y}-\mathbf{A}\mathbf{v}^{f}-\mathbf{a}^{d}\right)^{T}\mathbf{D}^{d}\left(\mathbf{y}-\mathbf{A}\mathbf{v}^{f}-\mathbf{a}^{d}\right)} e^{\mathbf{v}_{\mathcal{E}_{d}}^{d}{}^{T}\mathbf{c}^{e,d}+\mathbf{v}_{\mathcal{E}_{f}}^{f^{T}}\mathbf{c}^{e,f}}
\end{eqnarray}
We remark that the approximate distribution of the constrained fluxes does dependent on the coefficients $\mathbf{c}^{e}$, which, in a statistical mechanics picture, act as external fields on them, independently of $\boldsymbol{\gamma}_{n\times n}$ and $\boldsymbol{\gamma}_{m \times m}$. This is also retrieved in the true distribution in Eq. (\ref{eq:P-marginaliz})
of the main text. \\
Re-expressing the dependent and measured fluxes in terms of the free ones, and noticing that the following equivalences hold $\mathbf{v}_{\mathcal{E}_{f}}^{f^{T}}\mathbf{c}^{e,f} = \mathbf{v}^{f^{T}}\mathbf{c}^{f}$ and $(\mathbf{y}_{d}^{\rm exp}- \mathbf{A}^{\rm exp}_{d}\mathbf{v}^{f})^{T}\mathbf{c}^{d,e} = (\mathbf{y} - \mathbf{A}\mathbf{v}^{f})^{T}\mathbf{c}^{d}$, we can re-formulate the distribution of the constrained fluxes given the coefficients and the constant terms as
\begin{equation}
\label{eq:Full-Gauss-vf}
q\left(\mathbf{v}^{f};\mathbf{c}^{d},\mathbf{c}^{f}, \mathbf{y}\right)\propto e^{-\frac{1}{2}\left(\mathbf{v}^{f}-\boldsymbol{\mu}\right)^{T}\boldsymbol{\Sigma}^{-1}\left(\mathbf{v}^{f}-\boldsymbol{\mu}\right)}
\end{equation}
for
\begin{eqnarray}
\label{eq:par-Full-Gauss-vf}
\begin{cases}
\boldsymbol{\Sigma}^{-1} & =\left(\mathbf{D}^{f}+\mathbf{A}^{T}\mathbf{D}^{d}\mathbf{A}\right)\\
\boldsymbol{\mu} & =\boldsymbol{\Sigma}\left[\mathbf{D}^{f}\mathbf{a}^{f}+\mathbf{A}^{T}\mathbf{D}^{d}\left(\mathbf{y}-\mathbf{a}^{d}\right)+\mathbf{c}^{f}-\mathbf{A}^{T}\mathbf{c}^{d}\right]
\end{cases}
\end{eqnarray}
It is easy to see that the statistics of both the set of constrained and dependent fluxes $(\mathbf{v}^{d}, \mathbf{v}^f)$, according to the Gaussian density, can be computed as
\begin{equation}
\label{eq:Stat-dep-free}
\begin{array}{ccccccccccccc}
\left\langle v_{i}^{f}\right\rangle _{q} & = & \mu_{i} &  &  &  & \left\langle v_{i}^{f^{2}}\right\rangle _{q}-\left\langle v_{i}^{f}\right\rangle _{q}^{2} & = & \Sigma_{ii} &  &  &  & i=1,\ldots,n'\\
\left\langle v_{i}^{d}\right\rangle _{q} & = & \left[-\mathbf{A}\boldsymbol{\mu}+\mathbf{y}\right]_{i}\, &  &  &  & \left\langle v_{i}^{d^{2}}\right\rangle _{q}-\left\langle v_{i}^{d}\right\rangle _{q}^{2} & = & \left[\mathbf{A}\boldsymbol{\Sigma}\mathbf{A}^{T}\right]_{ii}\, &  &  &  & i=1,\ldots,m'
\end{array}
\end{equation}

Let us introduce the so-called `cavity' marginal distribution for flux $v^{f}_{i}$ (or $v^{d}_{i}$) obtained by (i) removing in Eq. (\ref{eq:Gauss-dep-free}) the univariate Gaussian factor associated with the target flux (here we multiply the full Gaussian by its inverse), (ii) marginalizing over the auxiliary fluxes, and (iii) marginalizing over all other fluxes except $v^{f}_{i}$ (or $v^{d}_{i}$), that is
\begin{eqnarray}
\label{eq:cavities1}
q^{\char`\\ i,f}\left(v^{f}_{i} ;\mathbf{c}^{e,d}, \mathbf{c}^{e,f},\mathbf{y}\right) & \propto & \int d\mathbf{v}^{f}_{/ v_{i}^{f}} d\mathbf{v}^{d} \int d\mathbf{v}^{e,d} d\mathbf{v}^{e,f} q\left(\mathbf{v}^{d},\mathbf{v}^{f},\mathbf{v}^{e,f},\mathbf{v}^{e,d};\mathbf{c}^{e,d}, \mathbf{c}^{e,f},\mathbf{y}\right)e^{\frac{\left(v^{f}_{i}-a^{f}_{i}\right)^{2}}{2d^{f}_{i}}} \\
\label{eq:cavities2}
q^{\char`\\ i,d}\left(v^{d}_{i} ;\mathbf{c}^{e,d}, \mathbf{c}^{e,f},\mathbf{y}\right) & \propto &\int d\mathbf{v}^{d}_{/ v_{i}^{d}} d\mathbf{v}^{f} \int d\mathbf{v}^{e,d} d\mathbf{v}^{e,f} q\left(\mathbf{v}^{d},\mathbf{v}^{f},\mathbf{v}^{e,f},\mathbf{v}^{e,d};\mathbf{c}^{e,d}, \mathbf{c}^{e,f},\mathbf{y}\right)e^{\frac{\left(v^{d}_{i}-a^{d}_{i}\right)^{2}}{2d^{d}_{i}}}
\end{eqnarray} 
Performing the Gaussian integration in Eqs. (\ref{eq:cavities1}), (\ref{eq:cavities2}), we get a univariate density for the target flux, which we parametrize by a mean $\mu^{\char`\\ i,\alpha}_{i}$ and a variance $\Sigma^{\char`\\ i,\alpha}_{ii}$ (for $\alpha = {d,f}$ ):
\begin{equation}
q^{\char`\\ i,\alpha}\left(v^{\alpha}_{i} ;\mathbf{c}^{e,d}, \mathbf{c}^{e,f},\mathbf{y}\right) \propto e^{-\frac{(v_{i}^{\alpha} - \mu^{\char`\\ i,\alpha}_{i})^{2}}{2 \Sigma^{\char`\\ i,\alpha}_{ii}}} 
\end{equation}

Let us also introduce the `tilted' distribution for flux $v^{f}_{i}$ (or $v^{d}_{i}$) which is defined as the `cavity' density times the exact prior of the considered flux, i.e.
\begin{eqnarray}
\label{eq:marginal-tilted}
q^{(i,\alpha)}\left(v_{i}^{\alpha};\mathbf{c}^{e,d}, \mathbf{c}^{e,f},\mathbf{y}\right) & \propto & q^{\char`\\ i,\alpha}\left(v^{\alpha}_{i} ;\mathbf{c}^{e,d}, \mathbf{c}^{e,f},\mathbf{y}\right) \psi_{i,\alpha}(v_{i}^{\alpha}) 
\end{eqnarray}

Intuitively, the marginal probability of flux $v_{i}^{f}$ (or $v_{i}^{d}$) is more accurate if computed from the `tilted' distribution than from the full Gaussian as it encodes the exact term of the prior involving the $i^{th}$ free or dependent flux. This is exploited by EP approximation where, in fact, we determine each pair of parameters $\left(a_{i}^{f},d_{i}^{f}\right)$ (or $\left(a_{i}^{d}, d_{i}^{d}\right)$) requiring that the `tilted' distribution is as close as possible to the multivariate Gaussian in Eq. (\ref{eq:Full-Gauss-vf}) (which it can be easily re-phrased in terms of both set of fluxes). Practically one can minimize the Kullback-Leibler distance between the two distributions ~\cite{braunstein17} realizing that this computation leads to the moment matching conditions
\begin{equation}
\label{eq:mom-match}
\begin{cases}
\left\langle v_{i}^{\alpha}\right\rangle _{q^{\left(i,\alpha\right)}} & =\left\langle v_{i}^{\alpha}\right\rangle _{q}\\
\left\langle v_{i}^{{\alpha}^{2}}\right\rangle _{q^{\left(i,\alpha\right)}} & =\left\langle v_{i}^{{\alpha}^{2}}\right\rangle _{q}
\end{cases}\,\alpha = {f,d} 
\end{equation}
to be solved with respect to the unknown parameters of the approximation $(a_{i}^{f}, d_{i}^{f})$ or $(a_{i}^{d}, d_{i}^{d})$.

At each iteration of the algorithm (and for fixed $\mathbf{c}$), we compute the l.h.s. of Eq. (\ref{eq:mom-match}) and we adjust the parameters $\left(a_{i}^{\alpha}, d_{i}^{\alpha}\right)$ for $\alpha = {f, d}$ appearing in the r.h.s. to ensure the two conditions. At the fixed point, we can extract an approximation of the marginal probability densities from the set of the `tilted' distributions and an approximate covariance matrix of the constrained fluxes, $\boldsymbol{\Sigma}$. The expectation values of the tilted densities depend on the single-variable priors. In this context, it turns out that the first and second moments we are interested in are those of a truncated Gaussian:
\begin{eqnarray}
\langle v_{i}^{\alpha}\rangle_{q^{\left(i,\alpha\right)}} &	= &	\mu_{i}^{\backslash i,\alpha}+\frac{\mathcal{\mathcal{N}}\left(\frac{v_{i,\alpha}^{{\rm min}}-\mu_{i}^{\backslash i,\alpha}}{\sqrt{\Sigma_{ii}^{\backslash i,\alpha}}}\right)-\mathcal{N}\left(\frac{v_{i,\alpha}^{{\rm max}}-\mu_{i}^{\backslash i,\alpha}}{\sqrt{\Sigma_{ii}^{\backslash i,\alpha}}}\right)}{\Phi\left(\frac{v_{i}^{{\rm max}}-\mu_{i}^{\backslash i,\alpha}}{\sqrt{\Sigma_{ii}^{\backslash i,\alpha}}}\right)-\Phi\left(\frac{v_{i}^{{\rm min}}-\mu_{i}^{\backslash i,\alpha}}{\sqrt{\Sigma_{ii}^{\backslash i,\alpha}}}\right)}\sqrt{\Sigma_{ii}^{\backslash i,\alpha}} \\
\langle v_{i}^{\alpha^{2}}\rangle_{q^{\left(i,\alpha\right)}}-\langle v_{i}^{\alpha}\rangle_{q^{\left(i,\alpha\right)}}^{2} &	= &	\Sigma_{ii}^{\backslash i,\alpha}\left[1+\frac{\frac{v_{i,\alpha}^{{\rm min}}-\mu_{i}^{\backslash i,\alpha}}{\sqrt{\Sigma_{ii}^{\backslash i,\alpha}}}\mathcal{N}\left(\frac{v_{i,\alpha}^{{\rm min}}-\mu_{i}^{\backslash i,\alpha}}{\sqrt{\Sigma_{ii}^{\backslash i,\alpha}}}\right)-\frac{v_{i,\alpha}^{{\rm max}}-\mu_{i}^{\backslash i,\alpha}}{\sqrt{\Sigma_{ii}^{\backslash i,\alpha}}}\mathcal{N}\left(\frac{v_{i,\alpha}^{{\rm max}}-\mu_{i}^{\backslash i,\alpha}}{\sqrt{\Sigma_{ii}^{\backslash i,\alpha}}}\right)}{\Phi\left(\frac{v_{i,\alpha}^{{\rm max}}-\mu_{i}^{\backslash i,\alpha}}{\sqrt{\Sigma_{ii}^{\backslash i,\alpha}}}\right)-\Phi\left(\frac{v_{i}^{{\rm min}}-\mu_{i}^{\backslash i,\alpha}}{\sqrt{\Sigma_{ii}^{\backslash i,\alpha}}}\right)}+\right. \\ & &\left.-\left(\frac{\mathcal{N}\left(\frac{v_{i}^{{\rm min}}-\mu_{i}^{\backslash i,\alpha}}{\sqrt{\Sigma_{ii}^{\backslash i,\alpha}}}\right)-\mathcal{N}\left(\frac{v_{i,\alpha}^{{\rm max}}-\mu_{i}^{\backslash i,\alpha}}{\sqrt{\Sigma_{ii}^{\backslash i,\alpha}}}\right)}{\Phi\left(\frac{v_{i,\alpha}^{{\rm max}}-\mu_{i}^{\backslash i,\alpha}}{\sqrt{\Sigma_{ii}^{\backslash i,\alpha}}}\right)-\Phi\left(\frac{v_{i}^{{\rm min}}-\mu_{i}^{\backslash i,\alpha}}{\sqrt{\Sigma_{ii}^{\backslash i,\alpha}}}\right)}\right)^{2}\right] \nonumber
\end{eqnarray}

We remark that, apparently, the update scheme requires to compute, at each iteration, the statistics of all the tilted distributions, and hence of all the possible cavity distributions. Fortunately, the expressions of the cavity parameters can be directly computed from the parameters of the full Gaussian density, i.e. by marginalization, reducing the computational time of a factor $N$ (see ~\cite{braunstein17} for more details). Besides, when dealing with the set of dependent and free variables, the statistics of the dependent set is retrieved from those of the free one, according to Eq. (\ref{eq:Stat-dep-free}) as already noticed in ~\cite{braunstein20}. The running time is therefore dominated by one matrix inversion per iteration aimed at computing the covariance matrix of the free fluxes, which scales as $O(n'^{3})$.
We report here the expression for the means and variances that ensure the moment matching condition in Eq. (\ref{eq:mom-match}) as a function of the cavity statistics, i.e.
\begin{equation}
\begin{cases}
d_{i}^{\alpha} = \left(\frac{1}{\left\langle v_{i}^{{\alpha}^{2}}\right\rangle _{q^{\left(i,\alpha\right)}} -\left\langle v_{i}^{\alpha}\right\rangle ^{2}_{q^{\left(i,\alpha\right)}}} - \frac{1}{\Sigma_{ii}^{\char`\\ i,\alpha}}\right)^{-1} \\
a_{i}^{\alpha} = d_{i}^{\alpha} \left[\left\langle v_{i}^{\alpha}\right\rangle _{q^{\left(i,\alpha\right)}} \left(\frac{1}{d_{i}^{\alpha}}+\frac{1}{\Sigma_{ii}^{\char`\\ i,\alpha}}\right) -\frac{\mu_{i}^{\char`\\ i,\alpha}}{\Sigma_{ii}^{\char`\\ i, \alpha}}\right]
\end{cases}
\end{equation}
where the latter are computed from the full Gaussian distribution (once the free and dependent fluxes are identified):
\begin{equation}
\begin{array}{ccc}
\begin{cases}
\Sigma_{ii}^{\char`\\ i,f}= & \frac{\Sigma_{ii}}{1-\Sigma_{ii}\frac{1}{d_{i}^{f}}}\\
\mu_{i}^{\char`\\ i,f}= & \frac{\mu_{i}-\frac{a_{i}^{f}\Sigma_{ii}}{d_{i}^{f}}}{1-\Sigma_{ii}\frac{1}{d_{i}^{f}}}
\end{cases} & \qquad & \begin{cases}
\Sigma_{ii}^{\backslash i,d}= & \frac{\left[\mathbf{A}\boldsymbol{\Sigma}\mathbf{A}^{T}\right]_{ii}}{1-\left[\mathbf{A}\boldsymbol{\Sigma}\mathbf{A}^{T}\right]_{ii}\frac{1}{d_{i}^{d}}}\\
\mu_{i}^{\backslash i,d}= & \frac{\left[-\mathbf{A}\boldsymbol{\mu}+\mathbf{y}\right]_{i}-\frac{a_{i}^{d}\left[\mathbf{A}\boldsymbol{\Sigma}\mathbf{A}^{T}\right]_{ii}}{d_{i}^{d}}}{1-\left[\mathbf{A}\boldsymbol{\Sigma}\mathbf{A}^{T}\right]_{ii}\frac{1}{d_{i}^{d}}}
\end{cases}\end{array}
\end{equation}

\subsubsection{Determining $\mathbf{c}$ and $\gamma$, given $\mathbf{a}$ and $\mathbf{d}$}

At difference with the parameters $(\mathbf{a}, \mathbf{d})$ which are associated with the approximation of the constrained fluxes distribution, the Lagrange multipliers $(\mathbf{c}, \boldsymbol{\gamma})$ must be set to ensure that the means and the variances (or an average variance for the unique $\gamma$ case) of the distribution of the auxiliary fluxes $\mathbf{v}^{e}$, match the empirical observations. 
Hence, considering the sub-division into free and dependent fluxes, we are interested in $q\left(\mathbf{v}^{e,d},\mathbf{v}^{e,f} ; \mathbf{c}^{e,d}, \mathbf{c}^{e,f}, \mathbf{y} \right)$ which can be computed marginalizing the joint distribution in  Eq. (\ref{eq:Gauss-free-exp})
with respect to the free fluxes belonging to the polytope, i.e.
\begin{eqnarray}
\label{eq:Gauss-aux-marg}
q\left(\mathbf{v}^{e,d},\mathbf{v}^{e,f} ; \mathbf{c}^{e,d}, \mathbf{c}^{e,f}, \mathbf{y} \right) \propto  e^{\mathbf{c}^{e,d^{T}}\cdot\mathbf{v}^{e,d}+\mathbf{c}^{e,f^{T}}\cdot\mathbf{v}^{e,f}}\int d\mathbf{v}^{f}e^{-\frac{1}{2}\left(\mathbf{v}^{f}-\mathbf{a}^{f}\right)^{T}\mathbf{D}^{f}\left(\mathbf{v}^{f}-\mathbf{a}^{f}\right)-\frac{1}{2}\left(\mathbf{y}-\mathbf{A}\mathbf{v}^{f}-\mathbf{a}^{d}\right)^{T}\mathbf{D}^{d}\left(\mathbf{y}-\mathbf{A}\mathbf{v}^{f}-\mathbf{a}^{d}\right)}\times\\\times e^{-\frac{1}{2}\left(\mathbf{y}_{d}^{\mathrm{exp}}-\mathbf{A}_{d}^{\mathrm{exp}}\mathbf{v}^{f}-\mathbf{v}^{e,d}\right)^{T}\boldsymbol{\gamma}_{m\times m}\left(\mathbf{y}_{d}^{\mathrm{exp}}-\mathbf{A}_{d}^{\mathrm{exp}}\mathbf{v}^{f}-\mathbf{v}^{e,d}\right)}e^{-\frac{1}{2}\left(\mathbf{v}_{\mathcal{E}_{f}}^{f}-\mathbf{v}^{e,f}\right)^{T}\boldsymbol{\gamma}_{n\times n}\left(\mathbf{v}_{\mathcal{E}_{f}}^{f}-\mathbf{v}^{e,f}\right)} \nonumber
\end{eqnarray}
For sake of simplicity, let us introduce the `extended' matrix $\boldsymbol{\gamma}_{n'\times n'}$, a diagonal matrix having non-zeros entries in the indices contained in $\mathcal{E}_{f}$, i.e.
\begin{equation}
\left[\boldsymbol{\gamma}_{n'\times n'}\right]_{j,j}	=	\begin{cases}
\gamma_{i} & \mathrm{for\ }j=\mathcal{E}_{f}\left(i\right),\,i=1,...,n\\
0 & \mathrm{otherwise}
\end{cases}
\end{equation}
and an `extended' vector $\mathbf{t}\in\mathbb{R}^{n'}$ which collects the auxiliary free fluxes (in the indices  $\mathcal{E}_{f}$) and taking zero values in correspondence to non-observed free fluxes:
\begin{equation}
t_{j}=\begin{cases}
\nu_{i}^{e,f} & \mathrm{for\ }j=\mathcal{E}_{f}\left(i\right),\,i=1,...,n\\
0 & \mathrm{otherwise}.
\end{cases}
\end{equation}
As a consequence, the following equivalence holds:
\begin{eqnarray}
\label{eq:Eq-t-gammap}
-\frac{1}{2}\left(\mathbf{v}_{\mathcal{E}_{f}}^{f}-\mathbf{v}^{e,f}\right)^{T}\boldsymbol{\gamma}_{n\times n}\left(\mathbf{v}_{\mathcal{E}_{f}}^{f}-\mathbf{v}^{e,f}\right) & = & -\frac{1}{2}\mathbf{v}^{f^{T}}\boldsymbol{\gamma}_{n'\times n'}\mathbf{v}^{f}+\mathbf{v}^{f^{T}}\boldsymbol{\gamma}_{n'\times n'}\mathbf{t}-\frac{1}{2}\mathbf{v}^{e,f^{T}}\boldsymbol{\gamma}_{n\times n}\mathbf{v}^{e,f}
\end{eqnarray}
Therefore, we can re-phrase Eq. (\ref{eq:Gauss-aux-marg}), writing the integrand in standard form, as
\begin{eqnarray}
\label{eq:Gauss-aux-marg-std}
q\left(\mathbf{v}^{e,d},\mathbf{v}^{e,f} ; \mathbf{c}^{e,d}, \mathbf{c}^{e,f}, \mathbf{y} \right) & \propto & e^{\mathbf{c}^{e,d^{T}}\cdot\mathbf{v}^{e,d}+\mathbf{c}^{e,f^{T}}\cdot\mathbf{v}^{e,f}}e^{-\frac{1}{2}\left(\mathbf{y}_{d}^{\mathrm{exp}}-\mathbf{v}^{e,d}\right)^{T}\boldsymbol{\gamma}_{m\times m}\left(\mathbf{y}_{d}^{\mathrm{exp}}-\mathbf{v}^{e,d}\right)} e^{-\frac{1}{2}\mathbf{v}^{e,f}{}^{T}\boldsymbol{\gamma}_{n\times n}\mathbf{v}^{e,f}} \times \\ & & \times e^{\frac{1}{2}\mathbf{h}^{T}\boldsymbol{\varLambda}^{-1}\mathbf{h}}  \int d\mathbf{v}^{f}e^{-\frac{1}{2}\left(\mathbf{v}^{f}-\mathbf{h}\right)^{T}\boldsymbol{\varLambda}^{-1}\left(\mathbf{v}^{f}-\mathbf{h}\right)} \nonumber
\end{eqnarray}
where
\begin{equation}
\label{eq:varL-h}
\begin{cases}
\boldsymbol{\varLambda}^{-1}= & \mathbf{D}^{f}+\mathbf{A}^{T}\mathbf{D}^{d}\mathbf{A}+\mathbf{A}_{d}^{\mathrm{exp^{T}}}\boldsymbol{\gamma}_{m\times m}\mathbf{A}_{d}^{\mathrm{exp}}+\boldsymbol{\gamma}_{n'\times n'}\\
\mathbf{h}= & \boldsymbol{\varLambda}\left[\mathbf{D}^{f}\mathbf{a}^{f}+\mathbf{A}^{T}\mathbf{D}^{d}\left(\mathbf{y}-\mathbf{a}^{d}\right)+\mathbf{A}_{d}^{\mathrm{exp^{T}}}\boldsymbol{\gamma}_{m\times m}\left(\mathbf{y}_{d}^{\mathrm{exp}}-\mathbf{v}^{e,d}\right)+\boldsymbol{\gamma}_{n'\times n'}\mathbf{t}\right]
\end{cases}
\end{equation}
Notice that the integration in Eq. (\ref{eq:Gauss-aux-marg-std}) returns a constant term which can be absorbed by the normalization constant, while the exponential term $e^{\frac{1}{2}\mathbf{h}^{T}\boldsymbol{\varLambda}^{-1}\mathbf{h}} $ depends on $(\mathbf{v}^{e,d}, \mathbf{v}^{e,f})$. Developing further the argument of the exponential, and recalling term-by-term the equivalence in Eq. (\ref{eq:Eq-t-gammap}), allows us to formulate the distribution of the free and dependent auxiliary fluxes as a multivariate Gaussian density
\begin{equation}
\label{Gauss-std-aux}
q\left(\mathbf{v}^{e,d},\mathbf{v}^{e,f};\mathbf{c}^{e,d},\mathbf{c}^{e,f}, \mathbf{y}\right)	\propto	e^{-\frac{1}{2}\left(\mathbf{v}^{e,d},\mathbf{v}^{e,f}\right)\mathbf{M}^{-1}\left(\begin{array}{c}
\mathbf{v}^{e,d}\\
\mathbf{v}^{e,f}
\end{array}\right)+\mathbf{v}^{e,d^{T}}\cdot\left(\mathbf{c}^{e,d}+\text{\ensuremath{\mathbf{u}^{e,d}}}\right)+\mathbf{v}^{e,f^{T}}\cdot\left(\mathbf{c}^{e,f}+\mathbf{u}^{e,f}\right)}
\end{equation}
where
\begin{eqnarray}
\mathbf{M}^{-1} & = &	\left(\begin{array}{cc}
\mathbf{U} & \mathbf{V}\\
\mathbf{V}^{T} & \mathbf{Y}
\end{array}\right) \\
\mathbf{U}&	= &	\boldsymbol{\gamma}_{m\times m}-\boldsymbol{\gamma}_{m\times m}\mathbf{A}_{d}^{\mathrm{exp}}\boldsymbol{\varLambda}\mathbf{A}_{d}^{\mathrm{exp^{T}}}\boldsymbol{\gamma}_{m\times m} \\
\mathbf{V}	& = &	\boldsymbol{\gamma}_{m\times m}\mathbf{A}_{d}^{\mathrm{exp}}[\boldsymbol{\varLambda}]_{1:n',\mathcal{E}_{f}^{\mathrm{}}}\boldsymbol{\gamma}_{n\times n} \\
\mathbf{Y} &	=	& \boldsymbol{\gamma}_{n\times n}-\boldsymbol{\gamma}_{n\times n}[\boldsymbol{\varLambda}]_{\mathcal{E}_{f}^{\mathrm{}},\mathcal{E}_{f}^{\mathrm{}}}\boldsymbol{\gamma}_{n\times n} \\
\mathbf{u}^{e,d} &	= &	\boldsymbol{\gamma}_{m\times m}\mathbf{y}_{d}^{\mathrm{exp}}-\boldsymbol{\gamma}_{m\times m}\mathbf{A}_{d}^{\mathrm{exp}}\boldsymbol{\varLambda}\left[\mathbf{D}^{f}\mathbf{a}^{f}+\mathbf{A}^{T}\mathbf{D}^{d}\left(\mathbf{y}-\mathbf{a}^{d}\right)+\mathbf{A}_{d}^{\mathrm{exp^{T}}}\boldsymbol{\gamma}_{m\times m}\mathbf{y}_{d}^{\mathrm{exp}}\right] \\
\mathbf{u}^{e,f} &	= &	\boldsymbol{\gamma}_{n\times n}[\boldsymbol{\varLambda}]_{\mathcal{E}_{f}^{\mathrm{}},1:n'}\left[\mathbf{D}^{f}\mathbf{a}^{f}+\mathbf{A}^{T}\mathbf{D}^{d}\left(\mathbf{y}-\mathbf{a}^{d}\right)+\mathbf{A}_{d}^{\mathrm{exp^{T}}}\boldsymbol{\gamma}_{m\times m}\mathbf{y}_{d}^{\mathrm{exp}}\right]
\end{eqnarray}
The symbol $[\boldsymbol{\varLambda}]_{x,y}$ denotes a sub-matrix of $\boldsymbol{\varLambda}$ (whose inverse is defined in Eq. (\ref{eq:varL-h})) in which we consider the rows associated with indices $x$ and the columns associated with indices $y$. \\
The constraints on the averages allows us to fix the values of the coefficients $\mathbf{c}$. More precisely, since the expectation values of $\left(\mathbf{v}^{e,d},\mathbf{v}^{e,f}\right)$, with respect to the distribution in Eq. (\ref{Gauss-std-aux}), must coincide with the experimental ones (properly divided in those associated with dependent and independent fluxes) $\left(\mathbf{v}^{\mathrm{xp,d}},\mathbf{v}^{\mathrm{xp,f}}\right)$,
\begin{equation}
\mathbf{M}\left(\begin{array}{c}
\mathbf{c}^{e,d}+\mathbf{u}^{e,d}\\
\mathbf{c}^{e,f}+\mathbf{u}^{e,f}
\end{array}\right)	=	\left(\begin{array}{c}
\mathbf{v}^{\mathrm{xp,d}}\\
\mathbf{v}^{\mathrm{xp,f}}
\end{array}\right)
\end{equation}
we can obtain, inverting the relation, an update equation for the coefficients $\mathbf{c}$, that is
\begin{equation}
\left(\begin{array}{c}
\mathbf{c}^{e,d}\\
\mathbf{c}^{e,f}
\end{array}\right)	=	\mathbf{M}^{-1}\left(\begin{array}{c}
\mathbf{v}^{\mathrm{xp,d}}\\
\mathbf{v}^{\mathrm{xp,f}}
\end{array}\right)-\left(\begin{array}{c}
\mathbf{u}^{e,d}\\
\mathbf{u}^{e,f}
\end{array}\right)
\end{equation}
Finally, we can proceed, using the same argument, with the determination of an update equation for $\boldsymbol{\gamma}_{n\times n}$ and $\boldsymbol{\gamma}_{m\times m}$. In this case, we seek the values of the Lagrange multipliers ensuring that the diagonal terms of the covariance matrix
\begin{equation}
\mathbf{M}	=	\left(\begin{array}{cc}
\mathbf{U} & \mathbf{V}\\
\mathbf{V}^{T} & \mathbf{Y}
\end{array}\right)^{-1}
\end{equation}
are equal to the experimental variances. In the following we treat the most general case but we recall that, when one value of $\gamma$ is sought, it is sufficient to require that the sum of the diagonal elements of $\mathbf{M}$ coincides with the sum of the experimental variances (being the mean obtained by dividing by the same number, that is the number of experimentally measured fluxes). \\
Notice that, as in this case, if the matrix to be inverted can be sub-divided in four blocks (in which those in the diagonal are squared), this property holds  ~\cite{bernstein09}
\begin{equation}
\mathbf{M}	=	\left(\begin{array}{cc}
\left(\mathbf{U}-\mathbf{V}\mathbf{Y}^{-1}\mathbf{V}^{T}\right)^{-1} & ...\\
... & \left(\mathbf{Y}-\mathbf{V}^{T}\mathbf{U}^{-1}\mathbf{V}\right)^{-1}
\end{array}\right)
\end{equation}
Therefore, to get the update equations we are seeking, it is sufficient to compute the diagonal blocks of $\mathbf{M}$ and to fix their diagonal elements to the experimental variances.
For sake of simplicity let us define and develop
\begin{eqnarray}
\mathbf{W}&=&\left(\mathbf{U}-\mathbf{V}\mathbf{Y}^{-1}\mathbf{V}^{T}\right)^{-1} \\
& = & \left[\mathbb{I}_{m\times m}-\mathbf{A}_{d}^{\mathrm{exp}}\boldsymbol{\varLambda}\mathbf{A}_{d}^{\mathrm{exp^{T}}}\boldsymbol{\gamma}_{m\times m}-\mathbf{A}_{d}^{\mathrm{exp}}[\boldsymbol{\varLambda}]_{1:n',\mathcal{E}_{f}}\boldsymbol{\gamma}_{n\times n}\left(\boldsymbol{\gamma}_{n\times n}-\boldsymbol{\gamma}_{n\times n}[\boldsymbol{\varLambda}]_{\mathcal{E}_{f},\mathcal{E}_{f}}\boldsymbol{\gamma}_{n\times n}\right)^{-1}\times\right. \nonumber \\ 
& & \times \left.\boldsymbol{\gamma}_{n\times n}[\boldsymbol{\varLambda}]_{\mathcal{E}_{f},1:n'}\mathbf{A}_{d}^{\mathrm{exp^{T}}}\boldsymbol{\gamma}_{m\times m}\right]^{-1}\boldsymbol{\gamma}_{m\times m}^{-1} \nonumber \\
\mathbf{T}&=&\left(\mathbf{Y}-\mathbf{V}^{T}\mathbf{U}^{-1}\mathbf{V}\right)^{-1} \\
 & = & \left[\mathbb{I}_{n\times n}-[\boldsymbol{\varLambda}]_{\mathcal{E}_{f}^{\mathrm{exp}},\mathcal{E}_{f}^{\mathrm{exp}}}\boldsymbol{\gamma}_{n\times n}-[\boldsymbol{\varLambda}]_{\mathcal{E}_{f},1:n'}\mathbf{A}_{d}^{\mathrm{exp^{T}}}\boldsymbol{\gamma}_{m\times m}\left(\boldsymbol{\gamma}_{m\times m}-\boldsymbol{\gamma}_{m\times m}\mathbf{A}_{d}^{\mathrm{exp}}\boldsymbol{\varLambda}\mathbf{A}_{d}^{\mathrm{exp^{T}}}\boldsymbol{\gamma}_{m\times m}\right)^{-1}\times\right. \nonumber \\
 & & \left.\times\boldsymbol{\gamma}_{m\times m}\mathbf{A}_{d}^{\mathrm{exp}}[\boldsymbol{\varLambda}]_{1:n',\mathcal{E}_{f}}\boldsymbol{\gamma}_{n\times n}\right]^{-1}\boldsymbol{\gamma}_{n\times n}^{-1} \nonumber
\end{eqnarray}
Hence, imposing
\begin{eqnarray}
\mathbf{W}_{i,i} &	= &	\sigma_{i}^{\mathrm{xp,d}}\qquad i=1,\ldots,m \\
\mathbf{T}_{i,i} &	= &	\sigma_{i}^{\mathrm{xp,f}}\qquad i=1,\ldots,n, \nonumber
\end{eqnarray}
where $(\boldsymbol{\sigma}^{\rm xp,d}, \boldsymbol{\sigma}^{\rm xp, f})$ are the experimental error associated with the corresponding dependent and free fluxes, we obtain
\begin{eqnarray}
\gamma_{n\times n}^{\left(i,i\right)} &	= & 	\frac{\left[\left[\mathbb{I}_{n\times n}-\boldsymbol{\varLambda}_{\mathcal{E}_{f}^{\mathrm{}},\mathcal{E}_{f}^{\mathrm{}}}\boldsymbol{\gamma}_{n\times n}-\boldsymbol{\varLambda}_{\mathcal{E}_{f}^{\mathrm{}},1:n'}\mathbf{A}_{d}^{\mathrm{exp^{T}}}\boldsymbol{\gamma}_{m\times m}\left(\boldsymbol{\gamma}_{m\times m}-\boldsymbol{\gamma}_{m\times m}\mathbf{A}_{d}^{\mathrm{exp}}\boldsymbol{\varLambda}\mathbf{A}_{d}^{\mathrm{exp^{T}}}\boldsymbol{\gamma}_{m\times m}\right)^{-1}\boldsymbol{\gamma}_{m\times m}\mathbf{A}_{d}^{\mathrm{exp}}\boldsymbol{\varLambda}_{1:n',\mathcal{E}_{f}^{\mathrm{}}}\boldsymbol{\gamma}_{n\times n}\right]^{-1}\right]_{i,i}}{\sigma_{i}^{\mathrm{xp,f}}} \nonumber \\
\gamma_{m\times m}^{\left(i,i\right)} &	= &	\frac{\left[\left[\mathbb{I}_{m\times m}-\mathbf{A}_{d}^{\mathrm{exp}}\boldsymbol{\varLambda}\mathbf{A}_{d}^{\mathrm{exp^{T}}}\boldsymbol{\gamma}_{m\times m}-\mathbf{A}_{d}^{\mathrm{exp}}\boldsymbol{\varLambda}_{1:n',\mathcal{E}_{f}^{\mathrm{}}}\boldsymbol{\gamma}_{n\times n}\left(\boldsymbol{\gamma}_{n\times n}-\boldsymbol{\gamma}_{n\times n}\boldsymbol{\varLambda}_{\mathcal{E}_{f}^{\mathrm{}},\mathcal{E}_{f}^{\mathrm{}}}\boldsymbol{\gamma}_{n\times n}\right)^{-1}\boldsymbol{\gamma}_{n\times n}\boldsymbol{\varLambda}_{\mathcal{E}_{f}^{\mathrm{}},1:n'}\mathbf{A}_{d}^{\mathrm{exp^{T}}}\boldsymbol{\gamma}_{m\times m}\right]^{-1}\right]_{i,i}}{\sigma_{i}^{\mathrm{xp,d}}} \nonumber
\end{eqnarray}

\subsubsection{Implementation details}
\label{sec:impl}
As already explored in ~\cite{braunstein17}, since the values taken by the fluxes may span several orders of magnitude, it is important to work with normalized fluxes within the EP approximation scheme, in particular to avoid the approximate variances $\mathbf{d}$ to take very heterogeneous values, ranging from zero to infinite. Therefore, we first compute a constant factor 
\begin{equation}
f = \max \left[ \max_{i} |v_{i}^{min}|, \max_{i} |v_{i}^{max}| \right] 
\end{equation}
and then we modify the lower and upper bounds of variability, together with the constant term of the stoichiometric equations, as
\begin{equation}
\mathbf{v}^{max}  \leftarrow  \frac{\mathbf{v}^{max}}{f}, \quad \mathbf{v}^{min} \leftarrow  \frac{\mathbf{v}^{min}}{f},  \quad \mathbf{b}  \leftarrow \frac{\mathbf{b}}{f} .
\end{equation}

Once a fixed point of the EP equations is reached, we can recover the average value of the fluxes in the proper range of variability, as well as the coefficients $\mathbf{c}$, by multiplying, and dividing respectively, the constant factor $f$. \\
We empirically found that the scheme with a unique $\gamma$ is preferable to the multiple $\gamma$-s scheme: the experimental errors often span several orders of magnitude which is then reflected in a very heterogeneous set of $\gamma$ Lagrange multipliers that, however, do not carry significant improvements in the fitting of the variance (with respect to the single $\gamma$ case). Furthermore, conversely to the unique $\gamma$ case, when multiple $\gamma$-s are inferred, the coefficients $\mathbf{c}$ vary a lot within the experiments rendering the aggregated analysis more difficult to interpret. \\
From an algorithmic point of view, the update of the parameters of the Gaussian approximation ($\mathbf{a}$, $\mathbf{d}$) and of the Lagrange multipliers ($\mathbf{c}$ and $\gamma$) is not performed synchronously: at each iteration of the main scheme, we update $(\mathbf{a}, \mathbf{d})$ and $\mathbf{c}$ while $\gamma$ is modified whenever the experimental averages are fitted. We then repeat the overall scheme up to the convergence associated with the $\gamma$ parameter, that is when the average experimental variance is fitted by the our model.
We monitor the convergence of the fixed point equations (for the normalized fluxes), at each iteration $t$, by computing
\begin{eqnarray}
\varepsilon_{\rm EP}^{t} & = & \max\left[ \max_{i}\varepsilon^{\rm mean}_{i}(t),\,\max_{i}\varepsilon^{\rm var}_{i}(t)\right] \\
\varepsilon_{\mathbf{c}}^{t} & = & \frac{1}{E} \sum_{i} | \langle  v^{e}_{i} \rangle_{{q}_{t}} - v^{\rm xp}_{i} | \\
\varepsilon_{\gamma}^{t} & = & | \frac{1}{E} \sum_{i} \langle v_{i}^{{e}^{2}} \rangle_{{q}_{t}} - \langle v_{i}^{e} \rangle_{{q}_{t}}^{2} - \frac{1}{E} \sum_{i} \sigma_{i}^{{\rm xp}^{2}} |
\end{eqnarray}
where $q_{t}$ is the approximated Gaussian distribution (we are omitting here the sub-division in free and dependent fluxes) at iteration $t$, and
\begin{eqnarray}
\varepsilon^{\rm mean}_{i}(t) & = & | \langle v_{i} \rangle_{{q}_{t}^{(i)}} - \langle v_{i} \rangle_{{q}_{t-1}^{(i)}} | \\
\varepsilon^{\rm var}_{i}(t) & = & |\langle v_{i}^{2} \rangle_{{q}_{t}^{(i)}} - \langle v_{i} \rangle^{2}_{{q}_{t}^{(i)}} - \langle v_{i}^{2} \rangle_{{q}_{t-1}^{(i)}} + \langle v_{i} \rangle^{2}_{{q}_{t-1}^{(i)}} |
\end{eqnarray}
being $q_{t}^{(i)}$ the tilted distribution of the $i^{\rm th}$ flux at time $t$.
We use as convergence tolerance $10^{-4}$, $10^{-5}$ for $\varepsilon_{\rm EP}$ and $\varepsilon_{\mathbf{c}}$, while for $\varepsilon_{\gamma}$ we use  $2.5\cdot10^{-5}$ for the experiments shown in ~\cite{nanchen06} and $9\cdot10^{-6}$ for those taken from ~\cite{schuetz12}.
These values are chosen to guarantee the convergence of the algorithm in each experiment presented in the two works (we notice that in few experiments a smaller threshold leads to divergent $\gamma$ which makes the computation of the covariance matrix $\mathbf{M}$ unfeasible), and also to allow for a reasonable fitting of the variances.
We show in Section \ref{sec:means} a set of plots presenting the average values of the constrained and of the noisy fluxes compared to the measured ones.

\subsection{Projections of the coefficients along individual flux directions}

The vector $\mathbf{c}$ of Lagrange multipliers appearing e.g. in Eq. (\ref{dist}) of the main test is defined in $\mathbb{R}^N$ with $N$ the number of fluxes. As fluxes lie in the feasible space $\mathcal{F}$, though, for the analysis one should get rid of components orthogonal to $\mathcal{F}$ by projecting $\mathbf{c}$ onto the feasible space. To be precise, the relevant information stored in $\mathbf{c}$ is encoded in the vector
\begin{equation}
    \mathbf{c}^{\mathrm{proj}}\equiv\hat{O}\mathbf{c}\in\mathbb{R}^{\mathrm{dim}(\mathcal{F})} 
    ~~,
    \label{eq:proj}
\end{equation} 
where $\hat{O}\in\mathbb{R}^{\mathrm{dim}(\mathcal{F})\times N}$ is an arbitrary orthonormal basis spanning $\mathcal{F}$. Such an operator is obtained by first performing a Gauss-Jordan decomposition of the stoichiometric matrix $\mathbf{S}$ and then by applying the Gram-Schmidt orthonormalization method. 
In Figure \ref{fig:ScProd} we show the value of the scalar product between the projected coefficients, for each inferred vector of coefficient $\mathbf{c}$, and the directions of each individual flux.
A value of the projection close to zero indicates that the mean of the corresponding flux is close to the mean flux pertaining to a homogeneous sampling of $\mathcal{F}$, whereas positive (resp. negative) values indicate that the flux is upregulated (resp. downregulated) compared to the unbiased mean. The color code in Figure \ref{fig:ScProd} mirrors the dilution rate used in each experiments while the shape of the points refer to the presence/absence of acetate and the database used. Flux names in the x-axis display different colors according to the sub-division driven by pathways.

\subsection{Dimensionality reduction}

We report here the details of the calculations associated with the Principal Component Analysis and additional results to complete the picture shown in Fig. \ref{four} of the main text. For each experiment $a$ we first project the inferred Lagrange multipliers to the feasible space $\mathcal{F}$ as in Eq. (\ref{eq:proj}) of the previous section and we divide each element by the norm of $\boldsymbol{c}^{\rm proj}_{a}$. Then, we compute a vector of means $\mathbf{m}\in\mathbb{R}^{\rm{dim}{\left(\mathcal{F} \right)}}$, averaged over the experiments, having elements
\begin{equation}
m_{i}=\frac{1}{E}\sum_{a}\left[\frac{\mathbf{c}^{\rm proj}_{a}}{||\mathbf{c}^{\rm proj}_{a}||}\right]_{i}
\end{equation}
used to build the standardized matrix of coefficients $\mathcal{C}\in\mathbb{R}^{\rm{dim}\left(\mathcal{F}\right)\times E}$ of entries
\begin{equation}
\mathcal{C}_{i,a} = \left[\frac{\mathbf{c}^{\rm proj}_{a}}{||\mathbf{c}^{\rm proj}_{a}||}\right]_{i} - m_{i}
\end{equation}
In a way, the columns of $\mathcal{C}$ collect the observations of all the components $i=1,\ldots,\rm{dim}\left(\mathcal{F}\right)$ of the projected coefficient $c_{i}^{\rm proj}$. Once the covariance matrix of the data $\mathcal{C}$ has been computed, we can calculate its eigenvalues and the associated eigenvectors. Let us sort the eigenvalues in descending order, and define a matrix $\mathbf{Q}\in\mathbb{R}^{\rm dim \left(\mathcal{F}\right)\times{\rm dim}\left(\mathcal{F}\right)}$ such that the $k$th column of $\mathbf{Q}$ contains the eigenvector associated with $k$th largest eigenvalue. Let us define $\mathbf{Q}_{1:\rm{dim}\left(\mathcal{F}\right),1}$ as the first principal component (PC1) and $\mathbf{Q}_{1:\rm{dim}\left(\mathcal{F}\right),2}$ the second principal component (PC2).
The projection of the original data $\mathcal{C}$ onto the bidimensional space spanned by (PC1, PC2) is reported in Figure \ref{fig:PCA}a where each point here represents an experiment. One sees that experiments organize in two clusters characterized by distinct acetate excretion profiles (note that PCA correctly separates the experiments from \cite{schuetz12} in two clusters). The first principal component projects the largest value on the acetate flux (ca. -0.74), confirming that carbon overflow is a key separator of phenotypic behaviour in carbon-limited {\it E. coli} growth. On the other hand, PC2 has the largest scalar product (ca. 0.50 in absolute value) against fluxes through reactions involved in acetate metabolism (e.g. isocitrate dehydrogenase ({\it icd}), 
phosphate acetyltransferase ({\it pta}) and acetate kinase ({\it ackA/B})) and the glyoxylate cycle (isocitrate lyase ({\it aceA/B})). A heatmap of the projection of the full set of eigenvectors is shown in Figure \ref{fig:PCA}b. \\

\begin{figure}
	\centering
	\includegraphics[width=\textwidth]{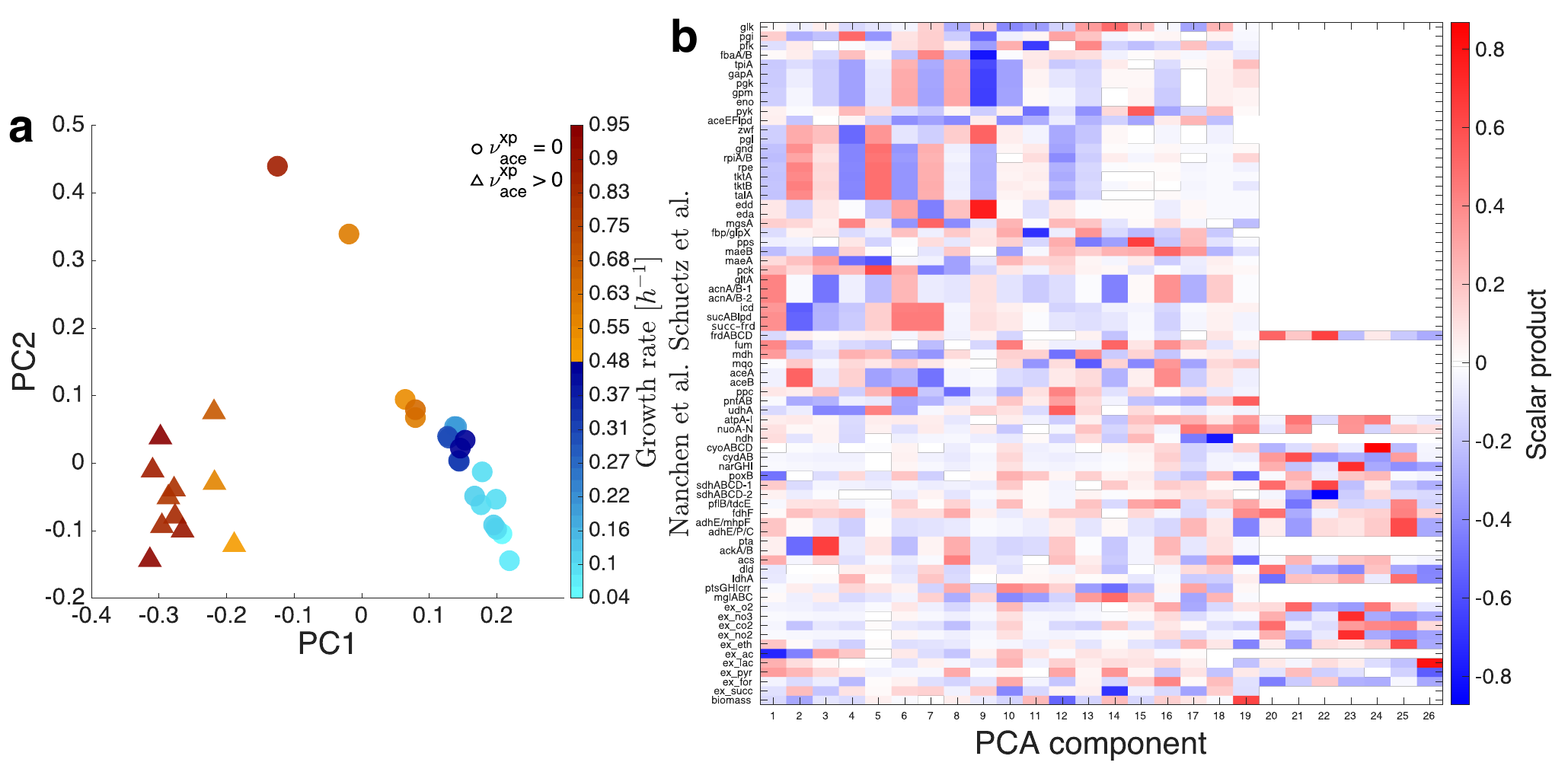} \\
	\caption{\label{fig:PCA}  \textbf{Principal Components Analysis} In this plot the show additional results of the PCA applied to the set of inferred coeffcients. In (a) we show the projection of the 33 experiments onto the first and second principal components of the coefficients covariance matrix. The color code mirrors both the glucose uptake for each experiment (as shown in the colorbar) and whether a point belongs to ~\cite{nanchen06} (blue palette) or ~\cite{schuetz12} (red palette); the markers are assigned in agreement with the value of the experimental acetate excretion. In (b) we show the heat-map of the projections of principal components of the inferred coefficients, i.e. the eigenvectors of the covariance matrix obtained from the normalized and projected Lagrange multipliers $\mathbf{c}$, along all the 73 fluxes of the models. }
\end{figure}

\subsubsection{Reconstruction of the coefficients}
To reconstruct the original coefficients, we can use a subset of the full set of eigenvectors as suggested by Figure \ref{four} of the main text. If we aim at using the first $k$ eigenvectors to reconstruct the original (normalized) projections of the $a$ experiment we can use
\begin{equation}
\mathbf{c}^{\rm PCA(k)}_{a} = \mathbf{m} + \left[ \mathcal{C}\mathbf{Q}_{1:\rm{dim}\left(\mathcal{F}\right),1:k}\mathbf{Q}^{T}_{1:\rm{dim}\left(\mathcal{F}\right),1:k}\right]_{1:\rm{dim}\left(\mathcal{F}\right),a} 
\end{equation}
Finally, we can re-project the approximate vector back to the fluxes space, obtaining
\begin{equation}
\tilde{\mathbf{c}}^{\rm PCA(k)}_{a} = \hat{O}^{T}\left( ||\mathbf{c}^{\rm proj}_{a}|| \mathbf{c}^{\rm PCA(k)}_{a} \right)
\end{equation}
to build an approximation of the flux probability density as
\begin{equation}
p\left(\mathbf{v};\tilde{\mathbf{c}}^{\rm PCA(k)}_{a}\right) \propto e^{\tilde{\mathbf{c}}^{{\rm PCA(k)}^{T}}_{a} \mathbf{v}} \mathbb{I}_{\mathcal{F}}\left(\mathbf{v}\right)
\label{eq:kPCAdist}
\end{equation}
Using the Expectation Propagation scheme, we can approximate the density in Eq. (\ref{eq:kPCAdist}) and get an estimate for the expected values of the constrained fluxes that have been measured, i.e. $\langle \mathbf{v}_{\mathcal{E}}\rangle_{\mathbf{c}^{\rm PCA(k)}_{a}}$ used in Eq. (\ref{eq:PCAest}) of the main text.

\subsection{Accessing the most informative fluxes}

Results discussed in the main text have made use of the full set of experimentally characterized fluxes in order to constrain the feasible space $\mathcal{F}$ and guide the inference procedure. However, because measured fluxes mostly belong to the central carbon pathways, they are likely to contain a significant amount of redundancy. An important question in this respect is whether one could learn the same information about flux distributions by employing a smaller number of fluxes to constrain $\mathcal{F}$, using the values of other measured fluxes as consistency checks. Importantly, non-zero coefficients can be understood as Lagrange multipliers associated with the most informative fluxes that are experimentally accessible. We have explored this issue by looking for a sparser representation of the vector $\mathbf{c}$ of Lagrange coefficients, i.e. one that, despite having a smaller number of non-zeros components, is still able to reproduce the mean values of all measured fluxes. This is achieved when the sparser and the original vectors have the same projection on the polytope $\mathcal{F}$ .

\begin{figure}[h]
\begin{center}
\includegraphics[width=1.0\textwidth]{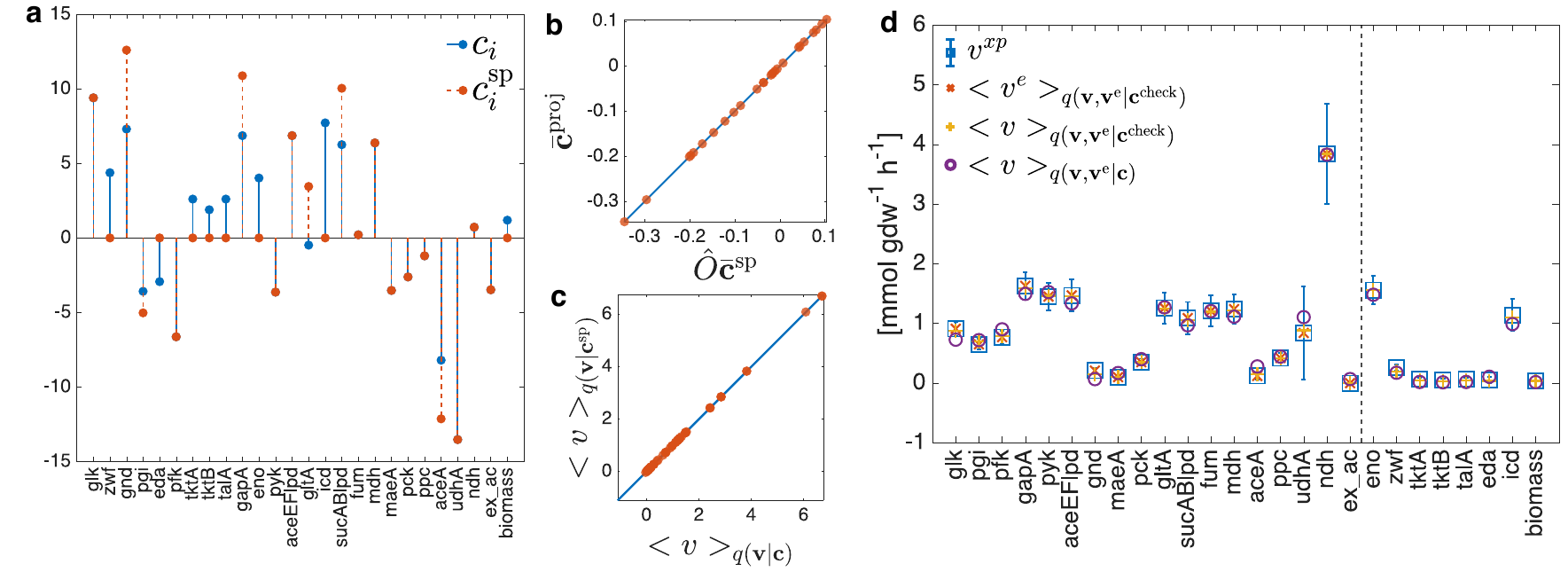}
\caption{\textbf{Compression of the coefficients} (a) Original coefficients inferred within the Expectation Propagation approximation (blue thick lines and points) associated with the 26 experimentally determined fluxes displayed in the x-axis. Using red dashed lines we plot the result of the compression; here only 18 fluxes out of 26 have to be fixed, and their coefficients refined, to get an equivalent probability density. In the (b) panels we show the projection of the (normalized) original and sparse coefficients into the sub-space $\mathcal{F}$: they perfectly match. As a consequence, the first moments derived from the original density $q(\mathbf{v};\mathbf{c})$ and those computed from the distribution $q(\mathbf{v};\mathbf{c}^{\rm sp})$ associated with the sparse coefficients are identical as displayed in panel (c). In panel (d) we show the performance of a synthetic experiment in which we perform the inference from scratch, fixing 18 measured fluxes (having non-zeros coefficients in panel (a)) out of the 26. They are displayed in the leftmost part of the x-axis, while the remaining 8 occupy the rightmost spots in the x-axis. For the fixed fluxes, the average values of the $\mathbf{v}^{e}$ (red points) perfectly overlap to the experimental averages $\mathbf{v}^{\rm xp}$ (in blue) and their constrained counterparts $<\mathbf{v}>$ (plotted in yellow) lie mostly within the experimental error bars as those obtained in the original inference (purple dots). For the rightmost fluxes we plot the averages of the constrained fluxes computed according to the original coefficients and those inferred in the synthetic experiment (yellow and purple spots respectively): although we have not used here any information of the experimental means, we retrieve almost the same averages. This confirms the results obtained from the compression of the original coefficients. \label{fig:CS}}
\end{center}
\end{figure}
This problem can be mathematically re-phrased as a compressed sensing problem solvable via Expectation Propagation (see the next section for the formal definition and the approximation details). As a representative example, Figure \ref{fig:CS}a shows the original coefficients and the compressed ones $\mathbf{c}^{\textrm{sp}}$ for one of the experiments in ~\cite{nanchen06}. One sees that, of the 26 degrees of freedom whose Lagrange multipliers are represented in blue, only 8 can be set to zero by the sparse procedure (red markers). A comparison between the original projections and the ones accomplished by the sparse vector is shown in Figure \ref{fig:CS}b, while a scatter plot showing the averages of all fluxes (measured and non-measured) obtained through the inference procedure against those computed from the sparse representation is instead reported in Figure \ref{fig:CS}c. Both metrics are perfectly in agreement, suggesting that no information is lost in compression, at least for mean values. \\
As a consistency check, we performed, for the data used in Figure \ref{fig:CS}a, the inference procedure from scratch by only constraining the mean values of fluxes corresponding to non-zero coefficients in the sparse representation of $\mathbf{c}$. We call the inferred coefficients obtained by this last procedure $\mathbf{c}^{\textrm{check}}$. In Figure \ref{fig:CS}d we show that we are able to reproduce the values of the experimentally measured fluxes that were not used to constrain $\mathcal{F}$ (corresponding to the rightmost 8 fluxes on the $x$-axis, separated by the black dashed line). We plot here the experimental means $\mathbf{v}^{xp}$ using blue squares to which we overlap the averages of the constrained fluxes $\mathbf{v}$ computed according to $\mathbf{c}$ and $\mathbf{c}^{\textrm{check}}$ using purple circles and yellow crosses, respectively. For the noisy ones we also show $\mathbf{v}^{\textrm{e}}$ that by construction coincide with $\mathbf{v}^{\textrm{xp}}$.  \\
Remarkably, the qualitative scenario just described extends to the whole ensemble of experiments, as it turns out that 18 degrees of freedom are generically needed to achieve a perfect reconstruction of mean values. We stress that we retrieve the same number of necessary degrees of freedom brought in light by the Principal Component Analysis of the main text.
These results underline a somewhat unexpected picture: despite the fact that metabolic networks are relatively modular from a functional viewpoint (leading to significant correlations between reactions belonging to the same pathway), the complexity of flux configurations found in experiments cannot be reduced to a few reactions representative of distinct pathways. In other words, cross-talk between pathways is an essential component of the system-level organization of metabolism. 


\subsubsection{Compression of the inferred coefficients: mathematical details}

The original vector $\mathbf{c}$ is a vector that has non-zero values only in correspondence of the Lagrange multiplier associated with the measured fluxes, more precisely we fix 25 or 26 averages, depending on the reference experiments, among the $N = 73$ fluxes. It offers an interpretation in terms on `preferential' directions in the polytope $\mathcal{F}$ when projected in the proper sub-space spanned by the ortho-normal basis; all the information is thus encoded in a dense vector $\mathbf{c}^{\rm proj}=\hat{O}\mathbf{c}$. Let us assume that we are given the projected vector which fully determine the probability density associated with it. The linear dependencies among fluxes give us a certain freedom in the assignment of non-zeros coefficients of $\mathbf{c}$ that satisfy the projection constraint: being  $\hat{O}$ a rectangular matrix, the inverse problem of determining a vector $\mathbf{c}$ that satisfies these equations is under-determined and allows for infinite solutions. Among these, we aim at determining the sparsest vector $\mathbf{c}^{\rm sp}$ that has the given projection $\mathbf{c}^{\rm proj}$ on the polytope and, concurrently, it has the least possible non-zeros components in correspondence to a sub-set a fluxes (we restrict the analysis to the measured fluxes in $\mathcal{E}$, but one can extend the procedure to all fluxes or any relevant sub-set, like internal or irreversible fluxes).\\
To determine $\mathbf{c}^{\rm sp}$ we solve the so-called compressed sensing problem (CS) on the original projection of the coefficients using the Expectation Propagation approximation derived in ~\cite{braunstein20}. Within a Bayesian framework we can write the probability of a sparse vector of coefficients, given the projection, using Bayes theorem, as:
\begin{eqnarray}
P\left(\mathbf{c}^{\rm sp};\mathbf{c}^{\rm proj}\right)	& \propto & 
\delta(\hat{O}\mathbf{c}^{\rm sp} = \mathbf{c}^{\rm proj})	
\prod_{i}P_{i}\left(c_{i}^{\rm sp}\right)
\end{eqnarray}
where $\delta(\hat{O}\mathbf{c}^{\rm sp} = \mathbf{c}^{\rm proj})$ is the likelihood function enforcing the projection constraint and $P_{i}(c_{i}^{\rm sp})$ is the prior probability associated with the the $i^{\rm th}$ component of the sparse vector. The latter serves to stress the sparsity constraint on the chosen sub-set of indices, in our case $\mathcal{E}$, and it is equal (in our formulation) to the so-called \textit{spike-and-slab} prior, or $L_{0}$ regularization,
\begin{eqnarray}
P_{i}\left(c_{i}^{\rm sp}\right) & = & \rho\delta_{c_{i}^{{\rm sp}},0}+\left(1-\rho\right)\mathcal{N}\left(0,\lambda\right) \qquad i \in \mathcal{E} \\
P_{i}\left(c_{i}^{\rm sp}\right) & = & \delta_{c_{i}^{\rm sp},0} \qquad i \notin \mathcal{E}
\end{eqnarray}
where $\rho$ is the fraction of zeros components among the sparse sub-vector associated with the observed fluxes $\mathcal{E}$ and $\mathcal{N}(0,\lambda)$ denotes a normal density of zero mean and standard deviation equals to $\lambda$. \\
From the \textit{a posteriori} distribution one would extract an estimator for the sparse coefficients computing their expectation values from $P\left(\mathbf{c}^{\rm sp};\mathbf{c}^{\rm proj}\right)$
\begin{eqnarray}
\langle c_{i}^{\rm sp} \rangle_{P} = \int d\mathbf{c}^{\rm sp} P\left(\mathbf{c}^{\rm sp};\mathbf{c}^{\rm proj}\right) c_{i}^{\rm sp}
\end{eqnarray}
but the computation of any observable from the posterior probability is unfeasible due to the intractability of the normalization (or partition function) of such density. To cope with the marginalization and the estimation of the first moments, we apply the Expectation Propagation approximation to 
$P\left(\mathbf{c}^{\rm sp};\mathbf{c}^{\rm proj}\right)$ where here each \textit{spike-and-slab} prior is approximated using an univariate Gaussian density, whose parameters are determined using the EP scheme (see ~\cite{braunstein20} for all the details).\\
Surprisingly, even setting $\rho \rightarrow 1$, we have found that at least 18 non-zeros components are necessary to fulfill the constraints on the projections, for the full set of coefficients, namely inferred from both the experimental conditions in ~\cite{nanchen06} and ~\cite{schuetz12}. The 18 chosen fluxes vary from experiment to experiment: to quantify the possible emergence of a pattern, we compute the co-occurrence of each pair of measured fluxes and we clusterize the empirical two-point frequencies matrix $\mathbf{F}$ of elements
\begin{eqnarray}
F_{i,j} & = & \frac{1}{33} \sum_{k=1}^{33} \mathbb{I}\left[ c_{i}^{sp,k} \neq 0\right] \mathbb{I}\left[ c_{j}^{sp,k} \neq 0\right] \qquad i \in \mathcal{E},\, j \in \mathcal{E}.
\end{eqnarray}
The analysis suggests that a single cluster comes up composed of \textit{glk}, \textit{gnd}, \textit{ex ac}, \textit{udhA}, \textit{ppc}, \textit{pfk}, \textit{aceEFlpd}, \textit{maeA}, \textit{pck}, \textit{mdh}, \textit{fum} and \textit{pyk} which appear within the compressed set of fluxes in more than 90 \% of the times. \\
We have also investigated the results of the compression when the non-zero coefficients can be assigned to all the fluxes or to a different sub-set composed of internal or irreversible fluxes. However, also in these cases, the procedure is not able to compress more and the least number of degrees of freedom remain unchanged or slightly increases.
Overall the compression analysis shows that the information content of the inferred coefficients cannot be encoded in few measured fluxes, confirming the analysis performed on the basis of the Principal Components, but at least 18 degrees of freedom  are required to retrieve the average values of all the fluxes, under the chosen reference experimental conditions.

\subsection{Fitting averages using HR-based Boltzmann machine learning}

For sake of completeness and to illustrate the computational difficulties behind the inference task, we will describe here a more straightforward and standard way to compute the optimal
coefficients $c_j$  based on the so-called Boltzmann Learning scheme. 

Upon considering the log-likelihood of the parameters given the empirical data, i.e. \begin{equation}
\mathcal{L}(\mathbf{c};\mathrm{data})=\frac{1}{K}\sum_{k=1}^K \log P(\mathrm{data};\mathbf{c})~~,
\end{equation}
one sees that
\begin{gather}
\frac{\partial\mathcal{L}}{\partial c_j}=v_{j}^{xp}-\avg{v_j}_{\mathbf{c}}~~,
\end{gather}
where
\begin{equation}\label{meanc}
\avg{v_j}_{\mathbf{c}}=\int_\mathcal{F} v_j\,p(\mathbf{v};\mathbf{c})\, d\mathbf{v}~~.
\end{equation}
This suggests that the optimal vector $\mathbf{c}$ can be found by an updating dynamics driven by the difference between the empirical mean and the mean computed using the current vector $\mathbf{c}$, i.e. via a Boltzmann learning such as 
\begin{equation}\label{BL}
c_j(\tau+\delta\tau)-c_j(\tau)=\left[v_{j}^{xp}-\avg{v_j}_{\mathbf{c}(\tau)}\right]\delta\tau~~.
\end{equation}

To compute the optimal values of the coefficients $c_j$ from Eq. (\ref{BL}) the following procedure can be defined:
\begin{enumerate}
\item initialize $c_j(0)=0$ for all $j\in\{1,\ldots,E\}$
\item at each time step $\tau$: compute $\avg{v_j}_{\mathbf{c}(\tau)}$ from Eq. (\ref{meanc}) by sampling the distribution $p(\mathbf{v};\mathbf{c}(\tau))$ e.g. via Hit and Run Monte Carlo ~\cite{ell}; then
\item find the index $j$ for which the difference $v_{j}^{xp}-\avg{v_j}_{\mathbf{c}(\tau)}$ is largest, update its value according to Eq. (\ref{BL}), and iterate.
\end{enumerate}
We do see that this amounts at simulating a dynamical system where the evaluation of the dynamical laws at each time-step requires  a sampling of the high-dimensional space of steady state of the given metabolic network.
To provide an example, for the {\it E. Coli} core network ~\cite{orth10}, the inference task has been performed for one experimental point and with $\delta \tau=10^{-3}$ we do find a plateau of the coefficients after roughly $\tau\sim 10^7$ time-steps. With our implementation of the Hit-and-Run Monte Carlo (that includes optimization through ellipsoidal rounding) the sampling time, for one network instance, is of the order of $10$ms on an
quadcore CPU running at $1.90$GHz, therefore the overall machine time required for the final  inference is of $2-3$h. The running time of the EP-based scheme presented in this work, aimed at inferring the approximate distribution of fluxes (together with the computation of the Lagrange multipliers), is overall of $5.7$s for the 33 experiments.

\section*{Supplementary Figures}

\subsection{Fitting quality}
\label{sec:means}
We show in Figures \ref{fig:fittingN} and \ref{fig:fittingS}, a comparison between the experimental data, i.e. the means $\mathbf{v}^{\rm xp}$  and the standard deviations $\boldsymbol{\sigma}^{\rm xp}$ (denoted as \textit{data}), and the statistics of the constrained and noisy fluxes $\mathbf{v}$ and $\mathbf{v}^{e}$. The plots confirm that the matching constraints on the averages of the noisy flux, expressed in Eq. (\ref{eq:constr_means}) are all satisfied, and show how close the expected values of the constrained fluxes are, given the $\gamma$ found within the EP approximation scheme. 
We also plot a set of fluxes, called $\mathbf{v}^{\rm opt}$ given by the closest configuration of constrained fluxes to the experimental means, satisfying the mass-balance condition and the boundaries of variability, that is
\begin{align}
\label{eq:opt-fluxes}
\mathbf{v}^{\rm opt} = \arg\min_{\substack{\mathbf{v}: \\ \mathbf{S}\mathbf{v}=\mathbf{b} \\ \mathbf{v}^{\rm min} \leq \mathbf{v} \leq \mathbf{v}^{\rm max}} } (\mathbf{v}^{\rm xp} - \mathbf{v}_{\mathcal{E}})^{T}(\mathbf{v}^{\rm xp} - \mathbf{v}_{\mathcal{E}})
\end{align}
By construction, these can be thought as the result of the inference procedure per $\gamma \rightarrow +\infty$ and therefore the average values of the constrained fluxes $\mathbf{v}$ cannot be closer to the experimental means than the `optimized' fluxes.  
To further check our approach, we perform a Monte-Carlo Hit-and-Run (HR) sampling  \cite{ell}, given the set of Lagrange multipliers $\mathbf{c}$, and we compute the sampling average $<\mathbf{v}>^{\rm HR}$ for the measured fluxes (green triangles in Figure \ref{fig:fittingN}). The scatter plot of all the average values of the fluxes according to EP and HR are shown in Figure \ref{fig:HR} for all the experiments.

\begin{figure}
	\centering
	\includegraphics[width=\textwidth]{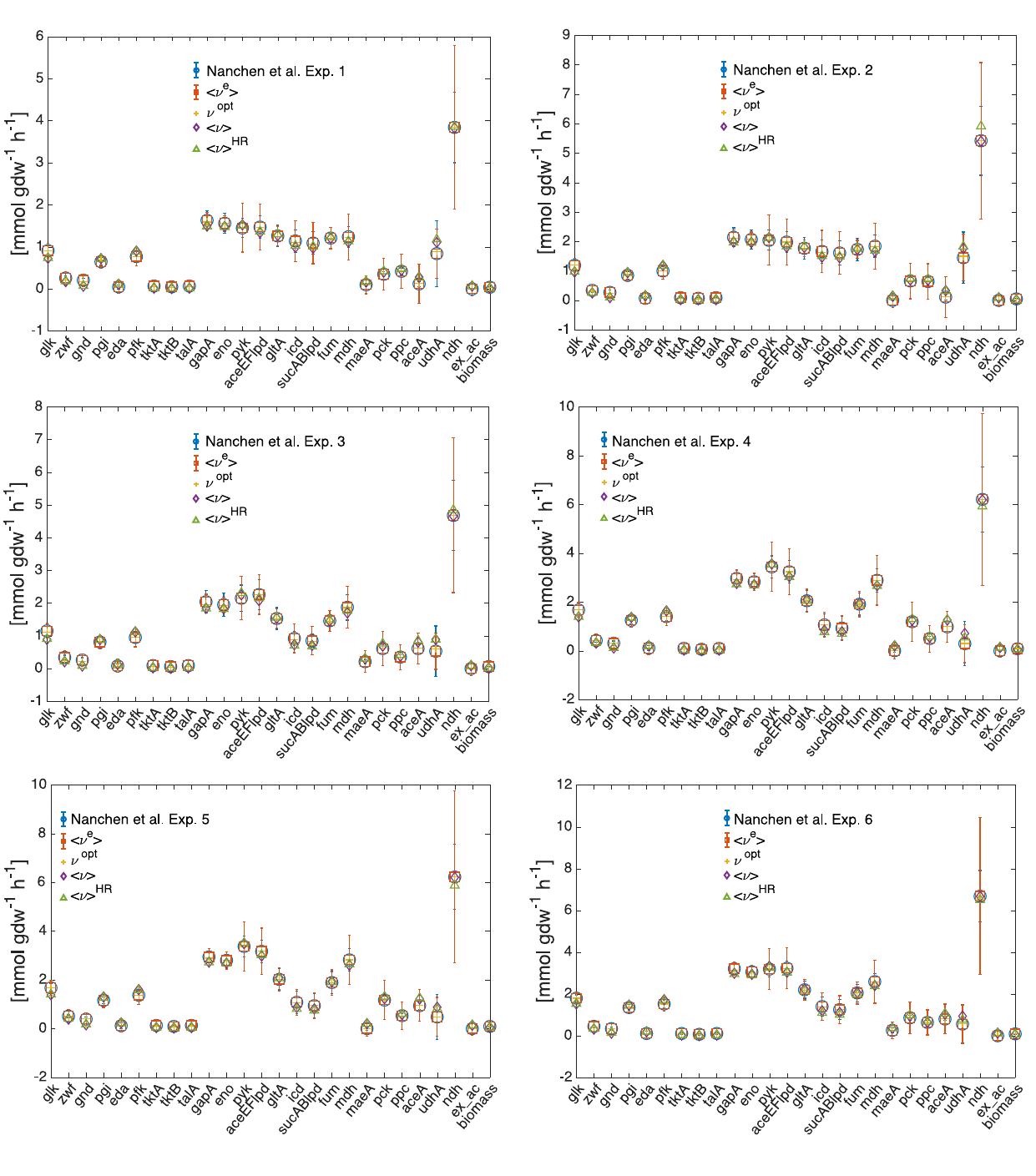} \\
\end{figure}

\begin{figure}
	\centering
	\includegraphics[width=\textwidth]{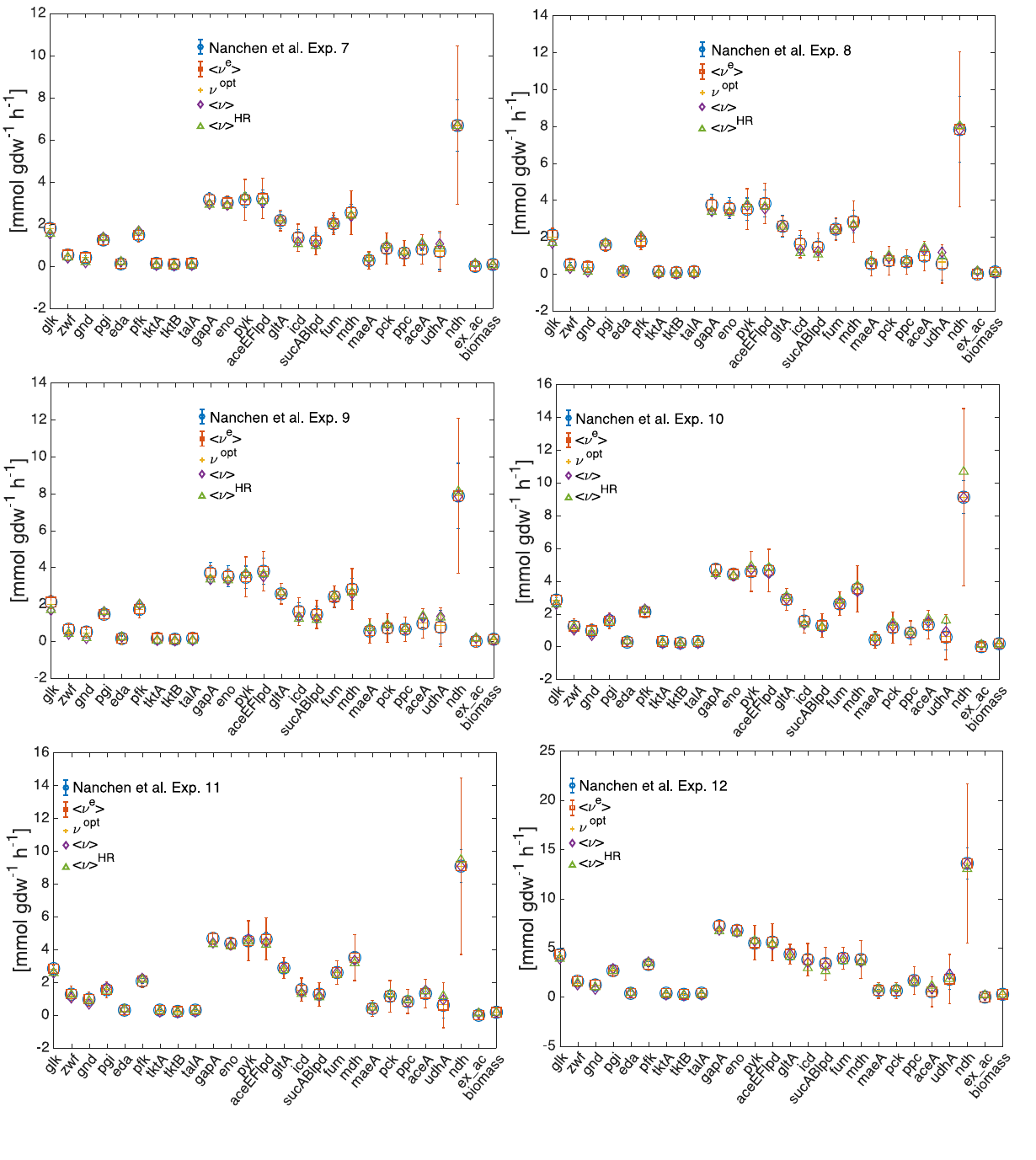} \\
\end{figure}

\begin{figure}
	\centering
	\includegraphics[width=\textwidth]{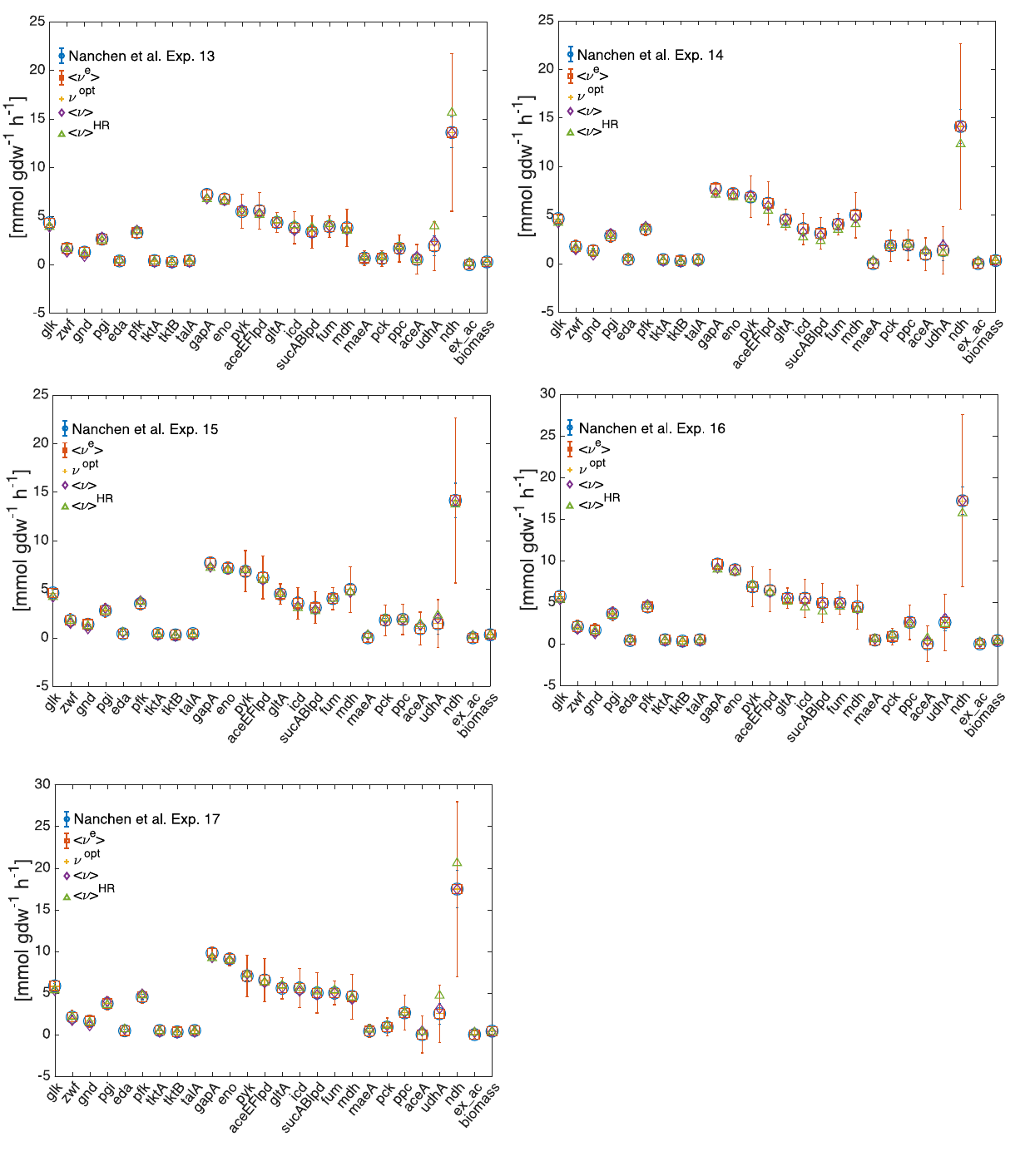}
	\caption{\label{fig:fittingN} \textbf{Fitting quality.} Plot of the experimental data for each of the 17 experiments in  ~\cite{nanchen06} (blue dots) together with the average value of the auxiliary fluxes $\mathbf{v}^{e}$ (red squares), the constrained fluxes $\mathbf{v}$ (purple diamonds), the optimized fluxes defined in Eq. (\ref{eq:opt-fluxes}) (yellow crosses) and the Hit-and-Run estimates (green triangles). All fluxes are reported in units of $mmol\,gdw^{-1}\,h^{-1}$ except the biomass synthetic rate which has dimension $h^{-1}$.}
\end{figure}

\begin{figure}
	\centering
	\includegraphics[width=\textwidth]{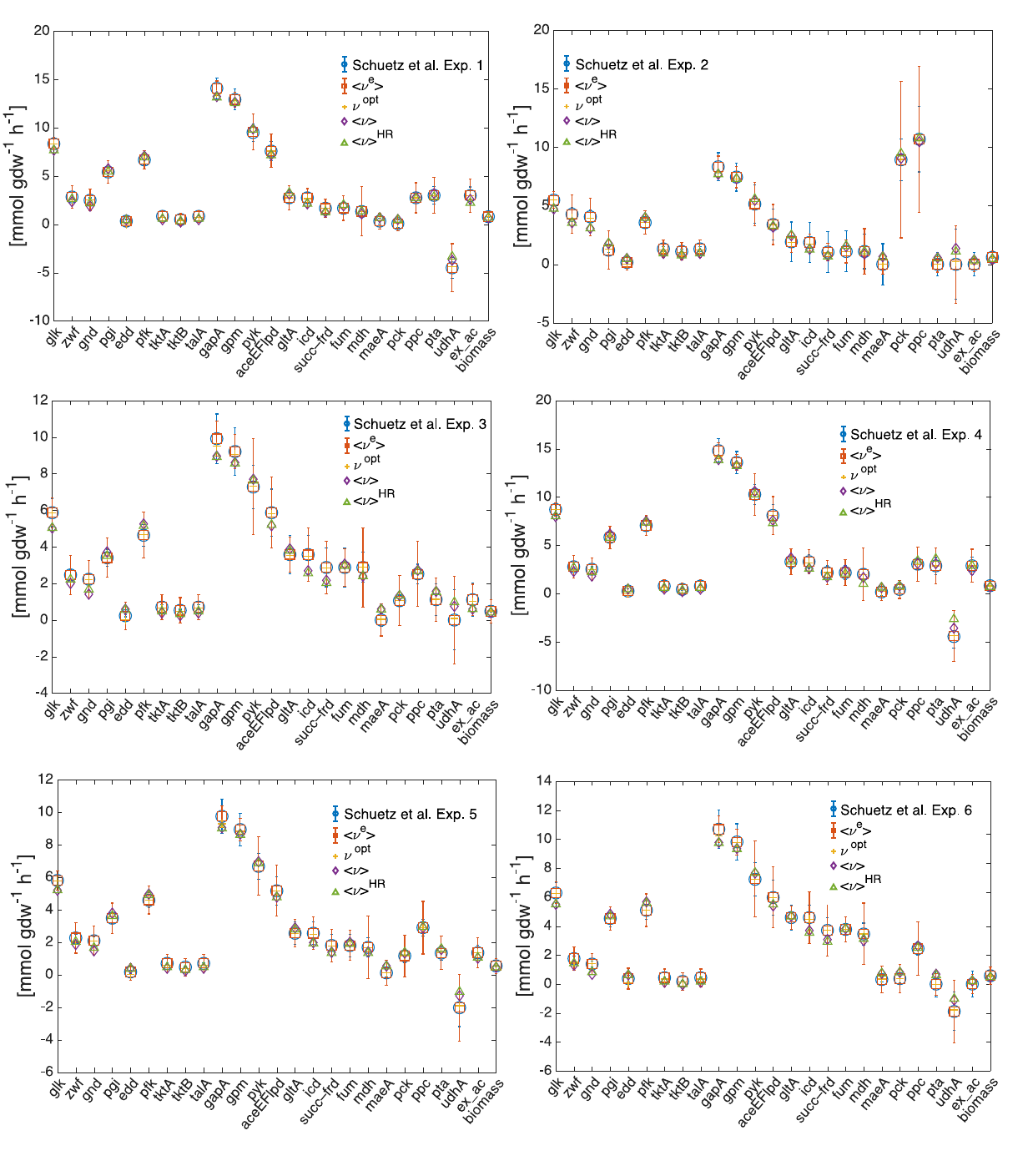}
\end{figure}

\begin{figure}
	\centering
	\includegraphics[width=\textwidth]{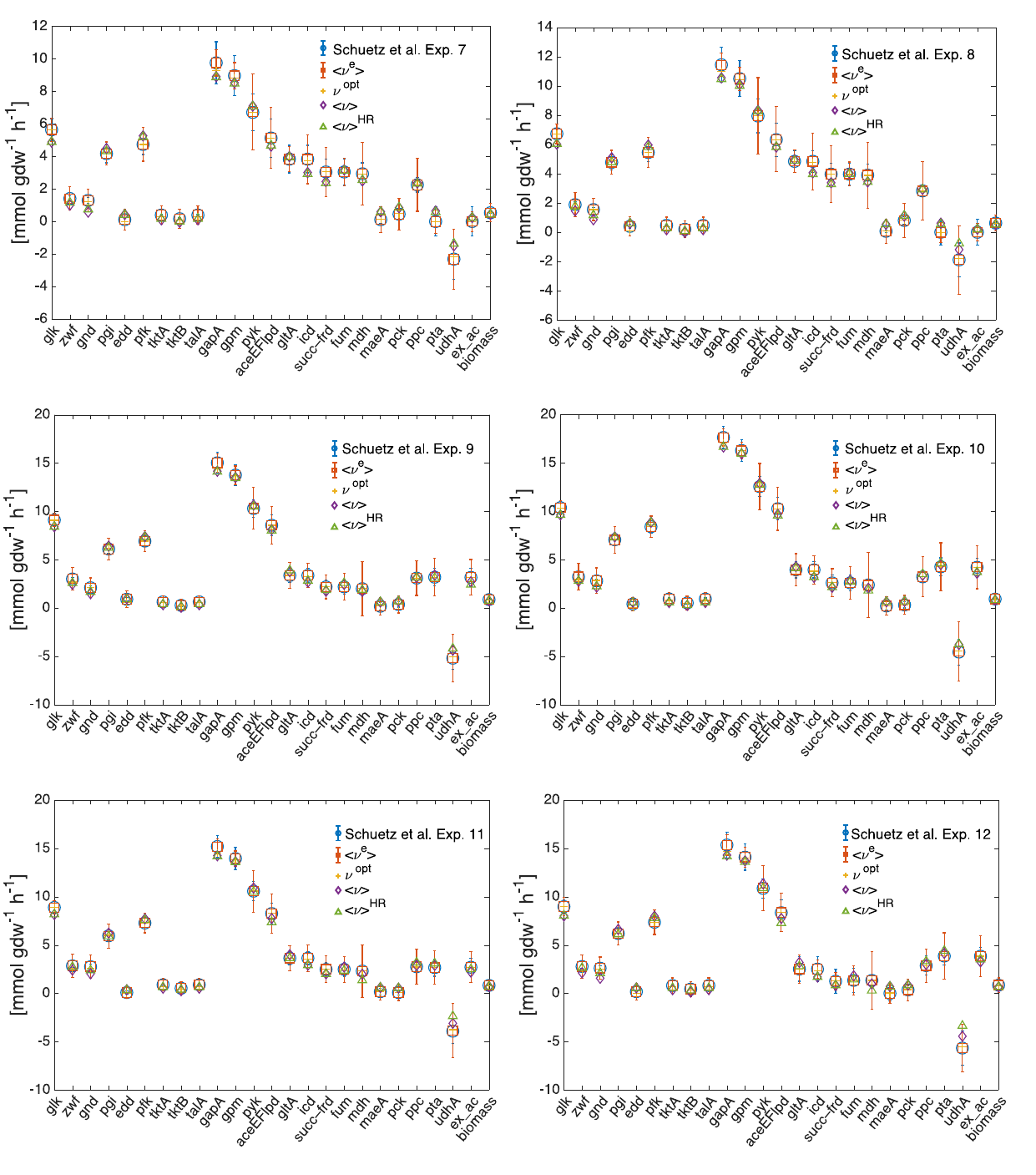}
\end{figure}

\begin{figure}
	\centering
	\includegraphics[width=\textwidth]{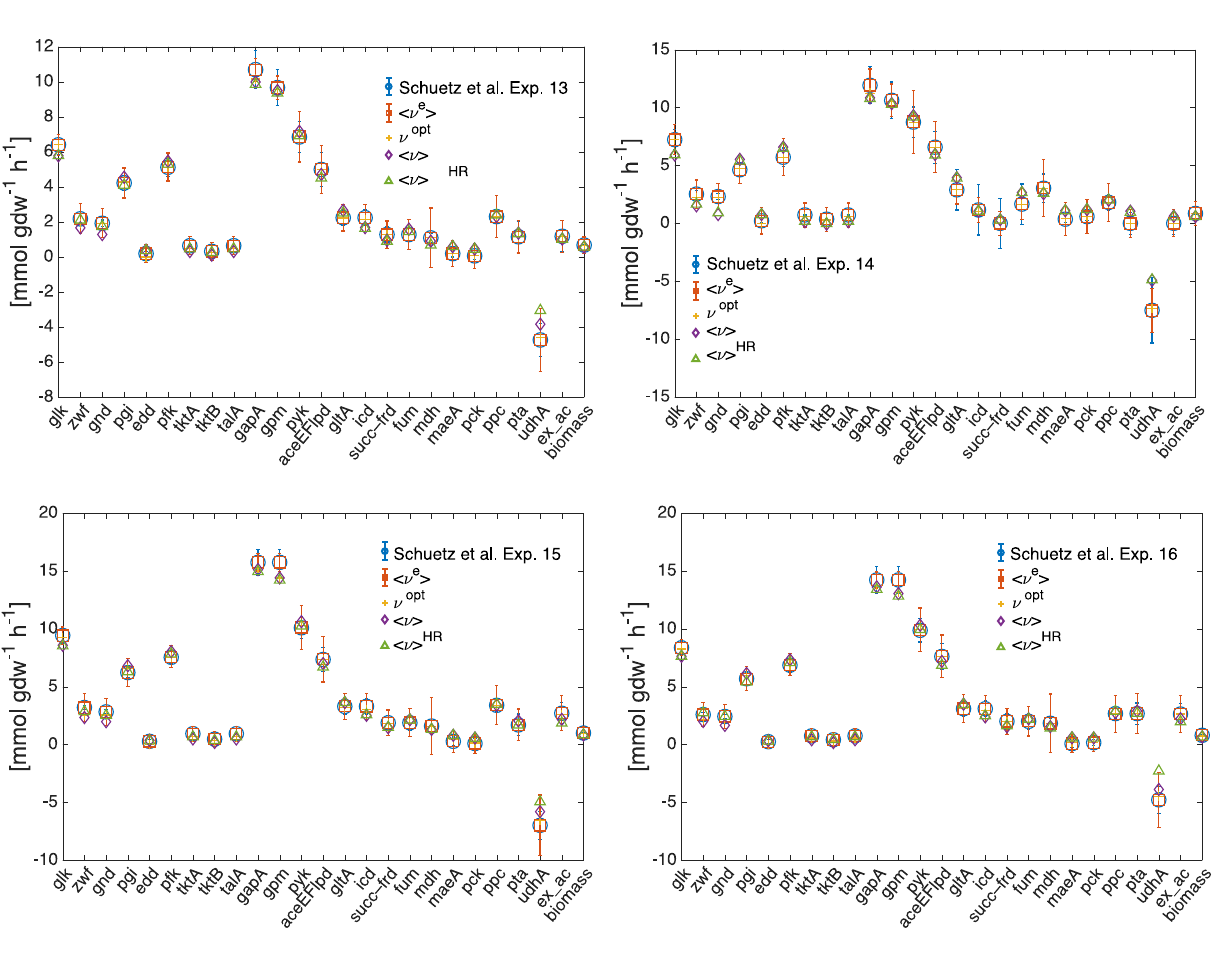}
	\caption{\label{fig:fittingS} \textbf{Fitting quality.} Plot of the experimental data for each of the 16 experiments in ~\cite{schuetz12} (blue dots) together with the average value of the auxiliary fluxes $\mathbf{v}^{e}$ (red squares), the constrained fluxes $\mathbf{v}$ (purple diamonds), the optimized fluxes defined in Eq. (\ref{eq:opt-fluxes}) (yellow crosses) and the Hit-and-Run estimates (green triangles). All fluxes are reported in units of $mmol\,gdw^{-1}\,h^{-1}$ except the biomass synthetic rate which has dimension $h^{-1}$.}
\end{figure}

\begin{figure}
\centering
\includegraphics[width=1.0\textwidth]{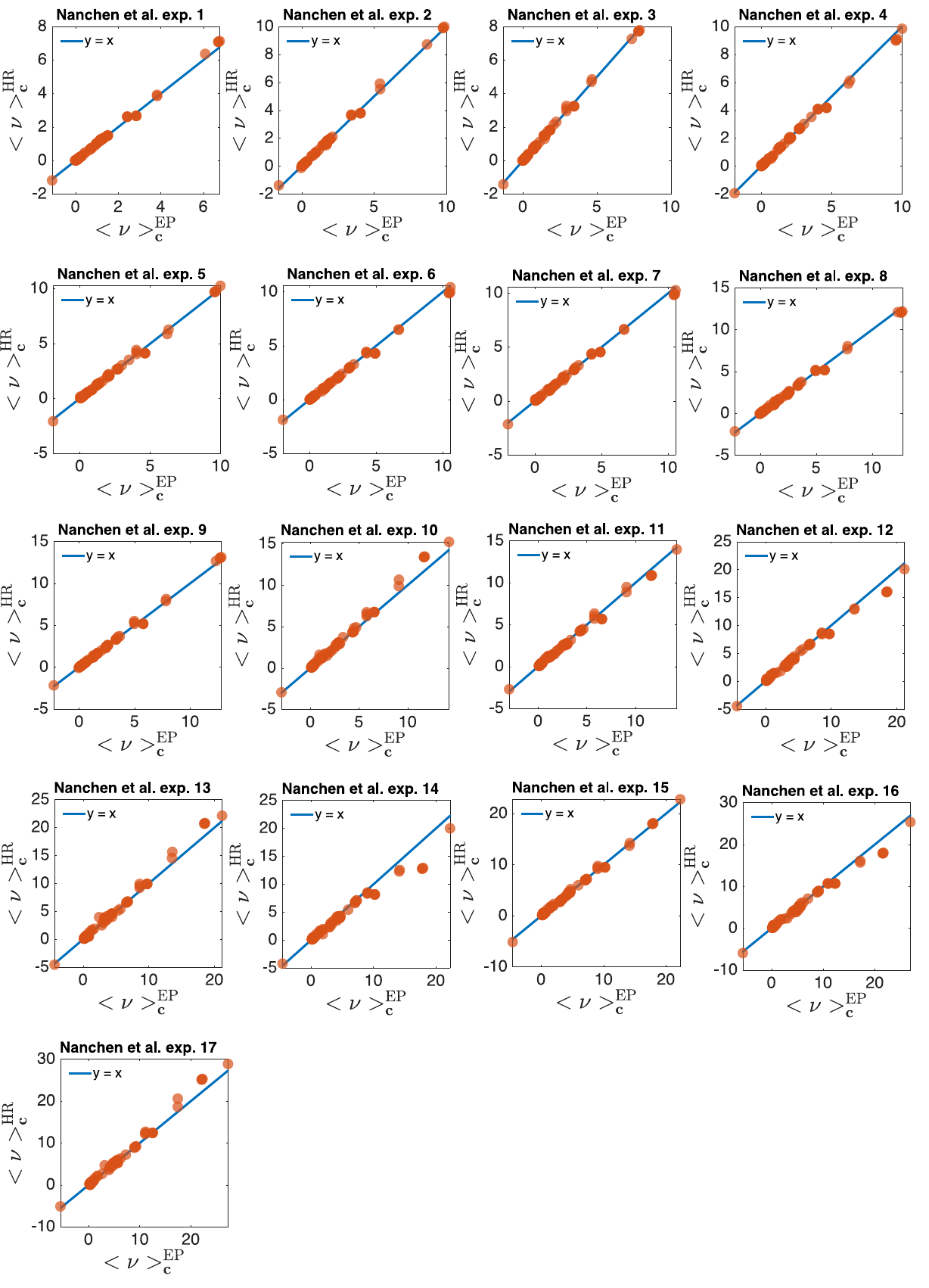}
\end{figure}

\begin{figure}
\includegraphics[width=1.0\textwidth]{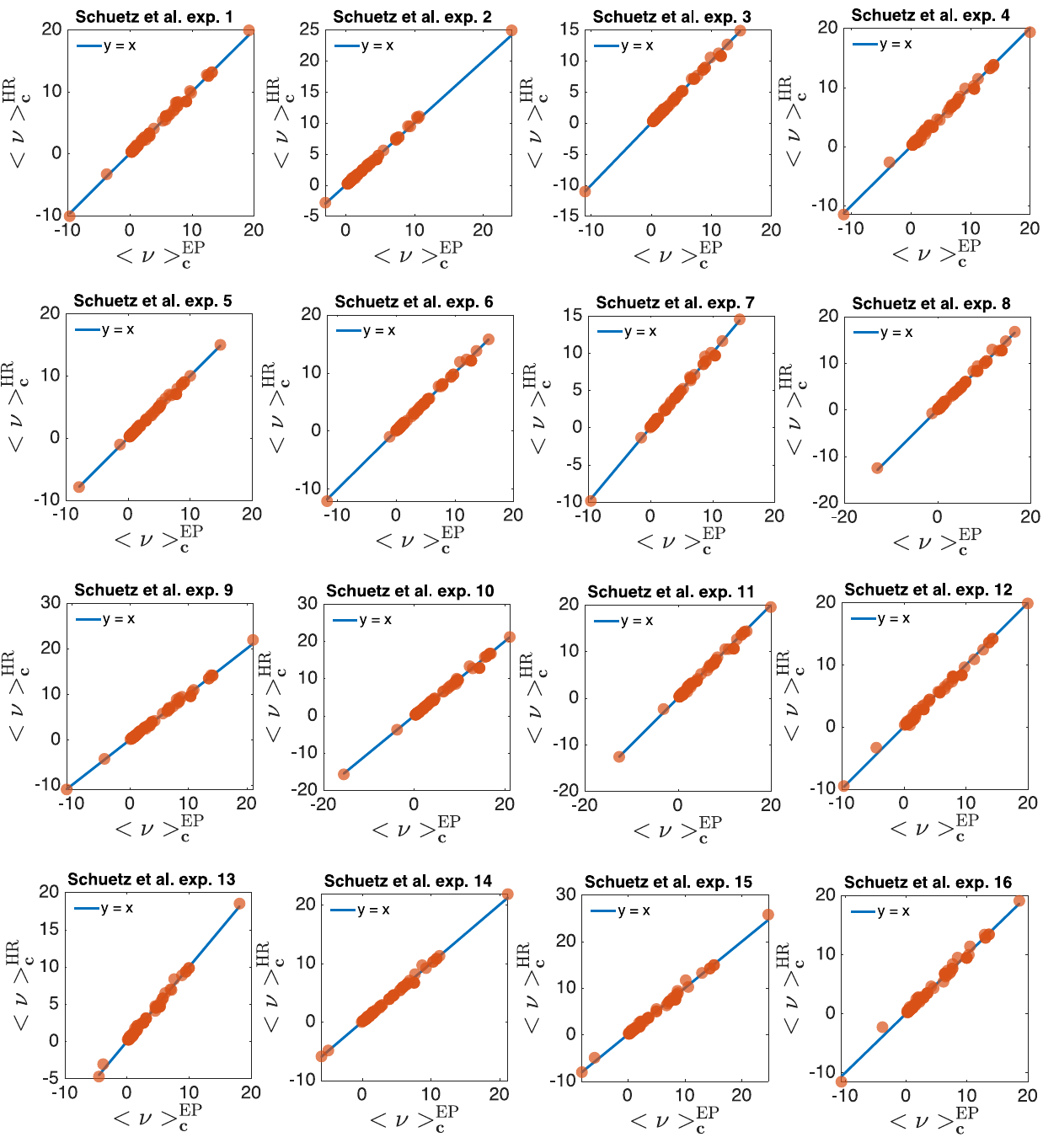}
\caption{\label{fig:HR} \textbf{Cross-check using Hit-and-Run Monte Carlo} For each experiment we show the scatter plot of the mean fluxes computed according to the Gaussian approximation provided by EP (x-axis) and the mean fluxes computed from a set of sampled configurations of the distribution in Eq. (\ref{dist}) (for given Lagrange multipliers $\mathbf
{c}$) obtained by Monte-Carlo Hit-and-Run.}
\end{figure}

\subsection{Correlation matrices}

In Figure \ref{fig:Pearson} we show the Pearson correlation coefficients between pairs of constrained fluxes, encoded in several $N \times N$ matrices, associated with the 33 experimental conditions. The computation of the Pearson correlation coefficients exploits the covariance matrices obtained at convergence of the Expectation Propagation scheme.

\begin{figure}
\centering
\includegraphics[height=0.60\textheight]{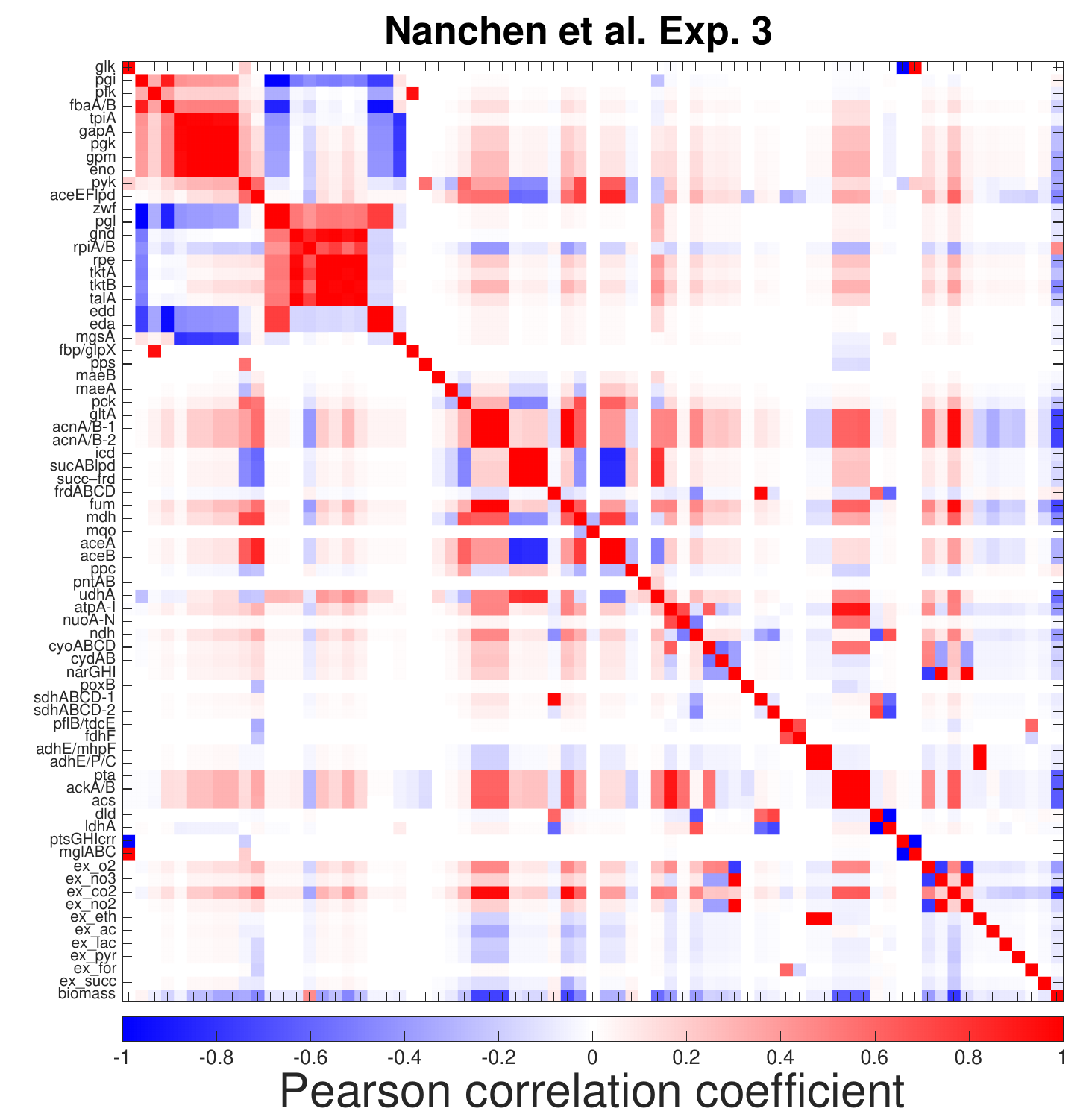}
\caption{\label{fig:Pearson} \textbf{Pearson correlation coefficient.} We plot here an example of the heatmap associated with the Pearson correlation coefficients of all the pairs of fluxes. The data displayed in the plot correspond to those of the third experiment in ~\cite{nanchen06}.}
\end{figure}

\subsection{Probability densities of the non-measured fluxes}

In Figure \ref{fig:marg} we plot the EP approximation of the marginal probability densities of the non-measured fluxes in all the experimental conditions. Specifically, we plot the tilted distribution associated with each constrained fluxes as in Eq. (\ref{eq:marginal-tilted}). The color code mirrors the dilution rate specified in each of the 33 experiments.

\subsection{Notation used for the biomass production rate}

In Table \ref{fig:table_bm} we show a summary of the symbols used to identify the biomass output within the manuscript.

\begin{table}
	\centering
	\begin{tabular*}{0.8\textwidth}{@{\extracolsep{\fill}}c|m{0.8\textwidth}}
		Symbol & Meaning\tabularnewline
		\hline 
		$\nu_{{\rm bm}}^{{\rm xp}}$ & Measured value of the biomass output according to the experimentally
		determined fluxes and the metabolic model. \newline This estimate
		is provided by the datasets in \cite{nanchen06} and \cite{schuetz12}.\tabularnewline
		\hline 
		$\nu_{{\rm bm}}^{{\rm max}}$ & Maximum attainable value of the biomass synthetic rate given the glucose
		consumption rate. This can be computed solving $\nu_{{\rm bm}}^{{\rm max}}=\max_{\mathbf{v}\in\mathcal{F}}\nu_{{\rm bm}}$
		where the feasible space encodes the NESS constraints for $\nu_{{\rm ex\,glc}}=\nu_{{\rm glc}}^{{\rm xp}}$,
		i.e. the value of the glucose uptake of the model is set equal to
		the experimental mean.\tabularnewline
		\hline 
		$\nu_{{\rm bm}}^{{\rm optm}}$ & Expectation value of the biomass production rate according to the
		optimal model. The latter is given by
		\[
		q\left(\mathbf{v};\beta\right)\sim e^{\beta\nu_{{\rm bm}}}\mathbb{I}\left[\mathbf{v}\in\mathcal{F}\right]
		\]
		where $\beta$ ensures that the expectation value $\left\langle \nu_{{\rm bm}}\right\rangle _{q\left(\mathbf{v};\beta\right)}$
		is equal to a given value $\bar{\nu}_{{\rm bm}}$. \newline Therefore $\left\langle \nu_{{\rm bm}}\right\rangle _{q\left(\mathbf{v};\beta\right)}=\nu_{{\rm bm}}^{{\rm optm}}$.\tabularnewline
		\hline 
		$\nu_{{\rm bm}}^{{\rm inf}}$ & Expectation value of the biomass production rate according to the
		inferred model 
		\[
		q\left(\mathbf{v},\mathbf{v}^{e};\mathbf{c}\right)\sim e^{\boldsymbol{c}^{T}\mathbf{v}^{e}-\frac{\gamma}{2}\sum_{i\in\mathcal{E}}\left(\nu_{i}-\nu_{i}^{e}\right)^{2}}\mathbb{I}\left[\mathbf{v}\in\mathcal{F}\right]
		\]
		where the Lagrange multipliers $\mathbf{c}$ and $\gamma$ are determined through Eqs. (\ref{eq:constr_means}) and (\ref{eq:constr_av_noise}) of the main text. \newline
		Therefore $\nu_{{\rm bm}}^{{\rm inf}}=\left\langle \nu_{{\rm bm}}\right\rangle _{q\left(\mathbf{v},\mathbf{v}^{e};\mathbf{c}\right)}$. 
		Being $\gamma>0$, $\nu_{{\rm bm}}^{{\rm inf}}\sim\nu_{{\rm bm}}^{{\rm xp}}$
		as shown in Figs. \ref{fig:fittingN} and \ref{fig:fittingS} \tabularnewline
	\end{tabular*}	
	\caption{Summary of the notation used to identified the biomass output under several conditions. \label{fig:table_bm}}
\end{table}

\begin{figure}
\centering
\includegraphics[width=1.0\textwidth]{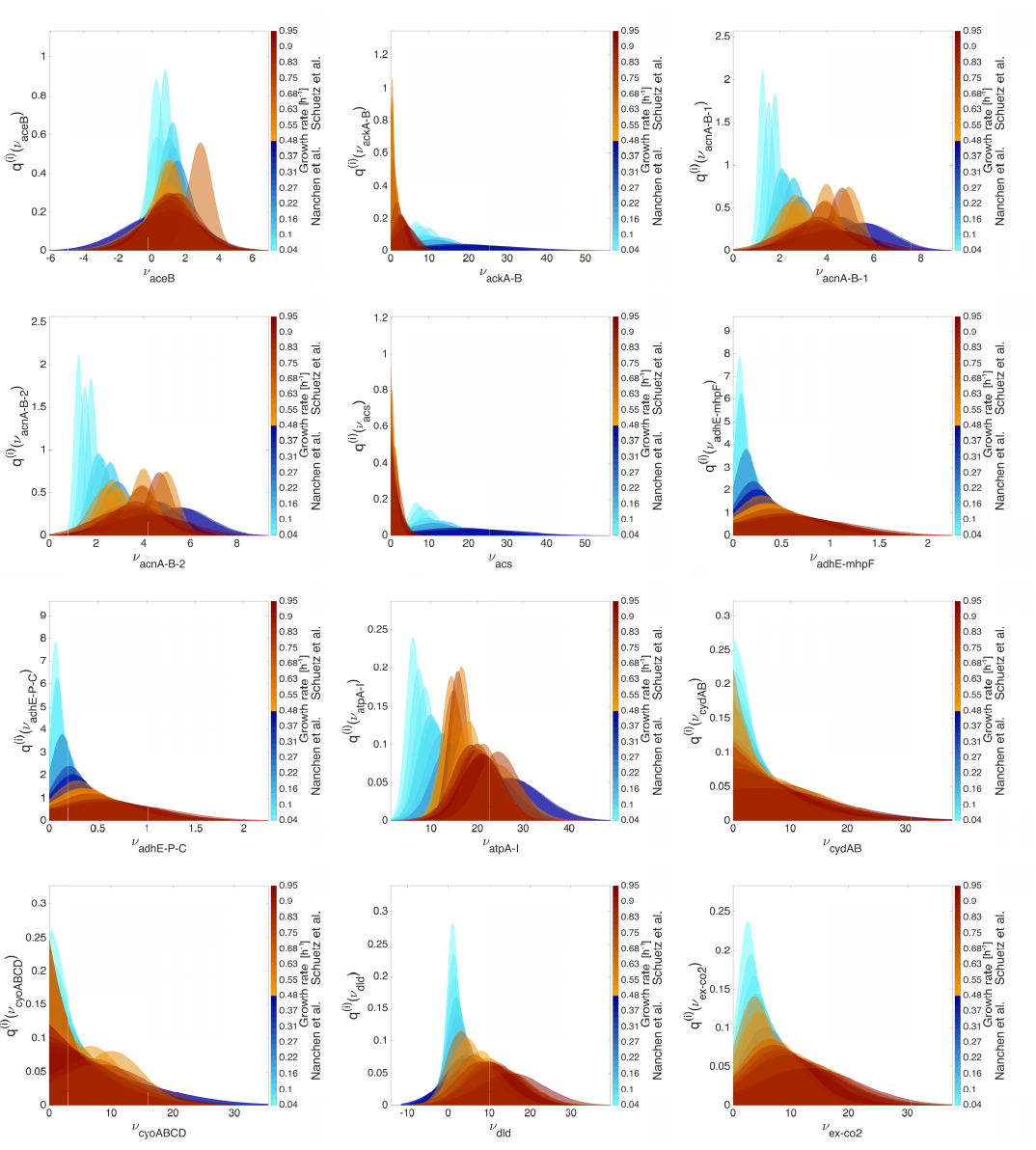}
\end{figure}

\clearpage

\begin{figure}
\centering
\includegraphics[width=1.0\textwidth]{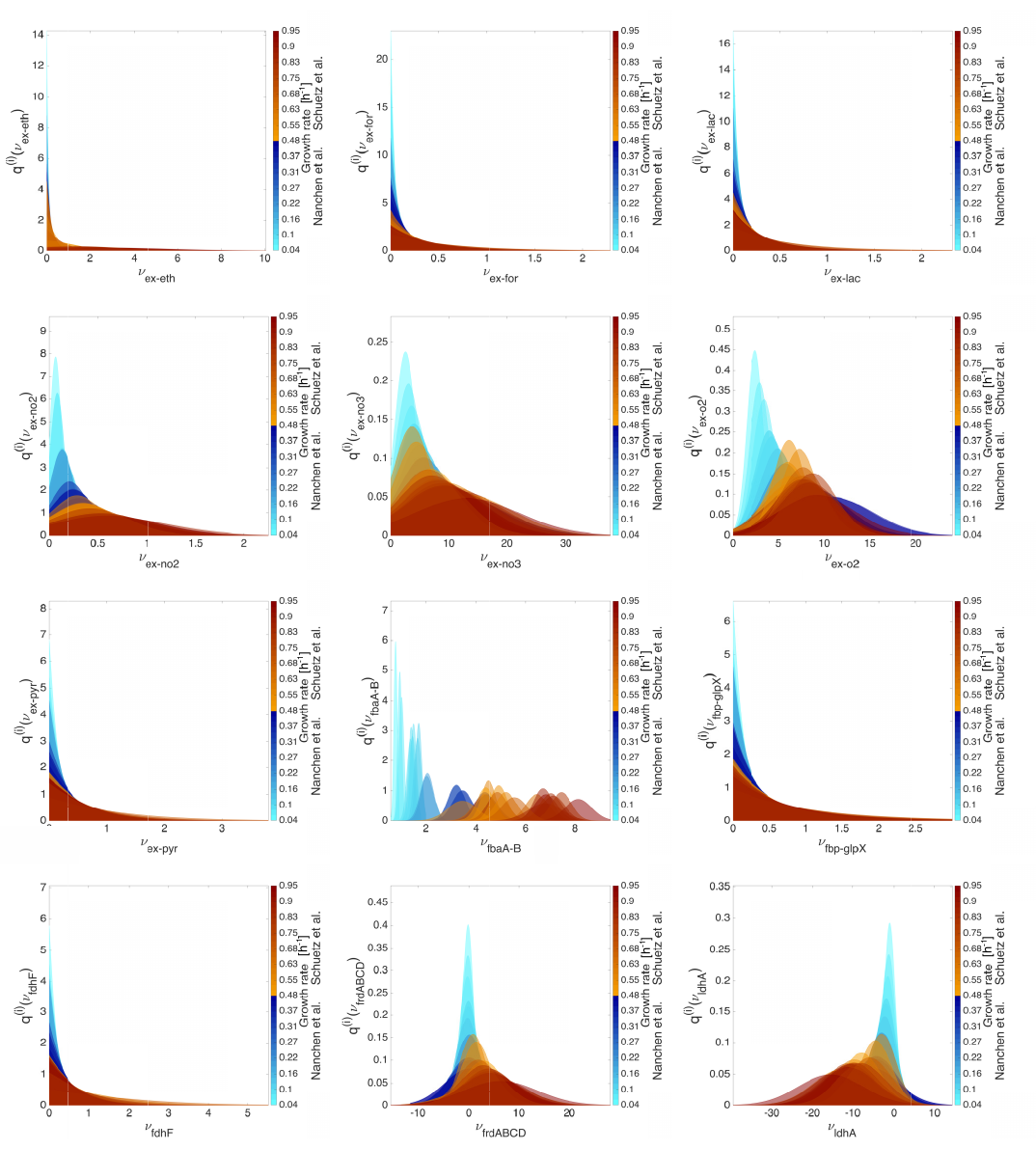}
\end{figure}

\clearpage

\begin{figure}
\centering
\includegraphics[width=1.0\textwidth]{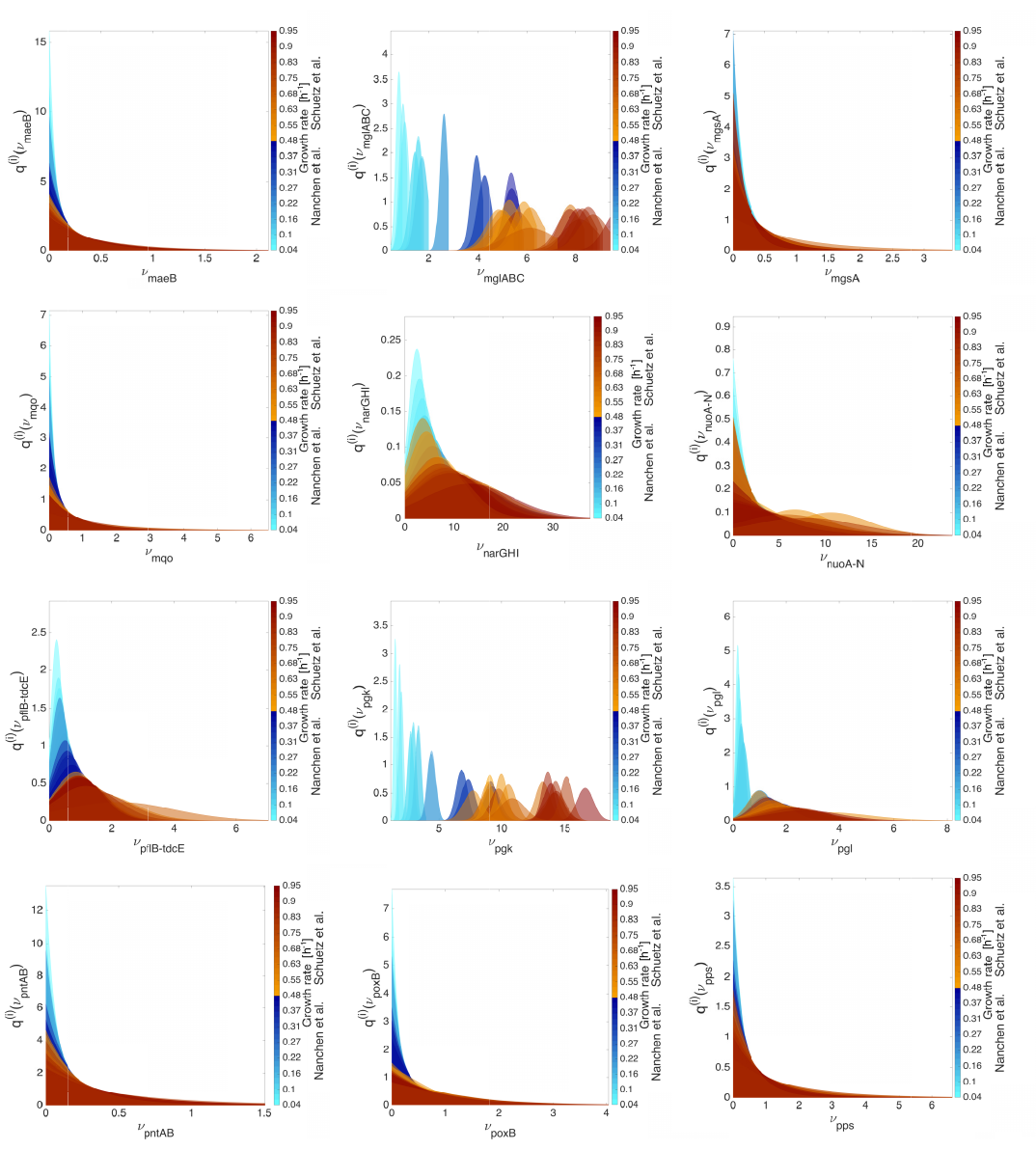}

\end{figure}

\clearpage

\begin{figure}
\centering
\includegraphics[width=1.0\textwidth]{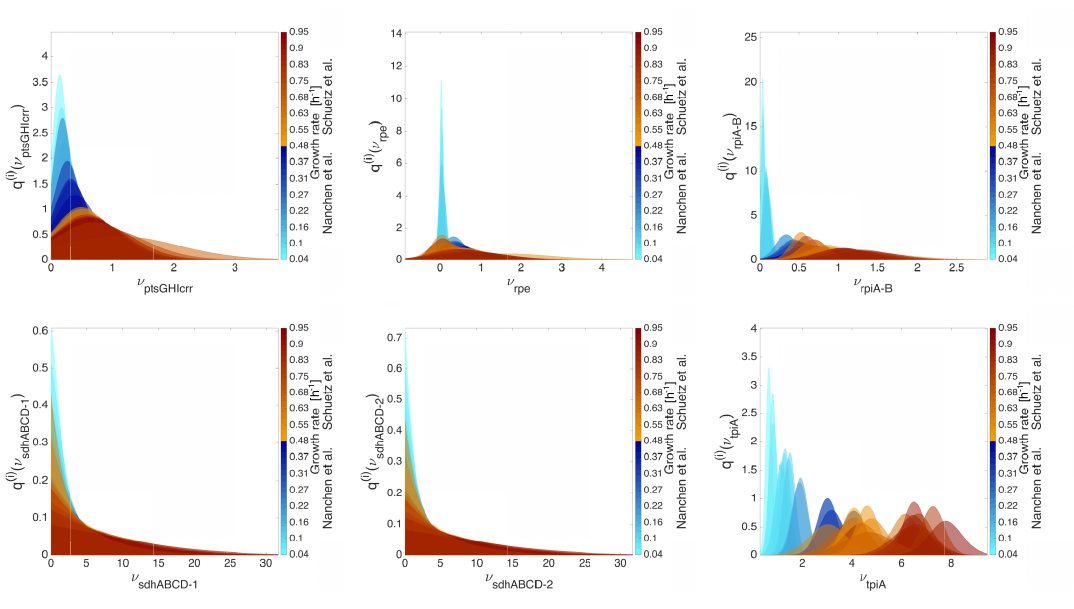}
\caption{\label{fig:marg} \textbf{Marginal probability densities} In these panels we show the approximation of the marginal probability densities revealed by Expectation Propagation. For each flux, we overlap the distributions obtained for all the 33 experiments, whose color is associated with the experimental growth rate. The unit of measurement of the fluxes in the x-axis is $mmol \, gdw^{-1} h^{-1}$.}
\end{figure}

\begin{figure}
\includegraphics[width=\textwidth]{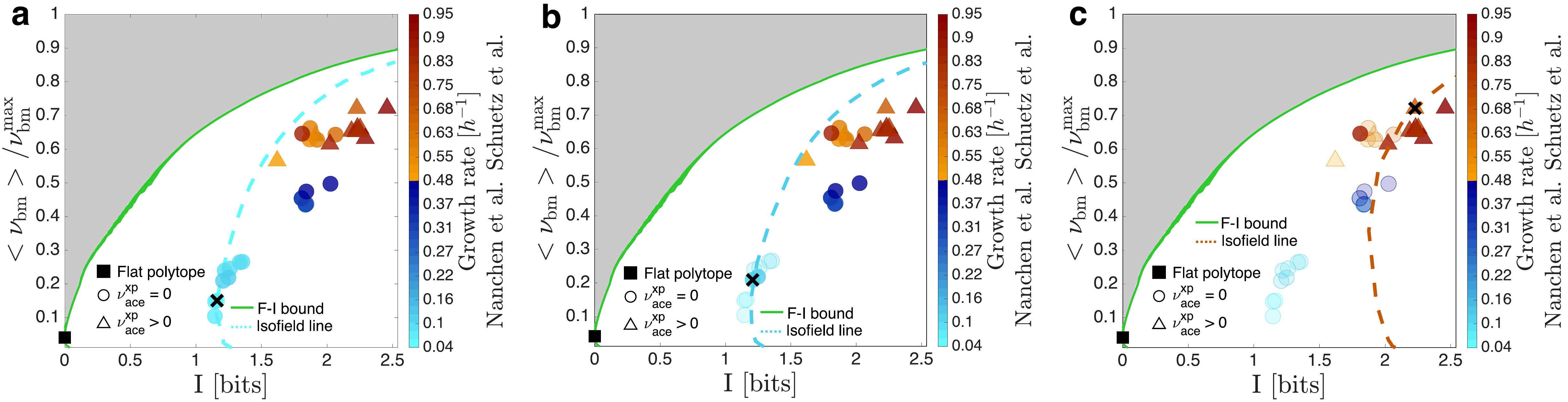}
\caption{\label{fig:Isofield} \textbf{Rate distortion curve and isofield lines.} In this plot we show the rate distortion curve and the scatter plot of the experimental data in the space $<v_{\rm bm}>/v^{\rm max}_{\rm bm}$ vs $I$ (as in Fig. 1d of the main text) together with the "isofield" lines associated with the first and the ninth experiments in ~\cite{nanchen06}, and the tenth experiment in ~\cite{schuetz12} in panels (a), (b) and (c) respectively. The cross indicates the reference population for which we construct the isofield line, while transparent markers are used when the glucose uptake of the corresponding experiments is smaller than that associated with the crossed point (i.e. they represent slower populations). Colored markers represent faster-growing populations which tend to lie on the right of the isofield line suggesting that they are more constrained than the reference population.}
\end{figure}

\begin{figure}
	\centering
	\includegraphics[width=\textwidth]{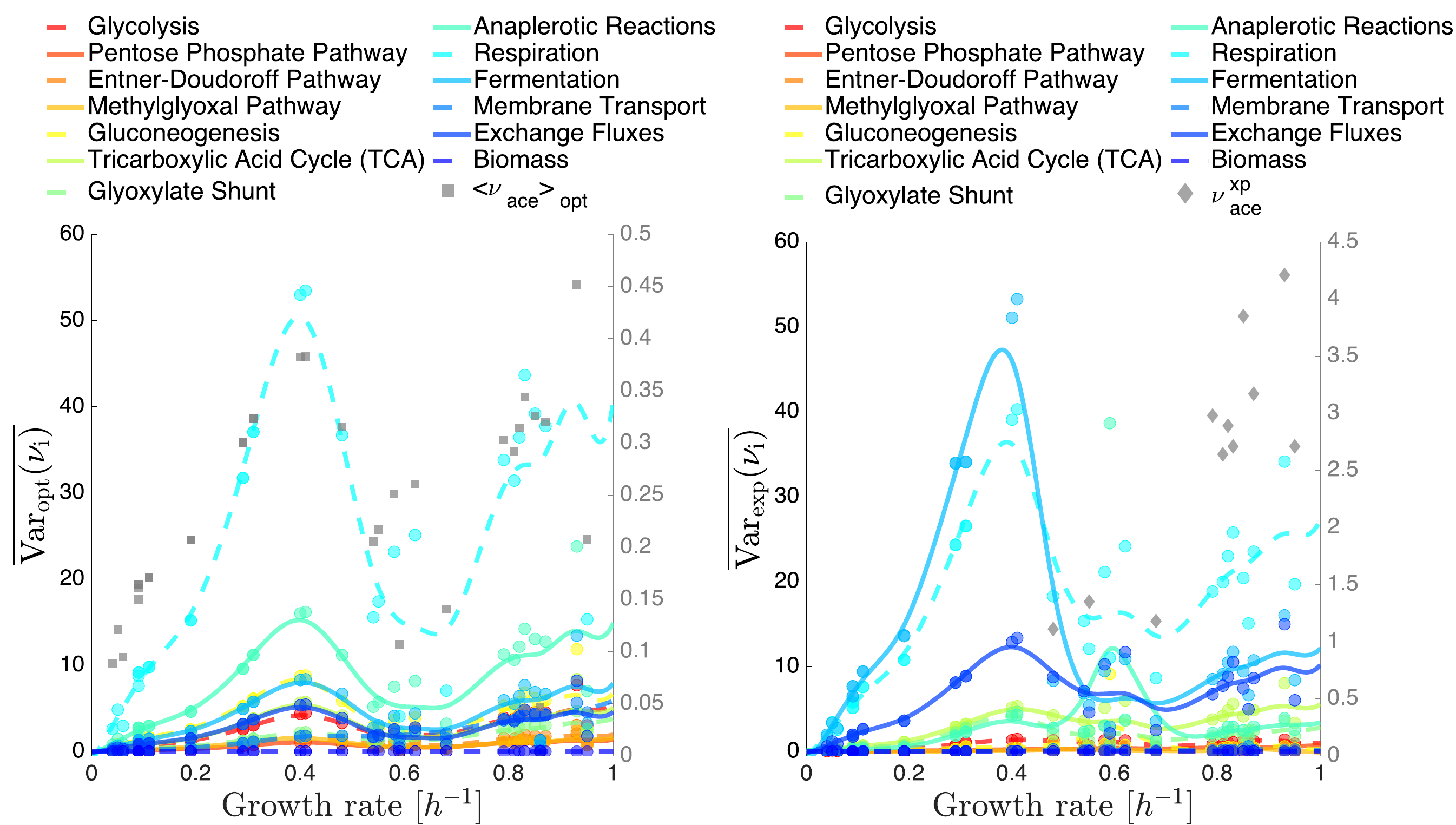} \\
	\caption{\label{fig:pathways} \textbf{Pathways variability.} Mean variance of fluxes through biochemically defined pathways as a function of the growth rate (empirical values) in optimal distributions (left) and in MaxEnt distributions inferred from experiments (right). Continuous lines are a guide for the eye. In both panels, the experimental value of the acetate excretion is reported by grey markers (right vertical axis), while the vertical dashed line in the right panel separates the growth rates of the two experimental data sets. (Optimal distributions are not influenced by this aspect.) }
\end{figure}

\begin{sidewaysfigure}
	 \centering
 \includegraphics[width=0.95\textwidth]{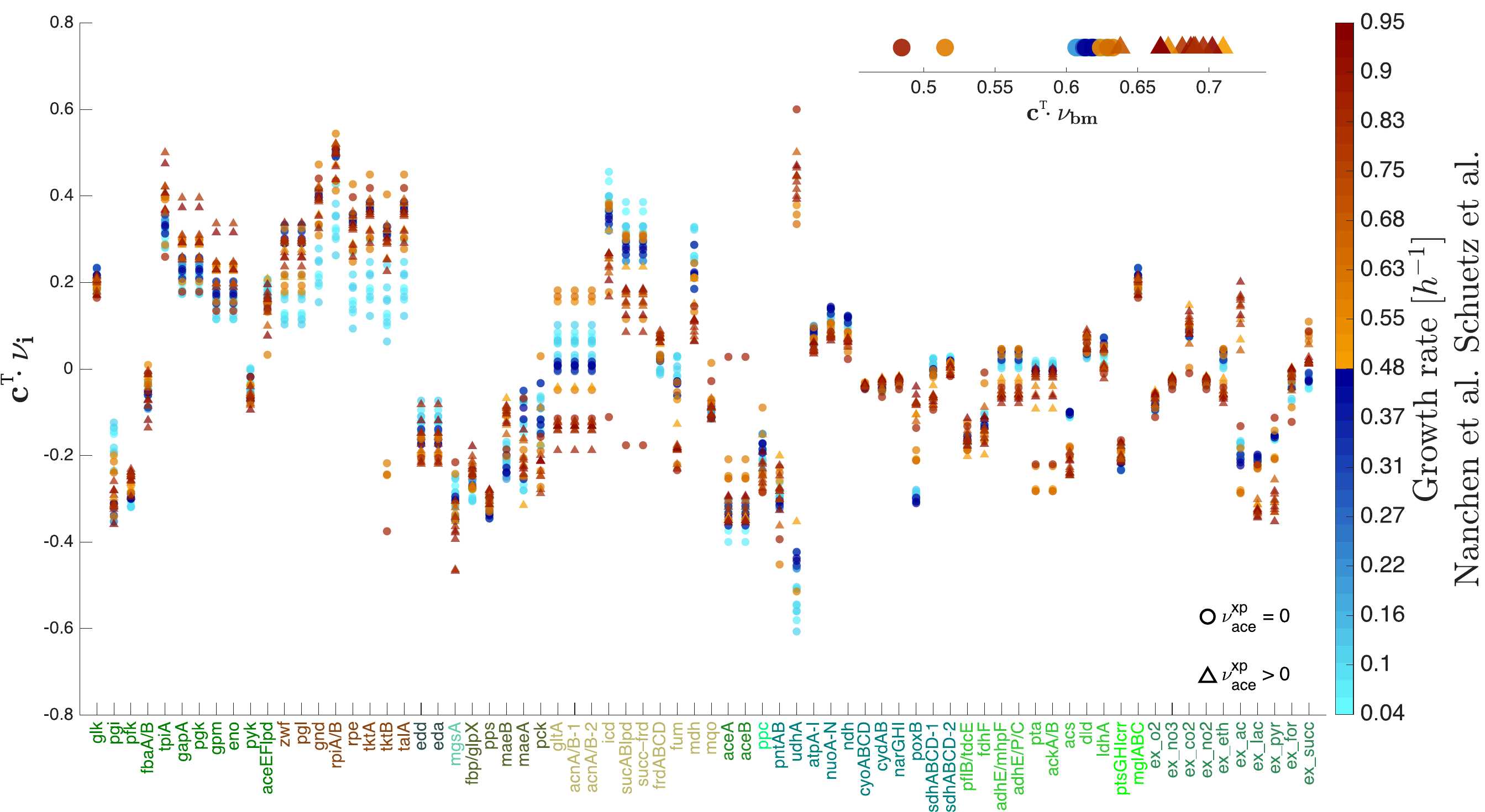}
 \caption{\textbf{Projections of the Lagrange multipliers.} In this plot we show the projections of the vector of coefficients $\mathbf{c}_{a}$ for experiment $a$ into the directions defined by individual fluxes. Positive (negative) values of the projection indicates that a flux is upregulated (downregulated) in experiment $a$ with respect to the reference value of the flux, i.e. its average value in the `flat` polytope $\mathcal{F}$. The color code mirrors both the glucose uptake of the experiments while the markers shape denote the presence or the absence) of the acetate excretion for circle and triangle points respectively.  \label{fig:ScProd}}
\end{sidewaysfigure}

\clearpage


\begin{thebibliography}{99}

\bibitem{lewis12}Lewis, N. E., Nagarajan, H., \& Palsson, B. O. (2012). Constraining the metabolic genotype-phenotype relationship using a phylogeny of in silico methods. Nature Reviews Microbiology, 10(4), 291-305.

\bibitem{feist10}Feist, A. M., \& Palsson, B. O. (2010). The biomass objective function. Current Opinion in Microbiology, 13(3), 344-349. 

\bibitem{dourado20}Dourado, H., \& Lercher, M. J. (2020). An analytical theory of balanced cellular growth. Nature Communications, 11(1), 1-14.

\bibitem{bruggeman20}Bruggeman, F. J., Planqu\'e, R., Molenaar, D., \& Teusink, B. (2020). Searching for principles of microbial physiology. FEMS Microbiology Reviews, 44(6), 821-844.

\bibitem{bordbar14}Bordbar, A., Monk, J. M., King, Z. A., \& Palsson, B. O. (2014). Constraint-based models predict metabolic and associated cellular functions. Nature Reviews Genetics, 15(2), 107-120.

\bibitem{scott10}Scott, M., Gunderson, C. W., Mateescu, E. M., Zhang, Z., \& Hwa, T. (2010). Interdependence of cell growth and gene expression: origins and consequences. Science, 330(6007), 1099-1102.

\bibitem{hui15}Hui, S., Silverman, J. M., Chen, S. S., Erickson, D. W., Basan, M., Wang, J., ... \& Williamson, J. R. (2015). Quantitative proteomic analysis reveals a simple strategy of global resource allocation in bacteria. Molecular Systems Biology, 11, 784.

\bibitem{flamholz13}Flamholz, A., Noor, E., Bar-Even, A., Liebermeister, W., \& Milo, R. (2013). Glycolytic strategy as a tradeoff between energy yield and protein cost. Proceedings of the National Academy of Sciences, 110(24), 10039-10044.

\bibitem{basan15}Basan, M., Hui, S., Okano, H., Zhang, Z., Shen, Y., Williamson, J. R., \& Hwa, T. (2015). Overflow metabolism in Escherichia coli results from efficient proteome allocation. Nature, 528(7580), 99-104.

\bibitem{basan20}Basan, M, et al. A universal trade-off between growth and lag in fluctuating environments. Nature 584.7821 (2020): 470-474.

\bibitem{utrilla16}Utrilla, J., O'Brien, E. J., Chen, K., McCloskey, D., Cheung, J., Wang, H., ... \& Palsson, B. O. (2016). Global rebalancing of cellular resources by pleiotropic point mutations illustrates a multi-scale mechanism of adaptive evolution. Cell systems, 2, 260-271.

\bibitem{mori17}Mori, M., Schink, S., Erickson, D. W., Gerland, U., \& Hwa, T. (2017). Quantifying the benefit of a proteome reserve in fluctuating environments. Nature Communications, 8, 1-8.

\bibitem{towbin17}Towbin, B. D., Korem, Y., Bren, A., Doron, S., Sorek, R., \& Alon, U. (2017). Optimality and sub-optimality in a bacterial growth law. Nature Communications, 8, 1-8.

\bibitem{erickson17}Erickson, D. W., Schink, S. J., Patsalo, V., Williamson, J. R., Gerland, U., \& Hwa, T. (2017). A global resource allocation strategy governs growth transition kinetics of Escherichia coli. Nature, 551(7678), 119-123.

\bibitem{schuetz12}Schuetz, R., Zamboni, N., Zampieri, M., Heinemann, M., \& Sauer, U. (2012). Multidimensional optimality of microbial metabolism. Science, 336(6081), 601-604.

\bibitem{shoval12}Shoval, O., Sheftel, H., Shinar, G., Hart, Y., Ramote, O., Mayo, A., ... \& Alon, U. (2012). Evolutionary trade-offs, Pareto optimality, and the geometry of phenotype space. Science, 336, 1157-1160.

\bibitem{mori19}Mori, M., Marinari, E., \& De Martino, A. (2019). A yield-cost tradeoff governs Escherichia coli's decision between fermentation and respiration in carbon-limited growth. NPJ Systems Biology and Applications, 5(1), 1-9.

\bibitem{kiviet14}Kiviet, D. J., Nghe, P., Walker, N., Boulineau, S., Sunderlikova, V., \& Tans, S. J. (2014). Stochasticity of metabolism and growth at the single-cell level. Nature, 514(7522), 376-379.

\bibitem{taheri15}Taheri-Araghi, S., Bradde, S., Sauls, J. T., Hill, N. S., Levin, P. A., Paulsson, J., ... \& Jun, S. (2015). Cell-size control and homeostasis in bacteria. Current Biology, 25(3), 385-391.

\bibitem{kennard16}Kennard, A. S., Osella, M., Javer, A., Grilli, J., Nghe, P., Tans, S. J., ... \& Lagomarsino, M. C. (2016). Individuality and universality in the growth-division laws of single E. coli cells. Physical Review E, 93, 012408.

\bibitem{demartino16}De Martino, D., Capuani, F., \& De Martino, A. (2016). Growth against entropy in bacterial metabolism: the phenotypic trade-off behind empirical growth rate distributions in E. coli. Physical Biology, 13(3), 036005.

\bibitem{demartino18}De Martino, D., Andersson, A. M., Bergmiller, T., Guet, C. C., \& Tkacik, G. (2018). Statistical mechanics for metabolic networks during steady state growth. Nature Communications, 9, 1-9.

\bibitem{obrien13}O'Brien, E. J., Lerman, J. A., Chang, R. L., Hyduke, D. R., \& Palsson, B. O. (2013). Genome-scale models of metabolism and gene expression extend and refine growth phenotype prediction. Molecular Systems Biology, 9, 693.

\bibitem{goelzer15}Goelzer, A., Muntel, J., Chubukov, V., Jules, M., Prestel, E., Noelker, R., ... \& Becher, D. (2015). Quantitative prediction of genome-wide resource allocation in bacteria. Metabolic Engineering, 32, 232-243.

\bibitem{mori16}Mori, M., Hwa, T., Martin, O. C., De Martino, A., \& Marinari, E. (2016). Constrained allocation flux balance analysis. PLoS Computational Biology, 12, e1004913.

\bibitem{reimers17}Reimers, A. M., Knoop, H., Bockmayr, A., \& Steuer, R. (2017). Cellular trade-offs and optimal resource allocation during cyanobacterial diurnal growth. Proceedings of the National Academy of Sciences, 114(31), E6457-E6465.

\bibitem{feist16}Feist, A.M. \& Palsson, B. O. (2016) What do cells actually want? Genome Biology, 17, 110.

\bibitem{dai17}Dai, Z., \& Locasale, J. W. (2017). Understanding metabolism with flux analysis: from theory to application. Metabolic Engineering, 43, 94-102.

\bibitem{demartino18h}De Martino, A., \& De Martino, D. (2018). An introduction to the maximum entropy approach and its application to inference problems in biology. Heliyon, 4(4), e00596.

\bibitem{mackay03}MacKay, D. J. (2003). Information theory, inference and learning algorithms. Cambridge University Press.

\bibitem{shrijver98}Schrijver, A. (1998). Theory of linear and integer programming. John Wiley \& Sons.

\bibitem{nanchen06}Nanchen, A., Schicker, A., \& Sauer, U. (2006). Nonlinear dependency of intracellular fluxes on growth rate in miniaturized continuous cultures of Escherichia coli. Applied  Environmental Microbiology, 72, 1164-1172.

\bibitem{wolfe05}Wolfe, A.J. (2005) The acetate switch. Microbiology and Molecular Biology Reviews, 69, 12-50.

\bibitem{buescher15}Buescher, J. M., Antoniewicz, M. R., Boros, L. G., Burgess, S. C., Brunengraber, H., Clish, C. B., ... \& Gottlieb, E. (2015). A roadmap for interpreting 13C metabolite labeling patterns from cells. Current Opinion in Biotechnology, 34, 189-201.

\bibitem{demartino13}De Martino, D., Capuani, F., Mori, M., De Martino, A., \& Marinari, E. (2013). Counting and correcting thermodynamically infeasible flux cycles in genome-scale metabolic networks. Metabolites, 3(4), 946-966.

\bibitem{gudmundsson10}Gudmundsson, S., \& Thiele, I. (2010). Computationally efficient flux variability analysis. BMC Bioinformatics, 11(1), 489.

\bibitem{opper00}Opper, M., \& Winther, O. (2000). Gaussian processes for classification: Mean-field algorithms. Neural computation, 12(11), 2655-2684.

\bibitem{minka01}Minka, T. P. Expectation propagation for approximate Bayesian inference. In Proceedings of the Seventeenth conference on Uncertainty in artificial intelligence, pp. 362-369 (Morgan Kaufmann Publishers Inc., 2001)

\bibitem{braunstein17}Braunstein, A., Muntoni, A. P., \& Pagnani, A. (2017). An analytic approximation of the feasible space of metabolic networks. Nature Communications, 8, 1-9.

\bibitem{ell} De Martino, D., Mori, M., \& Parisi, V. (2015). Uniform sampling of steady states in metabolic networks: heterogeneous scales and rounding. PloS one, 10(4), e0122670.

\bibitem{Kaneko} Furusawa, C., \& Kaneko, K. (2018). Formation of dominant mode by evolution in biological systems. Physical Review E, 97(4), 042410.


\bibitem{xavier14}Xavier, J. C., Patil, K. R., \& Rocha, I. (2014). Systems biology perspectives on minimal and simpler cells. Microbiology and Molecular Biology Reviews, 78, 487-509.

\bibitem{gerosa11}Gerosa, L., \& Sauer, U. (2011). Regulation and control of metabolic fluxes in microbes. Current Opinion in Biotechnology, 22(4), 566-575.

\bibitem{posfai06}Posfai, G., Plunkett, G., Feher, T., Frisch, D., Keil, G. M., Umenhoffer, K., ... \& Burland, V. (2006). Emergent properties of reduced-genome Escherichia coli. Science, 312, 1044-1046.

\bibitem{minton11}Minton, A. P., \& Rivas, G. (2011). Biochemical reactions in the crowded and confined physiological environment: physical chemistry meets synthetic biology. In The Minimal Cell (pp. 73-89). Springer, Dordrecht.

\bibitem{carlson04}Carlson, R., \& Srienc, F. (2004). Fundamental Escherichia coli biochemical pathways for biomass and energy production: identification of reactions. Biotechnology and Bioengineering, 85(1), 1-19.

\bibitem{trinh06}Trinh, C. T., Carlson, R., Wlaschin, A., \& Srienc, F. (2006). Design, construction and performance of the most efficient biomass producing E. coli bacterium. Metabolic Engineering, 8(6), 628-638.

\bibitem{trinh08}Trinh, C. T., Unrean, P., \& Srienc, F. (2008). Minimal Escherichia coli cell for the most efficient production of ethanol from hexoses and pentoses. Appl. Environ. Microbiol., 74(12), 3634-3643.

\bibitem{bialek12}Bialek, W. (2012). Biophysics: searching for principles. Princeton University Press.

\bibitem{burgard03}Burgard, A. P., \& Maranas, C. D. (2003). Optimization-based framework for inferring and testing hypothesized metabolic objective functions. Biotechnology and Bioengineering, 82(6), 670-677.

\bibitem{gianchandani08}Gianchandani, E. P., Oberhardt, M. A., Burgard, A. P., Maranas, C. D., \& Papin, J. A. (2008). Predicting biological system objectives de novo from internal state measurements. BMC Bioinformatics, 9(1), 43.

\bibitem{chiu08}Chiu, H. C., \& Segr\`e, D. (2008). Comparative determination of biomass composition in differentially active metabolic states. In Genome Informatics 2008: Genome Informatics Series Vol. 20 (pp. 171-182).

\bibitem{zhao16}Zhao, Q., Stettner, A. I., Reznik, E., Paschalidis, I. C., \& Segre, D. (2016). Mapping the landscape of metabolic goals of a cell. Genome Biology, 17(1), 109.

\bibitem{yang19}Yang, L., Saunders, M. A., Lachance, J. C., Palsson, B. O., \& Bento, J. (2019, July). Estimating Cellular Goals from High-Dimensional Biological Data. In Proceedings of the 25th ACM SIGKDD International Conference on Knowledge Discovery \& Data Mining (pp. 2202-2211).

\bibitem{knorr06}Knorr, A. L., Jain, R., \& Srivastava, R. (2007). Bayesian-based selection of metabolic objective functions. Bioinformatics, 23(3), 351-357.

\bibitem{miotto19}De Martino, A., Gueudr\'e, T. \& Miotto, M. (2019)  Exploration-exploitation tradeoffs dictate the optimal distributions of phenotypes for populations subject to fitness fluctuations. Physical Review E, 99, 012417.

\bibitem{zhao09}Zhao, Q., \& Kurata, H. (2009). Maximum entropy decomposition of flux distribution at steady state to elementary modes. Journal of Bioscience and Bioengineering, 107(1), 84-89.

\bibitem{zhao10}Zhao, Q., \& Kurata, H. (2010). Use of maximum entropy principle with Lagrange multipliers extends the feasibility of elementary mode analysis. Journal of Bioscience and Bioengineering, 110(2), 254-261.

\bibitem{demartino17}De Martino, D., Capuani, F., \& De Martino, A. (2017). Quantifying the entropic cost of cellular growth control. Physical Review E, 96(1), 010401.

\bibitem{cossio19}Fernandez-de-Cossio-Diaz, J. \& Mulet, R. (2019) Maximum entropy and population heterogeneity in continuous cell cultures. PLOS Computational Biology 15(2), e1006823.

\bibitem{pereiro21} Pereiro-Morej\'on, J. A., Fern\'andez-de-Cossio-D\'\i az, J., \& Mulet, R. (2021). Inferring metabolic fluxes in nutrient-limited continuous cultures: A Maximum Entropy Approach with minimum information. arXiv preprint arXiv:2109.13149.

\bibitem{tourigny20}Tourigny, D. (2020) Dynamic metabolic resource allocation based on the maximum entropy principle. Journal of Mathematical Biology, 80, 2395–2430.

\bibitem{masoero} De Martino, D., \& Masoero, D. (2016). Asymptotic analysis of noisy fitness maximization, applied to metabolism \& growth. Journal of Statistical Mechanics: Theory and Experiment, 2016(12), 123502.

~

{\bf Unique Supporting References:}

~

\bibitem{braunstein20}Braunstein, A., Muntoni, A. P., Pagnani, A., \& Pieropan, M. (2020). Compressed sensing reconstruction using expectation propagation. Journal of Physics A: Mathematical and Theoretical, 53(18), 184001.

\bibitem{saldida20}Saldida, J., Muntoni, A. P., De Martino, D., Hubmann, G., Niebel, B., Schmidt, A. M., ... \& Heinemann, M. (2020). Unbiased metabolic flux inference through combined thermodynamic and 13C flux analysis. bioRxiv - doi.org/10.1101/2020.06.29.177063

\bibitem{bernstein09}Bernstein, D. S. (2009). Matrix mathematics: theory, facts, and formulas. Princeton university press.

\bibitem{orth10}Orth, Jeffrey D. and Fleming, R. M. T. and Palsson, Bernhard Ø. (2010). Reconstruction and Use of Microbial Metabolic Networks: the Core \textit{Escherichia coli} Metabolic Model as an Educational Guide. EcoSal Plus 10.2.1.

\end{thebibliography}

\end{widetext}

\end{document}